\documentclass[aps,prd,amsmath,superscriptaddress,%
               nobibnotes,nofootinbib,preprintnumbers,showpacs,%
               onecolumn,12pt,tightenlines]{revtex4}

\usepackage{graphicx}

\usepackage{ifthen}

\usepackage{afterpage}

\newcommand{\ie}{\textit{i.e.}}
\newcommand{\eg}{\textit{e.g.}}

\newcommand{\rmd}{\ensuremath{\mathrm{d}}}

\newcommand{\erf}{\mathop{\mathrm{erf}}}

\newcommand{\gae}{%
  \ensuremath{\lower 2pt \hbox{%
    $\, \buildrel {\scriptstyle >}\over {\scriptstyle \sim}\,$}%
    }%
  }
\newcommand{\lae}{%
  \ensuremath{\lower 2pt \hbox{%
    $\, \buildrel {\scriptstyle <}\over {\scriptstyle \sim}\,$}%
    }%
  }

\newcommand{\refeqn}[2][eqn:]{Eqn.~(\ref{#1#2})}

\newcommand{\reftab}[2][tab:]{Table~\ref{#1#2}}

\newcommand{\reffig}[2][fig:]{Figure~\ref{#1#2}}
\newcommand{\Reffig}[2][fig:]{Figure~\ref{#1#2}}
\newcommand{\refsec}[2][sec:]{Section~\ref{#1#2}} 

\newcommand{\ifmulticol}[2]{%
  \ifthenelse{\lengthtest{1.9\columnwidth<\textwidth}}{#1}{#2}%
}

\newcommand{\insertfig}[1]{%
    \includegraphics[keepaspectratio,width=1.00\columnwidth,
                     height=0.45\textheight]{#1}
}

\newcommand{\dRdE}{\ensuremath{\frac{\rmd\! R}{\rmd\! E}}}

\newcommand{\vmin}{\ensuremath{v_\mathrm{min}}}
\newcommand{\vmp}{\ensuremath{\overline{v}_0}}
\newcommand{\vobs}{\ensuremath{v_\mathrm{obs}}}
\newcommand{\bvobs}{\ensuremath{\mathbf{v}_\mathrm{obs}}}
\newcommand{\vesc}{\ensuremath{v_\mathrm{esc}}}
\newcommand{\Nesc}{\ensuremath{N_\mathrm{esc}}}
\newcommand{\bu}{\ensuremath{\mathbf{u}}}  
\newcommand{\bv}{\ensuremath{\mathbf{v}}}  
\newcommand{\bV}{\ensuremath{\mathbf{V}}}  

\newcommand{\fpSI}{\ensuremath{f_{\mathrm{p}}}}
\newcommand{\fnSI}{\ensuremath{f_{\mathrm{n}}}}
\newcommand{\apSD}{\ensuremath{a_{\mathrm{p}}}}
\newcommand{\anSD}{\ensuremath{a_{\mathrm{n}}}}
\newcommand{\sigmaSI}{\ensuremath{\sigma_{\mathrm{SI}}}}
\newcommand{\sigmaSD}{\ensuremath{\sigma_{\mathrm{SD}}}}

\newcommand{\sigmapSI}{\ensuremath{\sigma_{\rm p,SI}}}

\newcommand{\sigmapSD}{\ensuremath{\sigma_{\rm p,SD}}}
\newcommand{\sigmanSD}{\ensuremath{\sigma_{\rm n,SD}}}
\newcommand{\mhat}{\ensuremath{\hat{m}}}
\newcommand{\apSDhat}{\ensuremath{\hat{a}_{\mathrm{p}}}}
\newcommand{\anSDhat}{\ensuremath{\hat{a}_{\mathrm{n}}}}
\newcommand{\sigmahat}{\ensuremath{\hat{\sigma}}}
\newcommand{\sigmapSIhat}{\ensuremath{\hat{\sigma}_{\rm p,SI}}}
\newcommand{\rhoDM}{\ensuremath{\rho_{0}}}
\newcommand{\eone}{\ensuremath{\hat{\boldsymbol{\varepsilon}}_1}}  
\newcommand{\etwo}{\ensuremath{\hat{\boldsymbol{\varepsilon}}_2}}  
\newcommand{\peone}{\ensuremath{\hat{\mathbf{e}}_1}}  
\newcommand{\petwo}{\ensuremath{\hat{\mathbf{e}}_2}}  
\newcommand{\alphabin}{\ensuremath{\alpha_{\mathrm{bin}}}}
\newcommand{\Nbin}{\ensuremath{N_{\mathrm{bin}}}}
\newcommand{\chisqmin}{\ensuremath{\chi_{\mathrm{min}}^2}}

\begin{document}


\preprint{FTPI--MINN--08/34}
\preprint{MCTP-08-59}


\title{Compatibility of DAMA/LIBRA dark matter detection with
       other searches}

\author{Christopher Savage}
\email[]{cmsavage@physics.umn.edu}
\affiliation{
 William I.\ Fine Theoretical Physics Institute,
 School of Physics and Astronomy,
 University of Minnesota,
 Minneapolis, MN 55455, USA}

\author{Graciela Gelmini}
\email[]{gelmini@physics.ucla.edu}
\affiliation{
 Department of Physics and Astronomy,
 UCLA,
 430 Portola Plaza,
 Los Angeles, CA 90095, USA}

\author{Paolo Gondolo}
\email[]{paolo@physics.utah.edu}
\affiliation{
 Department of Physics,
 University of Utah,
 115 S 1400 E \#201
 Salt Lake City, UT 84112, USA}

\author{Katherine Freese}
\email[]{ktfreese@umich.edu}
\affiliation{
 Michigan Center for Theoretical Physics,
 Department of Physics,
 University of Michigan,
 Ann Arbor, MI 48109, USA}

\date{\today}



\begin{abstract} 

The DAMA/NaI and DAMA/LIBRA annual modulation data, which may be
interpreted as a signal for the existence of weakly interacting dark
matter (WIMPs) in our galactic halo, are examined in light of null
results from other experiments: CDMS, XENON10, CRESST~I, CoGeNT,
TEXONO, and Super-Kamiokande (SuperK). We use the energy spectrum of the
combined DAMA modulation data given in 36 bins, and include the effect
of channeling.  Several statistical tools are implemented in our study:
likelihood ratio with a global fit and with raster scans in the WIMP
mass and goodness-of-fit (g.o.f.).  These approaches allow us to
differentiate between the preferred (global best fit) and allowed (g.o.f.)
parameter regions.
It is hard to find WIMP masses and couplings consistent with all
existing data sets; the surviving regions of parameter space are found
here.  For spin-independent (SI) interactions, the  best fit DAMA
regions are ruled out to the 3$\sigma$ C.L., even with channeling taken
into account.  However, for WIMP masses of $\sim$8~GeV 
some parameters outside these regions still yield a moderately
reasonable fit to
the DAMA data and are compatible with all 90\% C.L.\ upper limits from
negative searches, when channeling is included.  For spin-dependent
(SD) interactions with proton-only couplings, a range of masses
below 10~GeV is compatible with DAMA and other experiments, with and
without channeling, when SuperK indirect detection constraints are
included; without the SuperK constraints,
masses as high as $\sim$20~GeV are compatible.
For SD neutron-only couplings we find no parameters compatible with all
the experiments.
Mixed SD couplings are examined: \eg\ $\sim$8~GeV mass
WIMPs with $\anSD = \pm \apSD$ are found to be consistent with all
experiments.  In short, there are surviving regions at low mass for
both SI and SD interactions; if indirect detection limits are relaxed,
some SD proton-only couplings at higher masses also survive.

\end{abstract} 

\maketitle


\section{\label{sec:intro} Introduction}

Among the best motivated candidates for dark matter are Weakly
Interacting Massive Particles (WIMPs).  Direct searches for dark
matter WIMPs aim at detecting the scattering of WIMPs off of nuclei in
a low-background detector. These experiments measure the energy of the
recoiling nucleus, and are sensitive to a signal above a
detector-dependent energy threshold \cite{Jungman:1995df}.

The discovery of an annual modulation by the DAMA/NaI experiment
\cite{Bernabei:2003za}, confirmed by the new experiment
DAMA/LIBRA~\cite{Bernabei:2008yi} of the same collaboration, is the
only positive signal seen in any dark matter search.  The DAMA
collaboration has found an annual modulation in its data compatible
with the signal expected from dark matter particles bound to our
Galactic Halo \cite{Drukier:1986tm,Freese:1987wu}.  However, other
direct detection experiments, \eg\ 
CDMS~\cite{Akerib:2003px,Ahmed:2008eu},
CoGeNT~\cite{Aalseth:2008rx,Collar:2008pc},
COUPP~\cite{Behnke:2008zz,Collar:2008pc2},
CRESST~\cite{Angloher:2002in,Stodolsky:2004dsu},
KIMS~\cite{Lee.:2007qn},
TEXONO~\cite{Lin:2007ka,Wong:2008pc}, and
XENON10~\cite{Angle:2007uj,Angle:2008we}
have not found any signal from WIMPs. 
It has been difficult to reconcile a WIMP signal in DAMA with the
other negative results~\cite{previous}, for the case of canonical WIMP
masses $\sim 100$GeV with standard weak interactions motivated by
Supersymmetry (SUSY).  Yet, for light WIMPs, such compatibility was
possible.  Papers written prior to the latest DAMA/LIBRA results found
regions in cross-section and WIMP mass that reconciled all null
results with DAMA's positive signal, even assuming a standard halo
model: WIMPs with spin independent
interactions~\cite{Gelmini:2004gm,Gondolo:2005hh}
in the mass range 5--9~GeV, and WIMPs with spin dependent
interactions~\cite{Savage:2004fn} in the mass range 5--13~GeV were
found to be compatible with all existing data.
In addition, Ref.~\cite{Gelmini:2004gm,Gondolo:2005hh} studied not only
the case of
the standard halo model, but also models with dark matter streams;
they found that nonstandard velocity distributions could reconcile all
the existing data sets including DAMA.  Other dark matter explanations
invoke inelastic scattering of WIMPs \cite{TuckerSmith:2001hy}.

Since these studies of a few years ago, DAMA/LIBRA has reported new
experimental results.  First, of particular interest is the new
presentation of the modulation data in 36 separate energy bins, which
allows much more detailed investigations of the data.  Second,
in the past year another possible complication has come to light. The
DAMA collaboration has pointed out that ``channeling'' may affect the
interpretation of their data.  In general only a fraction (known as
the quenching factor) of the recoil energy deposited by a WIMP
is transferred to electrons and is converted into useful signal in
the DAMA detector (\eg\ ionization or scintillation); the remainder is
converted into phonons and heat and goes undetected.  Hence measured
energies must be corrected for this behavior to obtain the proper recoil
energy, by dividing by this quenching factor.  The DAMA detector is
composed of NaI, with quenching factors $Q_{Na} = 0.3$ and $Q_I = 0.09$.
Yet, recently, the DAMA collaboration pointed out that some fraction of
the nuclear recoils travel in straight paths along the crystal in such
a way as to lose very little energy to other atoms and to heat,
giving nearly all their energy to electrons; \ie, for these cases the
measured energy corresponds very nearly to the energy deposited by the
WIMP and thus have $Q \approx 1$.  As a consequence, the detector is
sensitive to lower mass WIMPs than previously thought.  The DAMA
threshold is 2~keVee (electron
equivalent). Using the above quenching factors this was thought to
correspond to 7~keV recoil energy for sodium and 22~keV recoil energy
for iodine; yet, due to channeling the actual threshold for some of
the events can be as low as 2~keV.  The new channeling studies revived
the possiblity of DAMA's compatibility with other data sets,
particularly at low masses.

Light neutralinos as WIMPs with masses as low as 2~GeV
\cite{Gabutti:1996qd} or 6~GeV \cite{Bottino:2002ry, Bottino:2008mf}
have been considered, even with the cross sections we find
necessary for WIMPs to be compatible with the DAMA signal and all
other negative searches (for spin independent
interactions)~\cite{Bottino:2008mf}.  In any event, in this paper we
proceed in a purely phenomenological way in choosing the WIMP mass and
cross section, with either spin-independent or spin-dependent cross
sections.  We do not attempt to provide an elementary particle model
to support the values of masses and cross sections.  As justification
of our approach, let us recall that there is no proven particle theory
of dark matter.  The candidates we are considering are stable neutral
particles which have very small cross sections with nucleons.
 Regarding their production in accelerators, they
would escape from the detectors without interacting.  Unless there is
a concrete specific model relating our neutral candidate to other
charged particles (which can, if fact, be observed) there is no way such
particles could be found in accelerators. The usual signature searched
for in accelerators, for example at LEP, Tevatron or LHC, is the
emission of a charged particle related to the neutral particle in
question.  For example, searching for ``neutralinos'' one puts bounds
on one of its cousins, a ``chargino'', or another relative, a
``slepton''. Without a detailed model there are no accelerator bounds 
on neutral dark matter candidates.

An alternative way to search for WIMP dark matter of relevance to this
paper is via indirect detection of WIMP annihilation in the Sun.  When
WIMPs pass through a large celestial body, such as the Sun or the
Earth \cite{indirectdet:solar,indirectdet:earth}, interactions can
lead to gravitational capture if enough energy is lost in the
collision to fall below the escape velocity.  Captured WIMPs soon fall
to the body's core and eventually annihilate with other WIMPs.  These
annihilations lead to high-energy neutrinos that can be observed by
Earth-based detectors such as Super-Kamiokande \cite{Desai:2004pq}
(SuperK), which currently provides the tightest indirect detection
bounds at light WIMP masses,
AMANDA \cite{Ahrens:2002eb,Ackermann:2005fr},
IceCube \cite{Ahrens:2002dv} and ANTARES \cite{Blanc:2003na}.
The annihilation rate depends on the capture rate of WIMPs, which is
in turn determined by the WIMP scattering cross section off nuclei in
the celestial body.  While the Earth is predominantly composed of
spinless nuclei, the Sun is mostly made of hydrogen, which has
spin. Thus the spin-dependent cross section of WIMPs off nucleons can
be probed by measuring the annihilation signals from WIMP annihilation
in the Sun.  Other indirect detection methods search for WIMPs that
annihilate in the Galactic Halo or near the Galactic Center where they
produce neutrinos, positrons, or antiprotons that may be seen in
detectors on the Earth
\cite{indirectdet:galactichalo,indirectdet:galacticcenter}. 
Of course, these
rely upon additional physics (WIMP annhilation) beyond what direct
detection experiments rely on (WIMP scattering).  Thermal WIMPs can
only achieve the correct relic density today if their annihilation
cross sections are fixed to be the value we use when computing the
SuperK limits; yet one can imagine
alternatives in which case the SuperK bounds should not be used.

In this paper, we reconsider the issue of compatibility of the new
DAMA/NaI and DAMA/LIBRA (hereafter DAMA) results with all other
experimental results.  We use the
36 bins of modulation data and take into account the channeling
effect.  We restrict our studies to the standard isothermal model for
the dark matter halo.  We carefully investigate both spin-independent
as well as spin-dependent couplings.  Our procedure is the
following. Assuming the standard dark halo model, we find the viable
region by making sure that we produce the correct amplitudes for the
DAMA modulation.  We then compare this region with the most stringent
current experimental constraints from negative searches.  We use three
different statistical methods to find the DAMA region:
the likelihood method with (i) a global fit and (ii) raster scans in
the WIMP mass, and (iii) goodness-of-fit.  These methods and the
different information derived from them are described in detail.

\section{\label{sec:Detection} Dark Matter Detection}

WIMP direct detection experiments seek to measure the energy deposited
when a WIMP interacts with a nucleus in the detector \cite{Goodman:1984dc}.
If a WIMP of mass $m$ scatters elastically from a nucleus of mass $M$,
it will deposit a recoil energy $E = (\mu^2v^2/M)(1-\cos\theta)$,
where $\mu \equiv m M/ (m + M)$ is the reduced mass, $v$ is the speed
of the WIMP relative to the nucleus, and $\theta$ is the scattering
angle in the center of mass frame.  The differential recoil rate per
unit detector mass for a WIMP mass $m$, typically given in units of
cpd/kg/keV (where cpd is counts per day), can be written as:
\begin{equation} \label{eqn:dRdE}
  \dRdE = \frac{\sigma(q)}{2 m \mu^2}\, \rho\, \eta(E,t)
\end{equation}
where $q = \sqrt{2 M E}$ is the nucleus recoil momentum, $\sigma(q)$ is
the WIMP-nucleus cross-section, $\rho$ is the local WIMP density,
and information about the WIMP velocity distribution is encoded into the
mean inverse speed $\eta(E,t)$,
\begin{equation} \label{eqn:eta}  
  \eta(E,t) = \int_{u > \vmin} \frac{f(\bu,t)}{u} \, \rmd^3u \, .
\end{equation}
Here 
\begin{equation} \label{eqn:vmin}
  \vmin = \sqrt{\frac{M E}{2\mu^2}}
\end{equation}
represents the minimum WIMP velocity that can result in a recoil energy
$E$ and $f({\bf u},t)$ is the (time-dependent) distribution of WIMP
velocities ${\bf u}$ relative to the detector.

To determine the number of expected recoils for a given experiment
and WIMP mass, we integrate Eqn.~(\ref{eqn:dRdE}) over the nucleus
recoil energy to find the recoil rate $R$ per unit detector mass:
\begin{equation} \label{eqn:rateone}
  R(t) = \int_{E_{1}/Q}^{E_{2}/Q}\rmd E \,
         \epsilon(QE) \frac{\rho}{2 m \mu^2} \, \sigma(q) \, \eta(E,t) .
\end{equation}
$\epsilon(QE)$ is the (energy dependent) efficiency of the experiment,
due, \eg, to data cuts designed to reduce backgrounds.  $Q$ is the
quenching factor relating the observed energy $E_{det}$ (in some cases
referred to as the electron-equivalent energy) with the actual recoil
energy $E_{rec}$: $E_{det} = Q E_{rec}$ (explained in more detail
below).  The energy range between $E_{1}$ and $E_{2}$ is that of
\textit{observed} energies for some data bin of the detector (where
experiments often bin observed recoils by energy).  The quenching
factor $Q$ depends on the nuclear target and the characteristics of
the detector.

The recoiling nucleus will transfer its energy to either electrons,
which may be observed as \eg\ ionization or scintillation in the
detector, or to other nuclei, producing phonons and heat.  Experiments
that do not measure the phonons/heat can only directly measure the
fraction of energy $Q$ that goes into the channel that is observed
(such as scintillation); we refer to this as the observed energy.
Observed energies for experiments that measure only from the
electron channel will be quoted in electron-equivalent energies
(keVee).  Note some detectors can calibrate their
energy scales to $E_{det} = E_{rec}$, in which case the quenching
factor can be ignored in the above formulations (such
calibrations may directly or indirectly involve a quenching factor).

Real experimental apparatus cannot determine the event energies with
perfect precision.
If the expected amount of energy in the channel an experiment observes
due to a nuclear recoil is $E^\prime$,
the \textit{measured} energy will be distributed about $E^\prime$.
The previous formulas apply only for a perfect energy
resolution and must be corrected for this finite resolution.
The observed rate over a measured energy range $E_1$--$E_2$,
\refeqn{rateone}, should be replaced with
\begin{equation} \label{eqn:rateER}
  R(t) = \int_{0}^{\infty}\rmd E \,
         \epsilon(QE) \, \Phi(QE,E_1,E_2) \, \frac{\rho}{2 m \mu^2}
         \, \sigma(q) \, \eta(E,t) .
\end{equation}
Here, $\Phi(E^\prime=QE,E_1,E_2)$ is a response function corresponding
to the fraction of events with an expected observed energy $E^\prime = QE$
(\ie\ the energy that would be observed with perfect energy resolution)
that will actually be measured between $E_1$ and $E_2$; recall $E$ in
the above equation is the recoil energy, of which only a fraction $Q$
is converted into a form that is potentially observable.
If the measured energies are normally distributed about $E^\prime$
with a (typically energy dependent) standard deviation
$\sigma(E^\prime)$,
\begin{equation} \label{eqn:normalRF}
  \Phi(E^\prime,E_1,E_2)
    = \frac{1}{2} \left[ 
        \erf\left( \frac{E_2 - E^\prime}{\sqrt{2} \sigma(E^\prime)} \right)
        - \erf\left( \frac{E_1 - E^\prime}{\sqrt{2} \sigma(E^\prime)} \right)
      \right] .
\end{equation}
In the limit $\sigma(E^\prime) \to 0$ (perfect energy resolution),
\refeqn{rateER} reduces to \refeqn{rateone}.

For detectors with multiple elements and/or isotopes, the total rate is
given by:
\begin{equation} \label{eqn:ratetot}
  R_{tot}(t) = \sum_i f_i R_i(t)
\end{equation}
where $f_i$ is the mass fraction and $R_i$ is the rate
(Eqn.~(\ref{eqn:rateone})) for element/isotope $i$.

The expected number of recoils observed by a detector is given by:
\begin{equation} \label{eqn:recoils}
  N_{rec} = M_{det} T R
\end{equation}
where $M_{det}$ is the detector mass and $T$ is the exposure time.

\subsection{\label{sec:SICS} Cross-Section}

The $\sigma(q)$ cross-section term in
Eqns.~(\ref{eqn:dRdE}) \&~(\ref{eqn:rateone}) is an effective
cross-section for scatters with a momentum exchange $q$.
The momentum exchange dependence appears in form factors that arise
from the finite size of the nucleus.
The total scattering cross-section generally has contributions from
spin-independent (SI) and spin-dependent (SD) couplings, with
\begin{equation} \label{eqn:CStot}
  \sigma = \sigmaSI + \sigmaSD ;
\end{equation}
these two cross-sections are described below.


\textit{Spin-independent (SI).}
For spin-independent WIMP interactions, we make the usual
assumption~\cite{Jungman:1995df} that the cross section $\sigma$ scales
with the square of the nucleus atomic number $A$ and is given by
\begin{equation} \label{eqn:SICS}
  \sigma = \sigma_{0} \, | F(E) |^2
\end{equation}
where $\sigma_{0}$ is the zero-momentum WIMP-nuclear cross-section and
$F(E)$ is a nuclear form factor,
normalized to $F(0) = 1$.  For purely scalar interactions,
\begin{equation} \label{eqn:scalar}
  \sigma_{0,\rm SI} = \frac{4 \mu^2}{\pi} [ Z \fpSI + (A-Z) \fnSI ]^2 \, .
\end{equation}
Here $Z$ is the number of protons, $A-Z$ is the number of neutrons,
and $\fpSI$ and $\fnSI$ are the WIMP couplings to the proton and nucleon,
respectively.  In most instances, $\fnSI \sim \fpSI$; the WIMP-nucleus
cross-section can then be given in terms of the WIMP-proton
cross-section as a result of Eqn.~(\ref{eqn:scalar}):
\begin{equation} \label{eqn:SICSproton}
  \sigma_{0,\rm SI} = \sigmapSI \left( \frac{\mu}{\mu_{\rm p}} \right)^2 A^2
\end{equation}
where the $\mu_{\rm p}$ is the proton-WIMP reduced mass, and $A$ is
the atomic mass of the target nucleus.
In this model, for a given WIMP mass, $\sigmapSI$ is the only free
parameter.

For the nuclear form factor we use the conventional Helmi
form~\cite{Jungman:1995df,SmithLewin},
\begin{equation} \label{eqn:SIFF}
  F(E) = 3 e^{-q^2 s^2/2} \, \frac{\sin(qr)- qr\cos(qr)}{(qr)^3} ,
\end{equation}
with $s=0.9$~fm, $r$ is an effective nuclear radius as described in
Ref.~\cite{SmithLewin},
and $q=\sqrt{2 M E}$.


\textit{Spin-dependent (SD).}
The generic form for the spin-dependent WIMP-nucleus cross-section
includes two couplings \cite{Engel:1991wq}, the WIMP-proton coupling
$a_p$ and the WIMP-neutron coupling $a_n$,
\begin{eqnarray} \label{eqn:SDCS}
  \sigma_{SD}(q) =
    \frac{32 \mu^2 G_F^2}{2 J + 1}
        \left[a_p^2 S_{pp}(q) + a_p a_n S_{pn}(q) 
      + a_n^2 S_{nn}(q) \right] .
\end{eqnarray}
Here, the quantities $a_p$ and $a_n$ are actually the axial
four-fermion WIMP-nucleon couplings in units of $2\sqrt{2} G_F$
\cite{Gondolo:1996qw,Tovey:2000mm,Gondolo:2004sc}.
The nuclear structure functions $S_{pp}(q)$, $S_{nn}(q)$, and
$S_{pn}(q)$ are functions of the exchange momentum $q$ and are
specific to each nucleus.  We take the structure functions for
Aluminum from Ref.~\cite{Engel:1995gw}; for Sodium, Iodine, and
Xenon from Ref.~\cite{Ressell:1997kx}; and for Silicon and Germanium
from Ref.~\cite{Ressell:1993qm}.  These and additional structure
functions may be found in the review of Ref.~\cite{Bednyakov:2006ux}.

\subsection{\label{sec:VelocityDist} Velocity Distribution}

The dark matter halo is likely to have a smooth, well mixed (virialized)
component that will lead to detectable recoils in direct detection
experiments; the simplest model of this component is the Standard
Halo Model (SHM) \cite{Freese:1987wu}.
In addition to this smooth component, the galaxy contains structure
from the galaxy formation process that has not yet virialized.  This
structure includes, \eg, tidal streams of dwarf galaxies in the process
of being absorbed by the Milky Way, such as the Sagittarius dwarf
galaxy \cite{Yanny:2003zu,Newberg:2003cu,Majewski:2003ux,Freese:2003na,
Freese:2003tt}.  Any such structure located about the Solar System
will also produce scattering events.

For each of these components, the smooth halo background and any
structure, we will use a Maxwellian distribution, characterized
by an rms velocity dispersion $\sigma_v$, to describe the WIMP speeds,
and we will allow for the distribution to be truncated at some escape
velocity $\vesc$,
\begin{equation} \label{eqn:Maxwellian}
  \widetilde{f}(\bv) =
    \begin{cases}
      \frac{1}{\Nesc} \left( \frac{3}{2 \pi \sigma_v^2} \right)^{3/2}
        \, e^{-3\bv^2\!/2\sigma_v^2} , 
        & \textrm{for} \,\, |\bv| < \vesc  \\
      0 , & \textrm{otherwise}.
    \end{cases}
\end{equation}
Here
\begin{equation} \label{eqn:Nesc}
  \Nesc = \erf(z) - 2 z \exp(-z^2) / \pi^{1/2} ,   
\end{equation}
with $z \equiv \vesc/\vmp$, is a normalization factor.  The most
probable speed,
\begin{equation} \label{eqn:vmp}
  \vmp = \sqrt{2/3} \, \sigma_v ,
\end{equation}
is used to generate unitless parameters such as $z$.
For distributions without an escape velocity ($\vesc \to \infty$),
$\Nesc = 1$.

The WIMP component (halo or stream) often exhibits a bulk motion
relative to us, so that
\begin{equation} \label{eqn:vdist}
  f(\bu) = \widetilde{f}(\bvobs + \bu) \, ,
\end{equation}
where $\bvobs$ is the motion of the observer relative to the rest frame
of the WIMP component described by \refeqn{Maxwellian}; this motion will
be discussed below.  For such a velocity
distribution, the mean inverse speed, \refeqn{eta}, becomes
\begin{equation} \label{eqn:eta2}
  \eta(E,t) =
    \begin{cases}
      \frac{1}{\vmp y} \, ,
        & \textrm{for} \,\, z<y, \, x<|y\!-\!z| \\
      \frac{1}{2 \Nesc \vmp y}
        \left[
          \erf(x\!+\!y) - \erf(x\!-\!y) - \frac{4}{\sqrt{\pi}} y e^{-z^2}
        \right] \, ,
        & \textrm{for} \,\, z>y, \, x<|y\!-\!z| \\
      \frac{1}{2 \Nesc \vmp y}
        \left[
          \erf(z) - \erf(x\!-\!y) - \frac{2}{\sqrt{\pi}} (y\!+\!z\!-\!x) e^{-z^2}
        \right] \, ,
        & \textrm{for} \,\, |y\!-\!z|<x<y\!+\!z \\
      0 \, ,
        & \textrm{for} \,\, y\!+\!z<x
    \end{cases}
\end{equation}
where 
\begin{equation} \label{eqn:xyz}
  x \equiv \vmin/\vmp \, , \quad
  y \equiv \vobs/\vmp \, , \quad \textrm{and} \quad
  z \equiv \vesc/\vmp \, ;
\end{equation}
recall $\vmin$ is given by \refeqn{vmin}.
Here, we use the common notational convention of
representing 3-vectors in bold and the magnitude of a vector in the
non-bold equivalent, \eg\ $\vobs \equiv |\bvobs|$.

Due to the motion of the Earth around the Sun, $\bvobs$ is time
dependent: $\bvobs = \bvobs(t)$.  We write this in terms of the Earth's
velocity $\bV_\oplus$ relative to the Sun as
\begin{equation} \label{eqn:vobs}
  \bvobs(t) = \bv_\odot
              + V_\oplus \left[
                  \eone \cos{\omega(t-t_1)} + \etwo \sin{\omega(t-t_1)}
                \right] \, ,
\end{equation}
where $\omega = 2\pi$/year, $\bv_\odot$ is the Sun's motion relative to
the WIMP component's rest frame, $V_\oplus = 29.8$ km/s is the Earth's
orbital speed, and $\eone$ and $\etwo$ are the directions of the Earth's
velocity at times $t_1$ and $t_1+0.25$ years, respectively.
With this form, we have neglected the ellipticity of the Earth's orbit,
although the ellipticity is small and, if accounted for, would give only
negligible changes in the results of this paper (see
Refs.~\cite{Green:2003yh,SmithLewin} for more detailed expressions).
For clarity, we have used explicit velocity vectors rather than the
position vectors $\peone$ and $\petwo$ used in
Refs.~\cite{Gelmini:2000dm,Freese:2003tt} and elsewhere (the position
vectors are more easily generalized to an elliptical orbit); the two
bases are related by $\eone = -\petwo$ and $\etwo = \peone$.

In Galactic coordinates, where $\hat{\mathbf{x}}$ is the direction to
the Galactic Center, $\hat{\mathbf{y}}$ the direction of disk rotation,
and $\hat{\mathbf{z}}$ the North Galactic Pole,
\begin{eqnarray}
  \label{eqn:eone}
    \eone &=& (0.9931, 0.1170, -0.01032) \, , \\
  \label{eqn:etwo}
    \etwo &=& (-0.0670, 0.4927, -0.8676) \, ,
\end{eqnarray}
where we have taken $\eone$ and $\etwo$ to be the direction of the
Earth's motion at the Spring equinox (March 21, or $t_1$) and Summer
solstice (June 21), respectively.

In this paper, we will take the halo to be a simple non-rotating
isothermal sphere \cite{Freese:1987wu}, also referred to as
the Standard Halo Model (SHM); more complicated halo models will be
examined in a future paper.  In this model, typical parameters of the
Maxwellian distribution for our location in the Milky Way are
$\sigma_{\mathrm{SHM}}$ = 270 km/s and $\vesc$ = 650 km/s, the
latter being the speed necessary to escape the Milky Way (WIMPs with
speeds in excess of this would have escaped the galaxy, hence the
truncation of the distribution in \refeqn{Maxwellian}).
Unlike the Galactic disk (along with the Sun), the halo has essentially
no rotation; the motion of the Sun relative to this stationary halo is
\begin{equation} \label{eqn:vsunSHM}
  \bv_{\odot,\mathrm{SHM}} = \bv_{\mathrm{LSR}} + \bv_{\odot,\mathrm{pec}}
    \, ,
\end{equation}
where $\bv_{\mathrm{LSR}} = (0,220,0)$ km/s is the motion of the
Local Standard of Rest and $\bv_{\odot,\mathrm{pec}} = (10,13,7)$ km/s
is the Sun's peculiar velocity.  The Earth's speed relative to the
halo, $\vobs(t)$, is maximized around June 1.  The local dark matter
density $\rhoDM$ is taken to be the estimated average density in the
local neighborhood, 0.3~GeV/cm$^3$.


\subsection{\label{sec:Modulation} Annual Modulation}

It is well known that the count rate in WIMP detectors will experience
an annual modulation as a result of the motion of the Earth around the
Sun described above \cite{Drukier:1986tm,Freese:1987wu}.  In some
cases, but not all, the count rate (\refeqn{dRdE}) has an approximate
time dependence
\begin{equation} \label{eqn:dRdEapprox}
  \dRdE(E,t) \approx S_0(E) + S_m(E) \cos{\omega(t-t_c)} ,
\end{equation}
where $t_c$ is the time of year at which $\vobs(t)$ is at its maximum.
$S_0(E)$ is the average differential recoil rate over a year and
$S_m(E)$ is referred to as the modulation amplitude (which may, in fact,
be negative).  The above equation is a reasonable approximation for the
SHM we are considering in this paper, but is not valid for all halo
models, particularly at some recoil energies for dark matter streams;
see Ref.~\cite{Savage:2006qr} for a discussion.  For the SHM,
\begin{equation} \label{eqn:SmSHM}
  S_m(E) = \frac{1}{2} \left[
             \dRdE(E,\,\textrm{June 1}) - \dRdE(E,\,\textrm{Dec 1})
           \right] .
\end{equation}
Experiments such as DAMA will often give the average amplitude over
some interval,
\begin{equation} \label{eqn:Sm}
  S_m = \frac{1}{E_2 - E_1} \int_{E_1}^{E_2} \rmd E \, S_m(E) .
\end{equation}

\subsection{\label{sec:Parameters} Parameter Space}

Many of the parameters that factor into the expected recoil rates for a
scattering detector are unknown, including the WIMP mass, four
WIMP-nucleon couplings (SI and SD couplings to each of protons and
neutrons), the local WIMP density, and the WIMP velocity distribution
in the halo.  In this paper, we shall fix the halo model to the SHM
and the local density to 0.3~GeV/cm$^3$.  In addition, we shall take
$\fpSI = \fnSI$ (equal SI couplings) so that there are only three
independent scattering couplings; the SI coupling will be given in
terms of the SI scattering cross-section off the proton, $\sigmapSI$.
The parameter space we examine will then consist of the four parameters
$m$, $\sigmapSI$, $\apSD$, and $\anSD$.

\section{\label{sec:Experiments} Null Experiments}

\begin{table*}
  \begin{tabular}{lcccccccc}
  \hline\hline
  Experiment & Element & Exposure & Energies & Quenching    & Efficiency
    & Constraint  & Ref. \\
             &         & [kg-day] & [keVee]  & Factor ($Q$) &
    & ($\dagger$) &      \\
  \hline
  DAMA
    & NaI
    & $2.99 \times 10^{5}$
    & 2--20
    & 0.3, 0.09 ($\ddag$)
    & 1
    & (various)
    & \protect\cite{Bernabei:2008yi} \\
  CDMS
    & Si
    & 6.58
    & 5--55
    & 1 ($\S$)
    & \refeqn{CDMSIeff}
    & Poisson
    & \protect\cite{Akerib:2003px} \\
    & Ge
    & 397.8
    & 10--100
    & 1 ($\S$)
    & \refeqn{CDMSIIeff}
    & Poisson
    & \protect\cite{Ahmed:2008eu} \\
  CoGeNT
    & Ge
    & 8.38
    & 0.23--4.1
    & 0.2
    & \refeqn{CoGeNTeff} 
    & BP
    & \protect\cite{Aalseth:2008rx,Collar:2008pc} \\
  CRESST I
    & Al$_2$O$_3$
    & 1.51
    & 0.6--20
    & 1 ($\S$)
    & 1
    & BP
    & \protect\cite{Angloher:2002in} \\
  TEXONO
    & Ge
    & 0.338
    & 0.2--8
    & 0.2
    & \refeqn{TEXONOeff}
    & MG
    & \protect\cite{Lin:2007ka,Wong:2008pc} \\
  XENON10
    & Xe
    & 316.4
    & 6.1--36.5  
    & 1 ($\S$)
    & \refeqn{XENON10eff}
    & MG
    & \protect\cite{Angle:2007uj,Angle:2008we} \\
  \hline\hline
  \end{tabular} 
  \caption{
    Experiments used in this study.
    Notes to the table:
      ($\dagger$) MG is the maximum gap method \cite{Yellin:2002xd},
        BP is a Poisson statistics based constraint for binned data as
        described in the text;
      ($\ddag$) some portion of recoils may undergo channeling with
        $Q \approx 1$ (see text);
      ($\S$) energies scaled to recoil energies.
  }
  \label{tab:ExpParam}
\end{table*}

Numerous experiments have searched for a dark matter signal, but all
these experimental results, apart from DAMA, are consistent with no
WIMP signal.  Here, we shall examine the constraints from these null
results; the positive DAMA signal will be addressed in the next
section.  We include here only a few experiments that provide some of
the strongest constraints at various points in the WIMP mass-coupling
parameter space.  We further restrict ourselves to experiments that
provide sufficient data for us to independently generate constraints.
Notably, this excludes COUPP \cite{Behnke:2008zz,Collar:2008pc2} for
which insufficient information on the data runs is provided to
both reproduce their results and extend them to arbitrary couplings.
COUPP is likely to provide strong constraints in areas of parameter
space that other direct detection experiments do not probe and could
significantly affect our results.  The experiments we include are
listed in \reftab{ExpParam}.

Most of the experiments observe some recoils; these recoils may be
due to background events (\eg\ neutrons from cosmic ray showers),
WIMP scatters, or a combination of the two.  While tighter constraints
may be found by modelling the background contribution,
we produce conservative constraints by analyzing the experimental
results \textit{without} background subtraction.  Additionally, some
of the data we use (as presented in the literature) is binned, whereas
analysis of the unbinned data (as available to the experimental groups)
generally allows for better constraints on the parameter space.
Some experiments are also able to combine and analyze their data sets
more effectively than we can.
Where other ambiguities arise, we err on the side of caution.
For these reasons, our constraints are generally more conservative
(by as much as a factor of $\sim$3 in some cases) than those provided
by the experimental groups themselves.  Readers should refer to the
original experimental publications for more thorough analyses and,
in some cases, the inclusion of background subtraction.

We will adopt the standard convention of displaying 90\% C.L.\ exclusion
regions for these null experiments.  For comparison, however, we will
also indicate the parameter space excluded at 3$\sigma$ and 5$\sigma$
for a few of the experiments.  We will first give a description
of the null experiments used in our analysis, then we will describe the
various statistical methods used to provide the constraints for these
experiments.

\subsection{\label{sec:ExptDescription} Descriptions}


\textit{CDMS.}
CDMS~II has a large exposure for its Germanium detectors,
397.8~kg-days, with a threshold of 10~keV \cite{Ahmed:2008eu}.
Observed energies are calibrated to recoil energies, so no quenching
factor is required. The efficiency of observing nuclear recoils after
data cuts is:
\begin{equation} \label{eqn:CDMSIIeff}
  \varepsilon(E) =
    \begin{cases}
      0.25 + 0.05 \frac{(E - 10\,\mathrm{keV})}{5\,\mathrm{keV}}
        & \textrm{for} \,\, 10\,\textrm{keV} < E < 15\,\textrm{keV} , \\
      0.30
        & \textrm{for} \,\, 20\,\textrm{keV} < E < 100\,\textrm{keV} .
    \end{cases}
\end{equation}

While newer data sets exist, we also include the Silicon data set from
CDMS-SUF \cite{Akerib:2003px} due to the low 5~keV threshold, which is
lower than the 7 \& 10~keV thresholds of the newer CDMS~II Si
data sets.  This CDMS-SUF set
involves 65.8~kg-days exposure on a 0.100~kg Si detector.  With an
80\% muon anti-coincident cut efficiency, 95\% nuclear recoil band
cut efficiency, and additional energy-dependent cuts, the total
efficiency can be approximated by\footnote{See also Fig.~(3) of
    Ref.~\cite{ArmelFunkhouser:2005zy}, but note that figure imposes
    a cut-off at 10~keV and an additional 75\% cut over 10--20~keV
    due to problems in one of the Ge detectors; such cuts do not
    apply in this case.}:
\begin{equation} \label{eqn:CDMSIeff}
  \varepsilon(E) = 0.80 \times 0.95 \times
    \begin{cases}
      0.10 + 0.30 \frac{(E - 5\,\mathrm{keV})}{15\,\mathrm{keV}}
        & \textrm{for} \,\, 5\,\textrm{keV} < E < 20\,\textrm{keV} , \\
      0.40 + 0.10 \frac{(E - 20\,\mathrm{keV})}{80\,\mathrm{keV}}
        & \textrm{for} \,\, 20\,\textrm{keV} < E < 100\,\textrm{keV} .
    \end{cases}
\end{equation}
Over a 5--100~keV recoil energy range, two events were detected at
55 \& 95~keV that are consistent with expected backgrounds.  In our
analysis, we will use zero observed events over a 5--55~keV recoil
range.  While it is not a statistically sound practice to
\textit{a postiori} select the bin to avoid observed events, we are
primarily interested in this data set to examine low mass
($\lae 30$ GeV) WIMPs, for which the observed events are at energies
too high to be compatible with WIMP scatters.  As such, using the
chosen bin will result in constraints comparable to more formal
statistical methods, such as S.~Yellin's maximum gap
method~\cite{Yellin:2002xd}, at low WIMP masses.  For higher mass WIMP
constraints based solely on this Silicon data, the two observed events
should be included over \eg\ a 5--100~keV range.  With inclusion of the
Ge data, however, the contribution of Silicon to the constraints at
higher WIMP masses is negligible.

For the combined CDMS Si+Ge data set, we use Poisson statistics to
provide a constraint based upon zero observed events.


\begin{table}
  \begin{tabular}{ccccc}
    \cline{1-2}\cline{4-5}\\[-3.0ex]
    \cline{1-2}\cline{4-5}
    Energy  & Events & \hspace{2em} & Energy  & Events \\\empty
    [keVee] &        &              & [keVee] &        \\
    \cline{1-2}\cline{4-5}
    0.338 - 0.371 & 4430
     & & 0.661 - 0.693 & 8 \\
    0.371 - 0.403 & 1104
     & & 0.693 - 0.725 & 5 \\
    0.403 - 0.435 & 140
     & & 0.725 - 0.757 & 18 \\
    0.435 - 0.467 & 29
     & & 0.757 - 0.790 & 15 \\
    0.467 - 0.499 & 12
     & & 0.790 - 0.822 & 12 \\
    0.499 - 0.532 & 12
     & & 0.822 - 0.854 & 5 \\
    0.532 - 0.564 & 19
     & & 0.854 - 0.886 & 12 \\
    0.564 - 0.596 & 9
     & & 0.886 - 0.919 & 16 \\
    0.596 - 0.628 & 12
     & & 0.919 - 0.951 & 9 \\
    0.628 - 0.661 & 8
     & & 0.951 - 0.983 & 8 \\
    \cline{1-2}\cline{4-5}\\[-3.0ex]
    \cline{1-2}\cline{4-5}
  \end{tabular} 
  \caption{
    CoGeNT binned data.
    Only the lowest energy bins are included here;
    an additional 
    98 bins over 0.983--4.14~keVee
    are not shown.
    Contraints at low WIMP masses mainly come from these lower energy
    bins.
  }
  \label{tab:CoGeNT}
\end{table}

\textit{CoGeNT.}
This experiment has a low exposure, 8.38~kg-days, on Germanium
detectors, but has a very low threshold of 0.23~keVee
\cite{Aalseth:2008rx,Collar:2008pc}.  Data is binned in observed
energies.
The CoGeNT collaboration has provided us with data in smaller
bins than given in their paper;
the lowest energy bins and the number of observed events in those bins
are presented in \reftab{CoGeNT} \cite{Collar:2008pc}.
From the number of events, exposure, and $\dRdE$ values in each
bin, we estimate the efficiency to be:
\begin{equation} \label{eqn:CoGeNTeff}
  \varepsilon(E^\prime) = 0.66 - \frac{E^\prime}{50\,\mathrm{keV}} ,
\end{equation}
where the efficiency is given as a function of \textit{observed} energy
($E^\prime = QE$).
We take the quenching factor to be $Q_{Ge} = 0.20$.
The widths of the bins are smaller than the energy resolution of the
CoGeNT detector, so the latter must be taken into account when
analyzing their results.
The energy resolution used by CoGeNT is:
\begin{equation} \label{eqn:CoGeNTER}
  \sigma(E^\prime) = \sqrt{(69.7\,\mathrm{eV})^2
                           + (0.98\,\mathrm{eV}) E^\prime} .
\end{equation}

There are significant background sources that contribute to the large
number of observed events.  The CoGeNT collaboration assumes specific
forms for this background and subtracts it during their analysis.
Here, we make no assumptions on the background and perform no background
analysis, so our constraint is more conservative.

For the binned CoGeNT data without background subtraction, we generate
constraints using the binned Poisson method described below.


\textit{CRESST~I.}
CRESST~I used Al$_2$O$_3$ detectors with an exposure of 1.51~kg-days
over an energy of 0.6--20~keV \cite{Angloher:2002in}.  The energy scale
is calibrated to the recoil energy.
Only scattering off of Aluminum is considered for the spin-dependent
scattering case (Oxygen is ignored).
The energy resolution used by CRESST~I is:
\begin{equation} \label{eqn:CRESSTIER}
  \sigma(E) = \sqrt{(0.220\,\mathrm{keV})^2 + (0.017 E)^2}
\end{equation}
Data is binned with a significant presence of backgrounds.
We generate the CRESST~I constraints using the binned Poisson method.


\textit{TEXONO.}
Another low exposure, low threshold Germanium experiment similar to
CoGeNT, TEXONO has an exposure of 0.338~kg-days over energies of
0.2--0.8~keVee \cite{Lin:2007ka,Wong:2008pc}.  The quenching factor is
taken to be $Q_{Ge} = 0.2$.  The combined efficiency after the
ACV (98.3\%), CRV (91.5\%), and PSD cuts (energy dependent, see
Figure~3 of Ref.~\cite{Lin:2007ka}) is estimated to be:
\begin{equation} \label{eqn:TEXONOeff}
  \varepsilon(E^\prime) = 0.983 \times 0.915 \times
    \begin{cases}
      \frac{(E^\prime - 0.1\,\mathrm{keV})}{0.25\,\mathrm{keV}}
        & \textrm{for} \,\, 0.1\,\textrm{keV} < E^\prime < 0.35\,\textrm{keV} , \\
      1
        & \textrm{for} \,\, 0.35\,\textrm{keV} < E^\prime .
    \end{cases}
\end{equation}
The constraints are generated using the unbinned maximum gap 
method \cite{Yellin:2002xd}, which examines the likelihood of
observing the gaps in energy between observed events.
TEXONO provides two intervals between event energies,
0.198--0.241~keVee and 1.39--1.87~keVee, that are used for the gaps
in this method; other gaps in the full 0.198--8~keVee range are ignored
(ignoring other intervals can lead to only more conservative constraints
in this type of analysis).

Some concerns have been raised regarding the claimed behavior of
TEXONO's detector response at very low energies that calls into
question their low energy results \cite{Avignone:2008xc}.  We shall
not address these concerns here.


\textit{XENON.}
XENON10 has a large exposure, 316.4~kg-days, on Xenon over
6.1--36.5~keV \cite{Angle:2007uj,Angle:2008we,Sorensen:2008ec};
energies are calibrated to recoil energies ($Q=1$).  That calibration
depends upon a quantity $\mathcal{L}_{\mathrm{eff}}$, described in
their papers, that was initially estimated to be 0.19 with large
uncertainties.  More recent results indicate it might be lower
\cite{Sorensen:2008ec} and we use a conservative value of 0.14.  Use
of this lower value requires all the nuclear recoil energy scales in
the XENON10 data to be multiplied by a factor of 0.19/0.14, \eg\ the
6.1--36.5~keV energy range is determined by multiplying the original
range of 4.6--26.9~keV by this factor.
We approximate the energy dependent efficiency by:
\begin{equation} \label{eqn:XENON10eff}
  \varepsilon(E) = 0.46 (1 - \frac{E}{135\,\mathrm{keV}})
\end{equation}
and use an energy resolution of \cite{Baudis:2008pc}
\begin{equation} \label{eqn:XENON10ER}
  \sigma(E) = (0.579\,\mathrm{keV}) \sqrt{E/\mathrm{keV}}
                     + 0.021 E .
\end{equation}
XENON10 observed 10 events, with an average background of 7 events
expected.  We use the maximum gap method to generate XENON10
constraints.



\textit{Super-Kamionkande.}
Aside from the direct detection experiments listed above, we shall
also show SuperK constraints from WIMP annihilation in the Sun
\cite{Desai:2004pq}.  SuperK determines these constraints by searching
for high energy neutrino products of the WIMP annihilation process
at the center of the Sun, where the annihilation rate is dependent upon
the capture rate of WIMPs and the capture rate depends upon the
scattering cross-section for WIMPs off of nuclei in the Sun.  Due to
the large proportion of Hydrogen, the neutrino flux is particularly
dependent upon the scattering cross-section off of the proton (both
SI and SD).  While there is a weak dependence on WIMP-neutron
scattering cross-section through the trace amounts of heavier elements,
we consider here only constraints for WIMP-proton scattering
and only in the SD case.  The SuperK collaboration also provides
SI constraints, but only for combined Earth and Solar WIMP
annihilations, which we do not include here.  Their SI constraints do
not affect our results.  The SD constraint is given at a 90\% C.L.
We have not performed our own reanalysis of the SuperK data and
simply present the constraints as determined by the SuperK
collaboration's own analysis (see Ref.~\cite{Hooper:2008cf} for
a reanalysis).

The SuperK constraints rely on the following assumptions:
(1) the WIMP/anti-WIMP abundance is not highly asymmetric, which would
suppress the annihilation rate;
(2) the annihilation cross-section is sufficiently high so that the
capture rate of WIMPs in the Sun (via scattering off of nuclei in the
Sun) is in equilibrium with the annihilation rate; and
(3) the WIMP does not annihilate predominantly into the light quarks,
which do not yield neutrinos in sufficient quantities and energies to
be observed.
For the Constrained Minimal Supersymmetric Standard Model,
these assumptions are mainly satisfied in the parameter space of
interest.  In general, however, these assumptions need not be satisfied,
so the SuperK constraints should be applied with caution.


\begin{figure}
  \insertfig{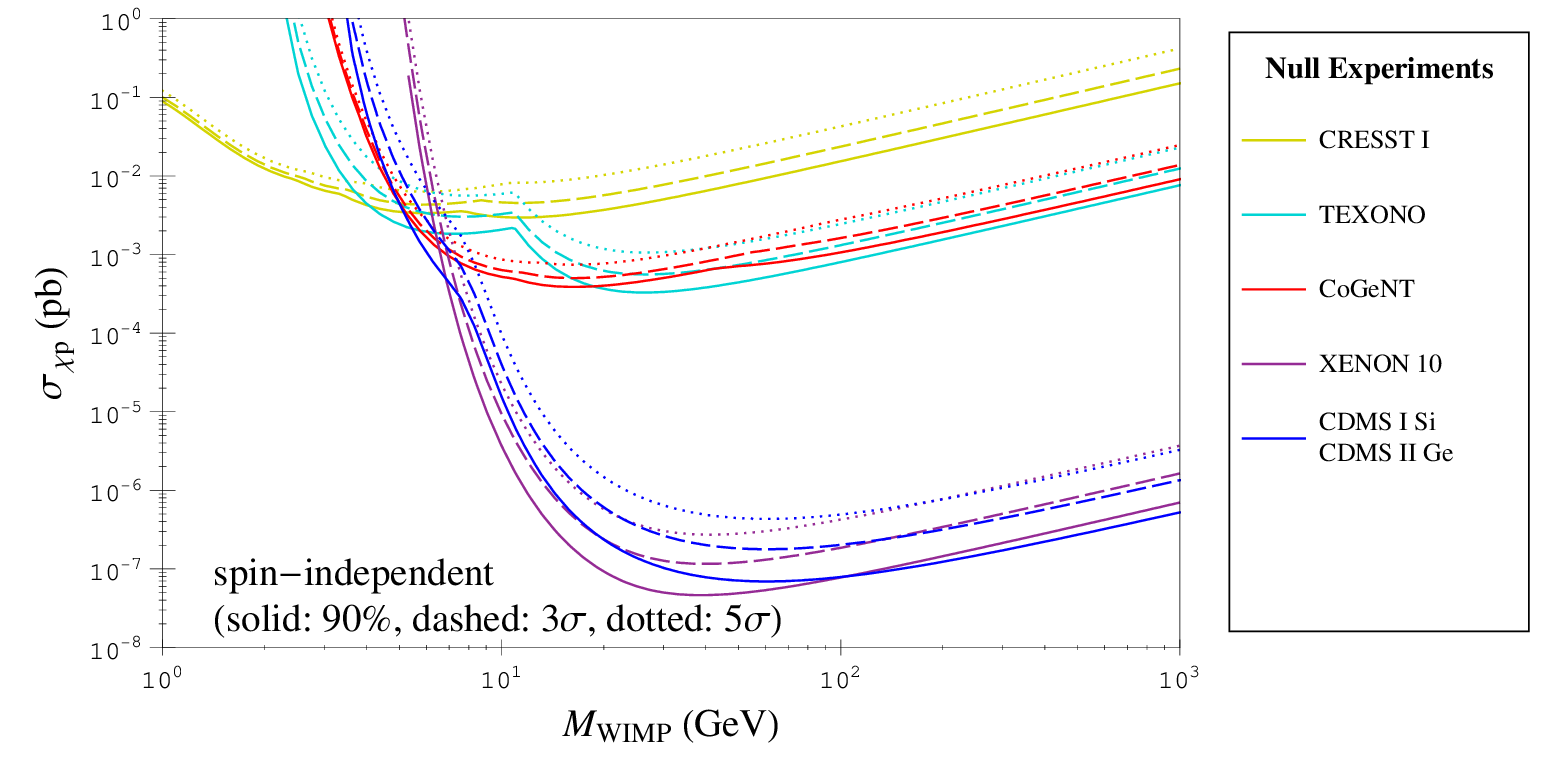}
  \caption{
    Contraints on spin-independent (SI) scattering cross-sections for
    various experiments with null results.
    Cross-sections below each line are excluded by the given experiment
    at the 90\% (solid), 3$\sigma$ (dashed), and 5$\sigma$ (dotted)
    confidence levels.
    The same coloring scheme will be used in later figures.
    }
  \label{fig:SIExpts}
\end{figure}

\textit{Constraints comparison.}
Constraints for these null experiments are shown in \reffig{SIExpts}
for the case of SI scattering.  We show here and in later figures
regions excluded at the 90\% level; for comparison, exclusion regions
at 3$\sigma$ and 5$\sigma$ are also shown here for a few of the null
experiments.  The regions above the curves are excluded.  The high
exposure experiments, CDMS and XENON10, provide very strong
constraints at high WIMP masses, but provide no constraints at low
WIMP masses.  Only the low threshold experiments---CoGeNT, CRESST~I,
and TEXONO---provide constraints at low WIMP masses (due to the fact
that light WIMPs only yield low energy collisions, which high
threshold experiments cannot detect).

\subsection{\label{sec:ExptConstraints} Constraints}

For the various null experiments, we define constraints in parameter
space at a certain exclusion level $1-\alpha$, typically 90\%, as the
parameters for which the probability of seeing the experimental
result is $\alpha$.  Parameters outside the corresponding contours
would yield the observed result with a probability less than $\alpha$.
We say the parameters within those contours are compatible within the
$1-\alpha$ level and parameters outside are excluded at the $1-\alpha$
level.  The probabilities are determined use different statistical
values, described here, in different cases.

\textit{Poisson.}
For data in a single bin, we use Poisson statistics to exclude
parameter space as follows: parameters that
predict an average number of events $\mu$ in that bin are excluded at a
level of $1-\alpha$ if the probability of seeing as few as the observed
number of events (zero in the case of CDMS Si+Ge) is less than
$\alpha$.  For zero observed events, that corresponds to upper limits
on $\mu$ of 2.3 at a 90\% exclusion level, 5.9 at $3\sigma$, and 14.4
at $5\sigma$.  Here, we use $N\sigma$ to indicate the probabilities
associated with the $\pm N\sigma$ bounds of a normal distribution.

\textit{Binned Poisson (BP).}
For binned data with unknown background, we exclude parameter space
using what we shall refer to as a ``binned Poisson'' (BP) technique,
defined as follows.  If $\alphabin$ is the probability of seeing below
a certain number of events in a bin (the lower tail of the Poisson
distribution), the probability $\alpha$ of seeing at least one bin
measurement falling below the $\alphabin$ lower tail of that bin is
related by:
\begin{equation} \label{eqn:BinnedPoisson}
  1 - \alpha = (1 - \alphabin)^{\Nbin} ,
\end{equation}
where $\Nbin$ is the number of bins.  For a desired exclusion level of
$1-\alpha$, we then require at least one bin to be at the bin's
exclusion level of $1-\alphabin$, where $\alphabin$ can be determined
from $\alpha$ using the above equation.  For example, for a desired
exclusion level of 90\% in two bins, the observed number of events in
at least one bin should be excluded at the 94.9\% level.  In other
words, if there is only a 5.1\% chance of seeing below a certain
number of events in each bin, there is a 10\% chance of the observed
number of events in at least one those bins to be below that number
of events.  In practice, the $1-\alpha$ exclusion contour is
determined by, at each WIMP mass, increasing the scattering
cross-section until the number of events $N_k$ observed in at least
one bin is in the $1-\alphabin$ lower tail of the Poisson distribution
with average $\mu_k$, where $\mu_k$ is the theoretical predicted
average number of events in that bin.  Background events make the
observed number of events (signal + background) less likely to fall
into the lower tail of a Poisson distribution based upon the expected
signal alone; bins with larger signal-to-noise will likely be the first
to reach the $1-\alphabin$ exclusion level (\ie\ too few events) and
define the overall constraint.
Then this technique determines the overall constraint from the bin that
would provide the strongest constraint, but takes the required
statistical penalty for choosing that bin, since that bin can simply
be one in which a statistical fluctuation produces a lower than
expected number of events.  See Ref.~\cite{Green:2001xy} for further
discussions.
This type of analysis is not optimal, but is simple and straightforward
to perform and works reasonably well for the large background cases
of CoGeNT and CRESST~I here.

\textit{Maximum Gap (MG).}
For unbinned data with unknown background, constraints are generated
using S.~Yellin's maximum gap (MG) method \cite{Yellin:2002xd}, which
examines the likelihood of observing the gaps in energy between
observed events.  This ubinned method generally provides a stronger
constraint that the previous (binned) methods for the case of an
unknown background.

\section{\label{sec:DAMA} DAMA}

\begin{table}
  \begin{tabular}{c@{\hspace{1em}}ccc@{\hspace{1em}}c}
    \cline{1-2}\cline{4-5}\\[-3.0ex]
    \cline{1-2}\cline{4-5}
    Energy  & Average $S_m$ & \hspace{2em} & Energy  & Average $S_m$ \\\empty
    [keVee] & [cpd/kg/keV]  &              & [keVee] & [cpd/kg/keV]  \\
    \cline{1-2}\cline{4-5}
    2.0  - 2.5  &  0.0162 $\pm$ 0.0048
     & & 11.0 - 11.5 & -0.0021 $\pm$ 0.0039 \\
    2.5  - 3.0  &  0.0287 $\pm$ 0.0054
     & & 11.5 - 12.0 & -0.0010 $\pm$ 0.0040 \\
    3.0  - 3.5  &  0.0250 $\pm$ 0.0055
     & & 12.0 - 12.5 &  0.0015 $\pm$ 0.0040 \\
    3.5  - 4.0  &  0.0140 $\pm$ 0.0050
     & & 12.5 - 13.0 &  0.0020 $\pm$ 0.0040 \\
    4.0  - 4.5  &  0.0101 $\pm$ 0.0045
     & & 13.0 - 13.5 & -0.0019 $\pm$ 0.0040 \\
    4.5  - 5.0  &  0.0118 $\pm$ 0.0041
     & & 13.5 - 14.0 &  0.0015 $\pm$ 0.0040 \\
    5.0  - 5.5  &  0.0039 $\pm$ 0.0041
     & & 14.0 - 14.5 & -0.0022 $\pm$ 0.0040 \\
    5.5  - 6.0  &  0.0030 $\pm$ 0.0039
     & & 14.5 - 15.0 & -0.0008 $\pm$ 0.0040 \\
    6.0  - 6.5  &  0.0059 $\pm$ 0.0038
     & & 15.0 - 15.5 & -0.0021 $\pm$ 0.0039 \\
    6.5  - 7.0  &  0.0011 $\pm$ 0.0036
     & & 15.5 - 16.0 & -0.0022 $\pm$ 0.0038 \\
    7.0  - 7.5  & -0.0001 $\pm$ 0.0036
     & & 16.0 - 16.5 &  0.0053 $\pm$ 0.0037 \\
    7.5  - 8.0  &  0.0004 $\pm$ 0.0036
     & & 16.5 - 17.0 &  0.0060 $\pm$ 0.0037 \\
    8.0  - 8.5  & -0.0014 $\pm$ 0.0037
     & & 17.0 - 17.5 &  0.0007 $\pm$ 0.0037 \\
    8.5  - 9.0  &  0.0039 $\pm$ 0.0037
     & & 17.5 - 18.0 &  0.0041 $\pm$ 0.0035 \\
    9.0  - 9.5  & -0.0034 $\pm$ 0.0037
     & & 18.0 - 18.5 &  0.0011 $\pm$ 0.0035 \\
    9.5  - 10.0 & -0.0071 $\pm$ 0.0038
     & & 18.5 - 19.0 & -0.0014 $\pm$ 0.0035 \\
    10.0 - 10.5 &  0.0086 $\pm$ 0.0038
     & & 19.0 - 19.5 &  0.0006 $\pm$ 0.0034 \\
    10.5 - 11.0 & -0.0018 $\pm$ 0.0039
     & & 19.5 - 20.0 &  0.0056 $\pm$ 0.0034 \\
    \cline{1-2}\cline{4-5}\\[-3.0ex]
    \cline{1-2}\cline{4-5}
    &&& \makebox[0in][r]{($\dagger$)\ }10.0 - 20.0 & 0.0011 $\pm$ 0.0008 \\
    \cline{4-5}\\[-3.0ex]
    \cline{4-5}
  \end{tabular} 
  \caption{
    DAMA/NaI + DAMA/LIBRA (DAMA) modulation amplitudes for the given
    observed energy bins.  Data is taken from Figure~9 of
    Ref.~\cite{Bernabei:2008yi}.
    ($\dagger$) The bins over 10-20~keV have been combined into a
    single bin for use with some of the analysis techniques discussed
    in the text.
  }
  \label{tab:DAMA}
\end{table}

The DAMA/NaI experiment \cite{Bernabei:2003za} used a large
NaI detector to search for the small modulation in the WIMP scattering
rate and became the first direct detection experiment to observe a
positive signal.  That signal has since been confirmed by the successor
to DAMA/NaI, DAMA/LIBRA \cite{Bernabei:2008yi} (also a NaI detector).
These two experiments remain the only direct detection experiments to
observe a signal.

\begin{figure}
  \insertfig{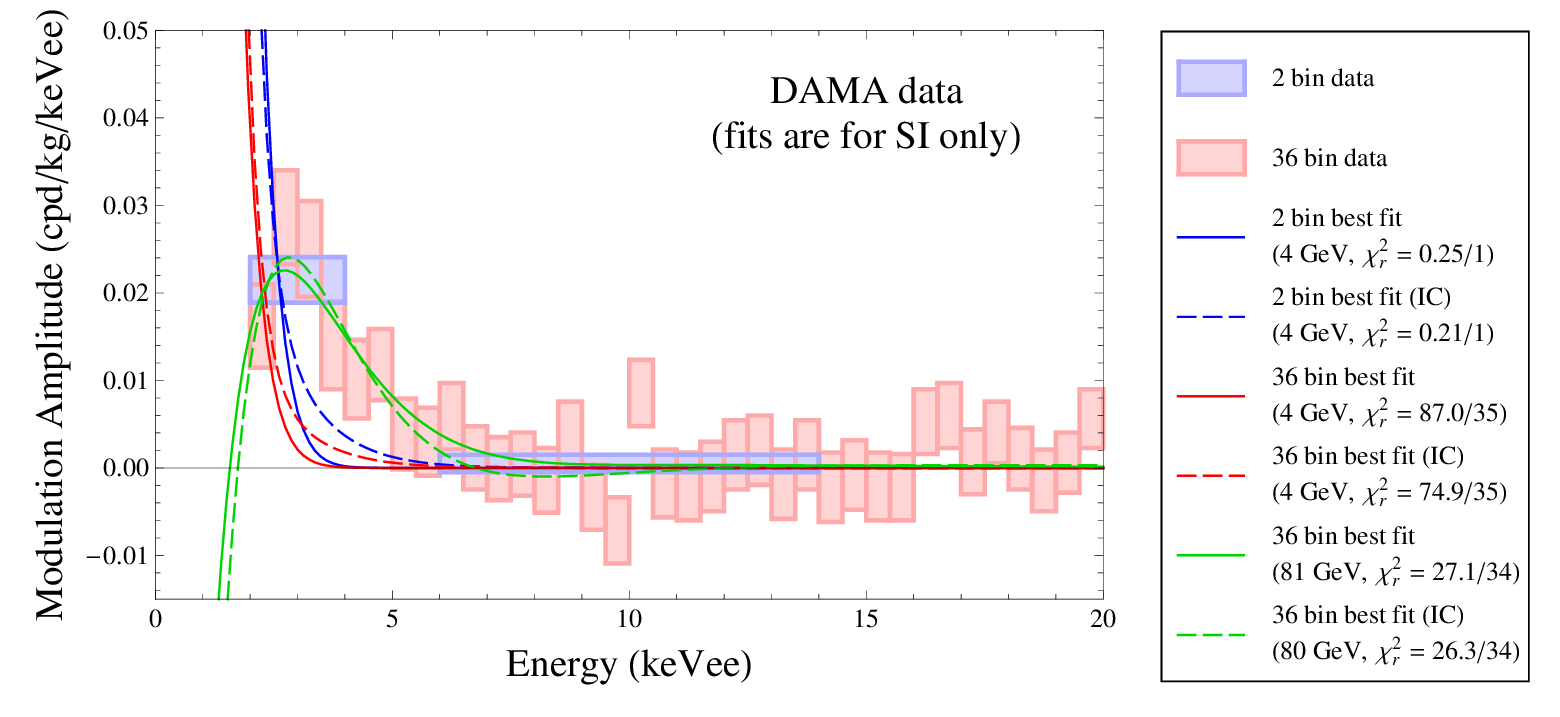}
  \caption{
    DAMA modulation amplitude data as well as best fit spectra for
    various cases.  Boxes represent the 1$\sigma$ bounds of the measured
    modulation amplitude for each bin: pink for the full 36 bin data
    set and blue for a 2 bin data set (2--4 \& 6--14~keVee).  Also
    shown are the $S_m$ spectra for the best fit SI cross-sections
    (in the SI only coupling case) at a WIMP mass of 4~GeV for the
    2 bin data (blue) and 36 bin data (red), with (dashed) and
    without (solid) inclusion of the ion channeling (IC) effect.  The
    $\chi^2$ values (over the degrees of freedom) are indicated in the
    legend.  For comparison, the best (SI only) fit at any mass,
    which occurs around 80 GeV, is shown (green).
    }
  \label{fig:DAMAFit}
\end{figure}

The original DAMA/NaI experiment released data for only two independent
bins, a low energy bin at 2--4, 2--5, or 2--6~keVee (these are not
independent) with a non-trivial modulation amplitude and a high
energy bin at 6--14~keVee with an amplitude consistent with zero.
The DAMA/LIBRA results, in combination with the earlier DAMA/NaI
results, have been released for 36 bins over a range of 2--20~keVee
(we shall henceforth use ``DAMA'' to refer to this combined
data set).  The modulation amplitudes for these bins, taken from
Figure~9 of Ref.~\cite{Bernabei:2008yi}, are given in \reftab{DAMA}.
The 1$\sigma$ bounds on the amplitudes for these bins are shown as
pink boxes in \reffig{DAMAFit}.  For comparison, modulation amplitudes
are shown in blue boxes for a 2 bin data set: $0.0215 \pm 0.0026$
cpd/kg/keVee over 2--4~keVee and $0.0005 \pm 0.0010$ over 6--14~keVee.

We use the quenching factors $Q_{Na} = 0.3$ and $Q_{I} = 0.09$ in our
calculations, with some exceptions as discussed in the following
section.

\subsection{Detector Effects}

Several physical and instrumental effects add complications
to the basic description of experimental recoil rates as described in
\refsec{Detection}.  These include a finite detector energy resolution
and two effects, ion channeling and the Migdal effect, that affect
how much of the deposited energy goes into the observed channels.

\subsubsection{Energy Resolution}

The DAMA/LIBRA apparatus, as with all experiments, cannot determine the
event energies with perfect precision.
If the expected amount of energy in the channel DAMA/LIBRA observes is
$E^\prime = Q E$, 
the \textit{measured} energy will be normally distributed about
this energy with a standard deviation \cite{Bernabei:2008yh},
\begin{equation} \label{eqn:DAMAER}
  \sigma(E^\prime) = (0.448\,\mathrm{keV}) \sqrt{E^\prime/\mathrm{keV}}
                     + 0.0091 E^\prime .
\end{equation}
While this equation technically applies only to the DAMA/LIBRA
apparatus, we use the same energy resolution for the combined data of
DAMA/NaI and DAMA/LIBRA.
We include this finite energy resolution in all our DAMA calculations.

\subsubsection{Ion Channeling}

Typically, only a small fraction $Q$ of the recoil energy, 0.3 for Na
and 0.09 for I, goes into a mode that DAMA measures (scintillation).
The rest is converted into, \eg, phonons/heat as the recoiling nucleus
collides with other nuclei and is not observed.  As pointed out by
the DAMA Collaboration \cite{Bernabei:2007hw}, nuclei that recoil
along the characteristic axes or planes of the crystal structure may
travel large distances without colliding with other nuclei.  For ions
such as recoiling Na$^+$ or I$^-$, the recoil energy in this case is
primarily transferred to electrons (which is observable) rather than
other nuclei (which becomes heat).  Some recoils that undergo this ion
channeling (IC) have $Q \approx 1$.

DAMA has used simulations to determine the energy-dependent fractions
of recoils off of Na ($f_{Na}$) and I ($f_{I}$) that will have
$Q \approx 1$ due to this IC effect.  These fractions, given by
Figure~4 of Ref.~\cite{Bernabei:2007hw}, are well approximated by
\begin{equation} \label{eqn:fNa}
  f_{Na} = 10^{-\sqrt{E / (6.9\,\mathrm{keV})}}
\end{equation}
and
\begin{equation} \label{eqn:fI}
  f_{I} = 10^{-\sqrt{E / (11.5\,\mathrm{keV})}} ,
\end{equation}
where these fractions are given as functions of the \textit{recoil}
energy.

We shall examine the DAMA results both with and without including the
IC effect and shall use ``IC'' in the figures to indicate those
analyses that include it.  We have not included this effect in other
experiments where it may also be applicable, such as CoGeNT and TEXONO
\footnote{Channeled events in experiments that measure phonons/heat,
    such as CDMS, would not be seen as any events with relatively
    high ionization/scintillation and low heat are characteristic of
    electromagnetic backgrounds (\eg\ $\gamma$-rays), which induce
    electron rather than nuclear recoils, and are thrown out.}.

\subsubsection{Migdal Effect}

In the 1940's, Migdal pointed out that the rapid change in velocity
of a recoiling nucleus can cause bound electrons to become excited or
ionized \cite{Migdal:1941}.  In the DAMA detector, the energy of
these ionized electrons goes into observable modes.  At the same
time, the \textit{apparent} recoil energy is reduced as the energy is
lost to the ionized electron almost immediately, before the nucleus
travels far enough to interact with other nuclei and electrons.
This Migdal effect (ME) can then affect the amount of observed energy
coming from a scattering event at a given energy.
Due to the complicated nature of this effect, we shall not show its
implementation here, but it results in a rather complicated response
function $\Phi$ that may be used in \refeqn{rateER}.
The implementation of the Migdal effect in an analysis of DAMA is
fully described in Ref.~\cite{Bernabei:2007jz}.
We do not include the Migdal effect in our analysis at this
time, but we plan to include it in a future paper.

\subsection{Analysis Techniques}

There are many possibilities for analyzing the DAMA results; the choice
of statistical tests depends upon how one wishes to interpret the
results.  Here, we will use three different tests to produce different
regions in parameter space: a likelihood ratio method with a global fit
of four parameters (which we refer to as the ``likelhood ratio'' method
from now on) to find the preferred (best fit) parameters to produce the DAMA
signal, a likelihood ratio method with a one parameter fit in raster
scans to find the preferred scattering cross-section to produce the
DAMA signal at each WIMP mass (which we call the ``raster scan''), and
a $\chi^2$ goodness-of-fit test (g.o.f.) to indicate
which parameters are compatible with the DAMA signal.  See
Ref.~\cite{Amsler:2008zz} for a short review of statistics or
Ref.~\cite{James:2006} for more extensive discussions.

\subsubsection{Likelihood Ratio}

\begin{table*}
  \addtolength{\tabcolsep}{0.25em}
  \begin{tabular}{lccccc}
    \hline\hline
     & Mass [GeV] & $\sigmapSI$ [pb]
     & $\apSD$ ($\sigmapSD$ [pb]) & $\anSD$ ($\sigmanSD$ [pb])
     & $\chisqmin$ (d.o.f.) \\
    \hline
    Global
      & 81.0 & $2.0 \times 10^{-5}$
      & 0.0 (0.0)       & 0.0 (0.0)
      & 27.1 (32) \\
    \hspace{2em} w/ channeling
      & 78.6 & $2.7 \times 10^{-5}$
      & 0.62 (0.13)     & 0.70 (0.17)
      & 25.4 (32) \\
    SI only
      & 81.0 & $2.0 \times 10^{-5}$
      & \textbf{0 (0)} & \textbf{0 (0)}
      & 27.1 (34) \\
    \hspace{2em} w/ channeling
      & 80.0 & $3.0 \times 10^{-5}$
      & \textbf{0 (0)} & \textbf{0 (0)}
      & 26.3 (34) \\
    SD only
      & 9.1  &\textbf{0}
      & 14.2 (59.4) & -176 (9020)
      & 29.3 (33) \\
    \hspace{2em} w/ channeling
      & 12.0 & \textbf{0}
      & 1.07 (0.35) & -12.8 (50.2)
      & 30.9 (33) \\
    SD proton only
      & 11.5 & \textbf{0}
      & 1.70 (0.88) & \textbf{0 (0)}
      & 30.8 (34) \\
    \hspace{2em} w/ channeling
      & 11.4 & \textbf{0}
      & 1.33 (0.53) & \textbf{0 (0)}
      & 31.0 (34) \\
    SD neutron only
      & 11.0 & \textbf{0}
      & \textbf{0 (0)} & 20.3 (125)
      & 31.0 (34) \\
    \hspace{2em} w/ channeling
      & 11.9 & \textbf{0}
      & \textbf{0 (0)} & 7.8 (18.6)
      & 30.9 (34) \\
    SD only ($\anSD = \apSD$)
      & 11.5 & \textbf{0}
      & 1.57 (0.75) & \textbf{1.57 (0.75)}
      & 30.8 (34) \\
    \hspace{2em} w/ channeling
      & 11.4 & \textbf{0}
      & 1.15 (0.40) & \textbf{1.15 (0.40)}
      & 31.0 (34) \\
    SD only ($\anSD = -\apSD$)
      & 11.6 & \textbf{0}
      & 1.88 (1.05) & \textbf{-1.88 (1.05)}
      & 30.8 (34) \\
    \hspace{2em} w/ channeling
      & 11.2 & \textbf{0}
      & 1.56 (0.73) & \textbf{-1.56 (0.73)}
      & 31.1 (34) \\
    \hline\hline
  \end{tabular} 
  \caption{
    Parameters at which the DAMA $\chi^2$ is minimized.
    The global minimum is found by allowing all parameters
    ($m$, $\sigmapSI$, $\apSD$, and $\anSD$) to vary, while the
    other cases allow only a subset of these parameters to vary
    freely with the other parameters fixed to zero or fixed in
    terms of the free parameters.
    Bold parameters in the table are those that have been fixed.
  }
  \label{tab:DAMAMin}
\end{table*}

To determine the most likely parameters for producing the DAMA signal,
we use the maximum likelihood method, based on the likelihood ratio
\begin{equation} \label{eqn:LR}
  \frac{L(S_{m,k}|m,\sigmapSI,\apSD,\anSD)}%
       {L(S_{m,k}|\mhat,\sigmapSIhat,\apSDhat,\anSDhat)} ,
\end{equation}
where $L$ is the likelihood function, $S_{m,k}$ are the observed
modulation amplitudes in each bin, and $\mhat$, $\sigmapSIhat$,
$\apSDhat$, and $\anSDhat$ are the values of the parameters
that maximize the likelihood for the observed $S_{m,k}$.  The
denominator of the above equation is the maximum likelihood value
$L_{\mathrm{max}}$.
Confidence regions in the parameters are determined by
\begin{equation} \label{eqn:LRCR}
  2\, \ln L(m,\sigmapSI,\apSD,\anSD)
  \ge 2\, \ln L_{\mathrm{max}} - 2\, \Delta \ln L ,
\end{equation}
where the value of $\Delta \ln L$ corresponds to the
confidence level (C.L.) of the confidence region.  Since each bin is
normally distributed, this equation may instead by given in terms of
the $\chi^2$,
\begin{equation} \label{eqn:ChiSqCR}
  \chi^2(m,\sigmapSI,\apSD,\anSD)
  \le \chisqmin + \Delta \chi^2 ,
\end{equation}
where
\begin{equation} \label{eqn:ChiSq}
  \chi^2(m,\sigmapSI,\apSD,\anSD)
    \equiv \sum_k \frac{(S_{m,k} - S_{m,k}^{\mathrm{Th}})^2}{\sigma_k^2} ,
\end{equation}
$\sigma_k$ is the uncertainty in $S_{m,k}$ (see \reftab{DAMA}), 
$S_{m,k}^{\mathrm{Th}} \equiv S_{m,k}^{\mathrm{Th}}(m,\sigmapSI,\apSD,\anSD)$
is the expected amplitude in a particular bin for the given parameters,
and $\chisqmin$ is the minimum value of $\chi^2$.
The $\chisqmin$ and the parameters at which it occurs is
given in \reftab{DAMAMin} in this global (four parameter) case both
with and without IC.  Also given are the minima in various specific
cases, such as SI only scattering.

The $\Delta \chi^2$ limit corresponding to different C.L.'s
depends upon the number of parameters that have been minimized
and, in the large data sample limit, is determined from a $\chi^2$
distribution with the degrees of freedom (d.o.f.) equal to the number
of minimized parameters.
For the four parameters considered here, $\Delta \chi^2$ =
7.8 (90\% C.L.), 16.3 (3$\sigma$), 34.6 (5$\sigma$), and
60.3 (7$\sigma$).

The confidence region for DAMA as described here is, in fact, a
4-dimensional region in the ($m,\sigmapSI,\apSD,\anSD$) parameter
space.  We shall be showing 2-dimensional slices of this larger
region for particular cases, such as the $\apSD = \anSD = 0$ slice
corresponding to SI only scattering.  This is \textit{not} equivalent
to fixing $\apSD = \anSD = 0$ and determining a 2-dimensional
confidence region by minimizing only over ($m,\sigmapSI$).  In the
latter case, a confidence region will always be defined in a
2-dimensional plane while, in the former, there may be no slice in
that plane if that plane represents a poor ``fit'' relative to the
overall parameter space.

Note, as will be discussed in \refsec{Results}, the confidence regions
obtained via this method yield the most \textit{preferred} (best fit) parameters
for producing the signal.  This method, in and of itself, does
\textit{not} imply parameters outside the confidence regions are
necessarily a bad fit to the data (or, conversely, that parameters
inside these regions are a good fit) and one should be careful using
these regions to compare with other experiments \footnote{In
    statistical parlance, the determination of the confidence
    region for this method is \textit{decoupled} from the
    goodness-of-fit.}.

\subsubsection{Raster Scan}

Since previous papers have examined the possibility of low mass WIMPs
as the source of the DAMA signal
\cite{Gelmini:2004gm,Gondolo:2005hh,Savage:2004fn}, we include
an analysis that gives the preferred scattering cross-sections at a
given WIMP mass.  If one specified how to determine the preferred
cross-sections for an arbitrary WIMP mass, the band formed by these
preferred cross-sections over all masses is what is referred to as a
raster scan.

The raster scans here involve only two parameters, $m$ and
$\sigma$, where the WIMP mass $m$ is the parameter we scan over and
$\sigma$ is some scattering cross-section, \eg\ the SI cross-section
(we shall be examining several different cases).  At a given  $m$,
the preferred values for $\sigma$ are determined using the likelihood
ratio method just described but leaving one parameter, which we call
here $\sigma$, as a variable and fixing the other two
couplings/cross-sections to some value, usually zero.  Thus
the $\chi^2$ is minimized over $\sigma$ at that particular mass,
$\chi_{\mathrm{RS,min}}^2(m) = \chi^2(m,\sigmahat)$.  The confidence
\textit{interval} in $\sigma$ at $m$ is then
\begin{equation} \label{eqn:RSChiSqCI}
  \chi^2(m,\sigma)
  \le \chi_{\mathrm{RS,min}}^2(m) + \Delta \chi_{\mathrm{RS}}^2 ,
\end{equation}
where limits on $\Delta \chi_{\mathrm{RS}}^2$, determined from a
$\chi^2$ distribution with 1 d.o.f. (large data sample limit), are
2.7, 9, 16, and 25 at C.L.'s of 90\%, 3$\sigma$, 5$\sigma$,
and  7$\sigma$, respectively.  The band of confidence \textit{intervals}
in $\sigma$ over all masses defines a confidence \textit{region} in
$m$ and $\sigma$ at the same C.L. as the intervals; we call this region
the raster scan region.

The above formulation is not technically correct when physical
boundaries are present, such as the requirement of a non-negative
cross-section, but it is approximately correct if the obtained
confidence intervals are not near the boundary.  As will be seen later,
the raster scans do yield confidence intervals that cross the physical
$\sigma = 0$ boundary; however, these occur only at high WIMP masses
that will not be of interest.  The raster scans over the masses of
interest are far enough from the boundary to be approximately correct.
Formally, one should use a method that can appropriately account for
such boundary conditions, such as that of Feldman and Cousins
\cite{Feldman:1997qc} \footnote{The result of using this more formal
    method will be to shift upwards the upper bounds on $\sigma$ at
    massses where the confidence interval is near the $\sigma = 0$
    boundary, but the shift will be less than a factor of 2.}.

The raster scan itself, by definition, provides no constraint on the
WIMP mass and the raster scan band includes WIMP masses at which no
scattering cross-section is capable of reasonably producing the DAMA
signal.  One may determine which WIMP masses provide poor fits to the
DAMA data by applying a $\chi^2$ goodness-of-fit test at the best fit
cross-section $\sigmahat$ at the mass to be examined.  The $\chi^2$
values at these minima should follow a $\chi^2$ distribution with
35 d.o.f.\ (36 bins with a fit to 1 parameter ($\sigma$)).  The $\chi^2$
for 35 d.o.f.\ is expected to fall below 46.1, 62.8, and 91.7 with
probabilities of 90\%, 99.73\% (3$\sigma$), and 99.999943\% (5$\sigma$),
respectively; then the $\chi^2$ should exceed those values only 10\%,
0.27\% and 5.7$\times 10^{-5}$\% of the time, respectively.

\subsubsection{Goodness-of-Fit}

To conservatively indicate parameters that are compatible with the DAMA
data set, we also indicate regions at which the $\chi^2$ falls within
a given level using a simple $\chi^2$ goodness-of-fit (g.o.f.) test on
the data.  In contrast to the previous two analyses methods, there
is no fit to the parameters here.  The g.o.f.\ regions are defined as
those parameters for which
\begin{equation} \label{eqn:GOFChiSq}
  \chi^2(m,\sigmapSI,\apSD,\anSD) \le \chi_{\mathrm{GOF}}^2 ,
\end{equation}
where $\chi_{\mathrm{GOF}}^2$ is the value at which the $\chi^2$
cumulative distribution function (CDF) for \eg\ 36 d.o.f.\ is equal to
the desired level of compatibility.  That is, for a desired
compatibility level of $1 - \alpha$, there is a probability $\alpha$
that $\chi^2$ will exceed $\chi_{\mathrm{GOF}}^2$.  Alternatively, we
can say parameters outside of the region are excluded at the
$1 - \alpha$ level.

For the SHM, the expected DAMA modulation amplitudes for energies over
10~keVee are negligibly small (statistically equal to zero) over the
parameter space examined in this paper.  The DAMA modulation amplitude
data is then expected to randomly vary about zero in the higher energy
bins, which increases the $\chi^2$ and dilutes the power of the
g.o.f.\ test.  To improve the ability of the g.o.f.\ test to exclude
some parameters, we combine all DAMA bins over 10-20~keVee into a
single bin ($0.0011 \pm 0.0008$ cpd/kg/keVee), resulting in a total of
17 data bins used for this test.  The previous two analyses (likelihood
ratio and raster scan) are not affected by the numerous high energy
bins varying about zero and use the full 36 bin data set.
For the 17 d.o.f.\ $\chi^2$ distribution, values of 24.8, 37.7, and 61.6
are excluded at the 90\%, 3$\sigma$, and 5$\sigma$ levels, respectively.

While the g.o.f.\ regions defined here are, in fact, confidence
regions---indicating the likely parameters to produce the DAMA signal
in our theoretical framework---we do not use these regions in that
manner.  The determination of a confidence region \textit{assumes}
that the theoretical framework is correct, \ie\ there exists some
choice of parameters that is correct.  That assumption may not be
valid if, \eg, the standard halo model is not a reasonable
approximation of the actual halo.
Instead, we will more conservatively only use this g.o.f.\ test to
\textit{exclude} parameters outside of the corresponding regions.
That is, parameters outside of the DAMA g.o.f.\ regions are
incompatible with the DAMA signal at the given level.

\subsection{Total Rate}

While we mainly examine the DAMA experiment by analyzing the modulation
amplitude $S_m$, DAMA can additionally constrain parameter space using
the total rate ($S_0$ in \refeqn{dRdEapprox}) as shown in Figure~1 of
Ref.~\cite{Bernabei:2008yi}.
The DAMA detectors do not strongly discriminate between WIMP scatters
and background events, so an unknown and possibly large portion of the
total observed events may be due to backgrounds (the backgrounds are
presumed not to vary with time and do not contribute to the modulation).
In the same manner as the various null experiments, we use the total
number of events observed by DAMA in 0.25~keVee width bins over
2-10~keVee to constrain the parameter space by excluding regions that
would predict more events than observed.
While DAMA shows data for several bins below 2~keVee, we conservatively
use only the data above their 2~keVee threshold as the behavior of
their detector at low energies may not be well understood.
Due to the large number of
observed events (in excess of 10,000 in each bin), statistical
fluctuations are relatively small and any choice of parameters that
predicts an $S_0$ value exceeding the measured rate by more than a few
percent is incompatible with the data to a very high level.

As with the null experiments, we will show regions excluded at the
90\% level due to the total rate for DAMA.  Constraints will be show
with and without the channeling effect.
Note that some choices of parameters that produce the correct
modulation signal may actually predict a total rate larger than
observed and, hence, those parameter choices are incompatible with
the full DAMA data set.

\section{\label{sec:Results} Results}

Here, we apply the statistical techniques described in the previous
section to the DAMA data and make comparisons with the null experiments.
We begin by examining how the increase in DAMA data bins affects the
results, then we examine the various DAMA detector effects, and finally
we compare the various experiments.

\subsection{DAMA Binning}

The DAMA/NaI data was only presented for two independent bins, so our
earlier analyses (performed prior to DAMA/LIBRA) used only one or two
bins.
To demostrate how the increase in the number of DAMA data bins from 2
to 36 has considerably increased the ability to constrain parameter
space, we first examine a 2 bin data set.  Here, we use the 2 bins from
the combined DAMA/NaI + DAMA/LIBRA data set as given in the previous
section (2--4 \& 6-14~keVee); this may be directly compared with the
36 bin DAMA results, which are derived from the same raw data set.  The
bins and the 1$\sigma$ bounds on the amplitudes are shown as blue boxes
in \reffig{DAMAFit}.  The low energy (2--4~keVee) bin has a positive
amplitude, while the high energy (6--14~keVee) bin is consistent with
no modulation.

\begin{figure}
  \insertfig{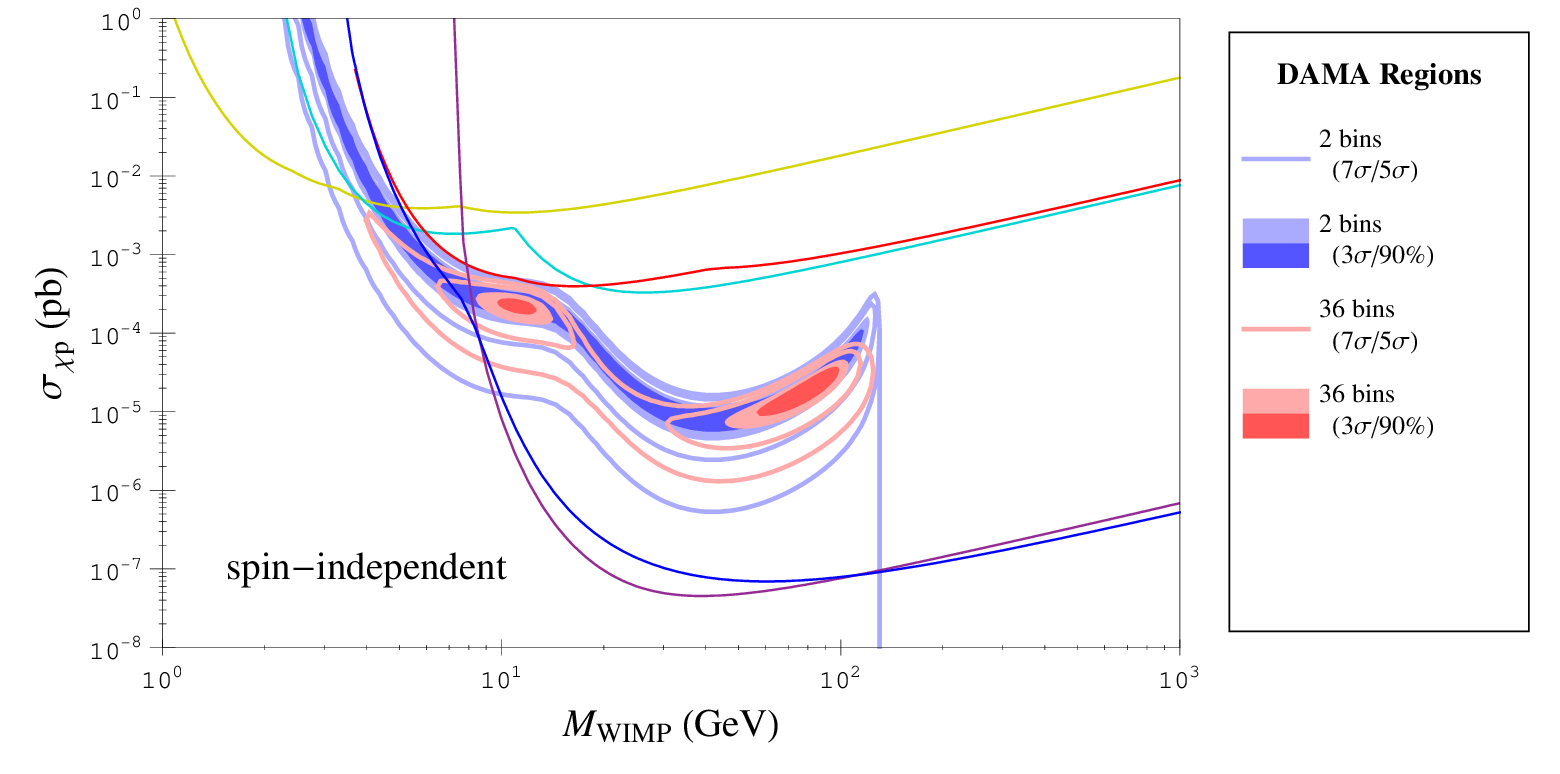}
  \caption{
    Most likely parameters to produce the DAMA signal as determined
    using 2 bins (2--4 \& 6-14~keVee; blue) and 36 bins (pink) for
    the case of SI scattering.  The
    dark/light solid regions correspond to 90\%/3$\sigma$ confidence
    regions; these regions are surrounded by 5$\sigma$ and 7$\sigma$
    contours.  The 2 bin data set provides very little constraint on
    the WIMP mass, while the 36 bin constrains the WIMP mass to two
    regions around 12~GeV and 80~GeV, corresponding to predominantly
    scattering off of Na and I, respectively.
    Null experiment constraints as described in \reffig{SIExpts}
    are shown at 90\% exclusion levels.
    }
  \label{fig:Binning}
\end{figure}

\begin{figure}
  \insertfig{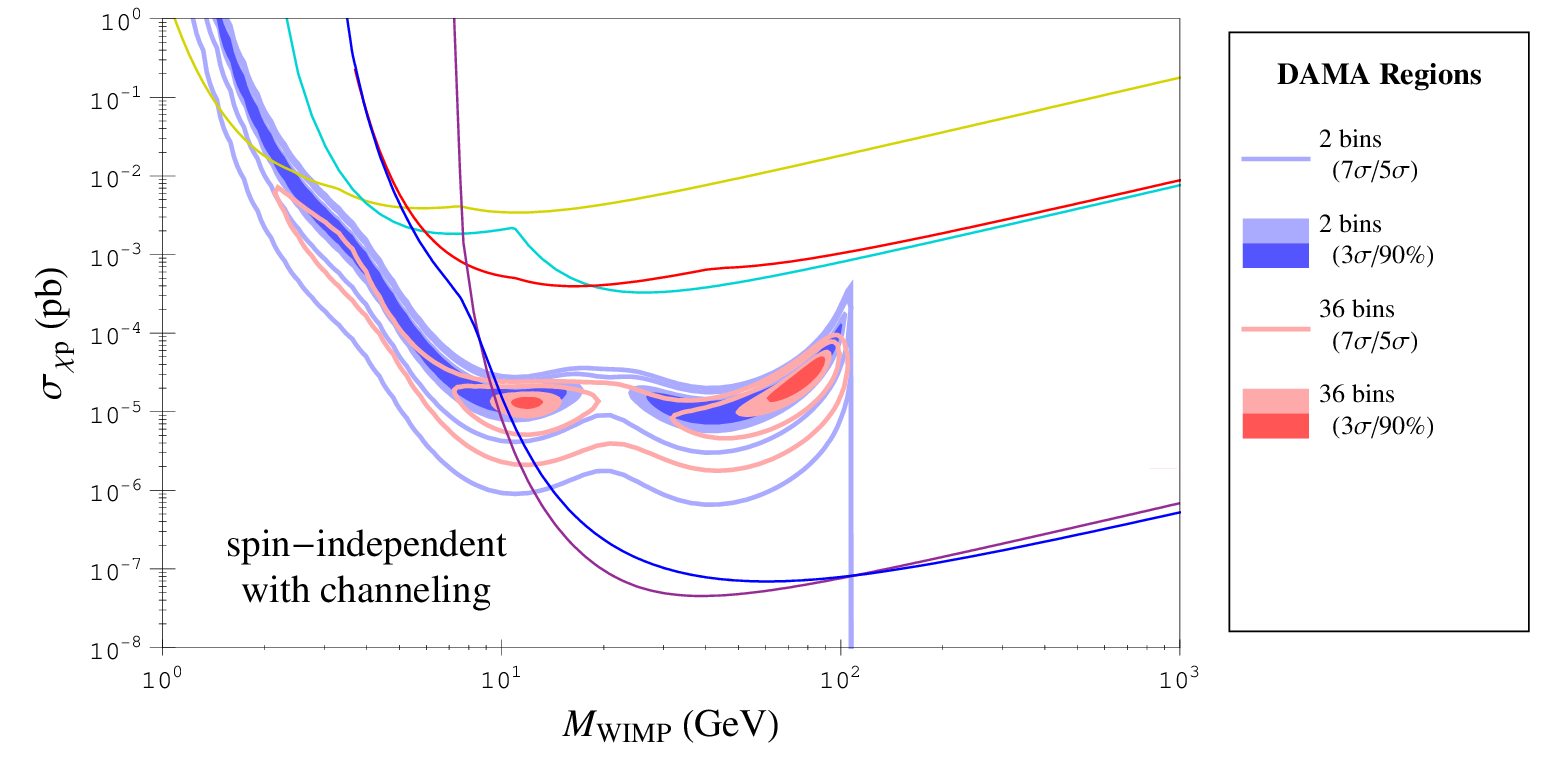}
  \caption{
    Same as \reffig{Binning}, but including the channeling effect
    for DAMA.
    The 36 bin low mass region around 12~GeV is predominantly due to
    scatters off of I that undergo the channeling effect, as opposed
    to scatters off of Na as in the no ion channeling case shown in
    \reffig{Binning}; ordinary Na scatters and channeled I scatters
    coincidentally provide good fits at similar WIMP masses.
    }
  \label{fig:BinningIC}
\end{figure}


In the SHM, there are very few recoils in NaI above 6~keVee, so the
expected modulation amplitude in the high energy bin will nearly
always be small.  In that case, the upper bin provides essentially no
constraint on parameter space and there is effectively only one bin
(the low energy bin) that yields constraints.  Since the amplitude is
proportional to the scattering cross-section, as long as a positive
modulation is predicted for the low energy bin at some WIMP mass,
one can always find a cross-section that yields an excellent fit to
the data.  Only at very low WIMP masses, where the recoils are all
of low energy and fall below the 2~keVee threshold so there is zero
predicted modulation, and at very high WIMP masses, where the
modulation amplitude is negative so that a good fit requires an
unphysical negative cross-section, we are unable to fit the 2 bin data.
Good fits can be found over a broad range of WIMP masses,
$\sim$2--100~GeV, as indicated in \reffig{Binning} and
\reffig{BinningIC}; the latter of which includes the channeling effect.
Effectively, analysis of the 2 bin data requires only a fit to the
\textit{amplitude} of the modulation spectrum, not the
\textit{shape} of the spectrum, and only weakly constrains the WIMP
mass. We used here two data points and four parameters, thus the value
of the minimum $\chi^2$, which is zero for several points in parameter
space, does not provide an indication of the goodness of he fit.

The 36 bin data set breaks the full energy range into much smaller
bins; the positive signal is now found in six separate bins over
2--5~keVee.  To provide a good fit to the data in this case, the
predicted spectra must not only have the correct \textit{amplitude},
it must have the correct \textit{shape} to produce the proper signal
in all six of those bins.  [The higher energy bins are consistent with
zero modulation, but again that is typically expected for the spectra
in the SHM.]  Since the shape of the spectrum depends significantly on
the WIMP mass, the 36 bin data set is able to much better constrain
the WIMP mass than the 2 bin data set, as shown in
Figures~\ref{fig:Binning} \&~\ref{fig:BinningIC}.  The reason for the
difference can more clearly be seen by the spectra in \reffig{DAMAFit}.
At a low WIMP mass of 4~GeV, the (no channeling) spectra at the best fit
SI cross-section is shown in blue for 2 bins, with a $\chi^2$ of 0.25 on
1 d.o.f., and in red for 36 bins, with a $\chi^2$ of 87.0 on 35 d.o.f.;
spectra including the channeling effect are shown in dashed lines with
the same colors (with corresponding $\chi^2$ given in the legend).
At such a low WIMP mass, the spectra are very sharp and drop off
rapidly as light WIMPs are not capable of inducing high energy recoils.
The fit to the 2 bin data is, in fact, nearly a perfect fit
to the low energy bin as the very large amplitude just above the
2~keVee threshold and small amplitude over the rest of the 2--4~keVee
bin gives the correct average modulation over that bin.  On the other
hand, the fit to the 36 bin data is very poor: it overestimates the
modulation amplitude in the first bin, but underestimates it for the
next five.  The \textit{shape} of the spectra required by the 36 bin
data cannot be matched at this WIMP mass.  For comparison, the spectrum
is shown in green for the best fit mass, near 80~GeV, which has a
$\chi^2$ of 27.1 on 34 d.o.f.  This spectrum is a very good fit to
both the 2 bin and 36 bin data sets.

Of note is the fact that the first (lowest energy) bin actually has a
lower amplitude than the next few bins.  At low enough energies, the
modulation actually reverses phase and the amplitude becomes negative.
The lower first bin may be an indication that the DAMA threshold is
just above the energy at which the modulation reduces phase.  Since
that energy is dependent upon the WIMP mass, the lowest bin makes an
important contribution to constraints on the WIMP mass.

For 36 bins, there are two mass regions that allow a good fit to the
data: around 10--15~GeV and 60--100~GeV, as seen in \reffig{Binning}.
In the lower mass region, scattering off of Na produces a spectrum that
matches the data, while in the higher mass region, scattering off of I
produces the appropriate spectrum.  In those two regions, scattering
is also predominantly off of Na and I, respectively.

The 36 bin data will be used for the remainder of the paper.

\subsection{DAMA Detector Effects}

\textit{Energy resolution.}
The finite energy resolution serves to smear out the modulation
spectrum.  It leads to a mild improvement in sensitivity to low
mass WIMPs, but is not as significant an effect as the two discussed
below.  The finite energy resolution is included in all our results.

\textit{Ion channeling (IC).}
The channeling effect reduces DAMA's recoil energy threshold.
Whereas a scattering event with the minimum observed energy of 2~keVee
is conventionally assumed to be the quenched energy of a 6.7~keV recoil
off of Na ($Q_{Na} = 0.3$) or a 22.2~keV recoil off of I
($Q_{I} = 0.09$), it could also be due to a channeled event with a true
recoil energy of 2~keV.  This lower recoil energy threshold provides
greater sensitivity to low mass WIMPs.  This is apparent when comparing
the 2 bin result of \reffig{BinningIC} (channeling) with
\reffig{Binning} (no channeling), as a much smaller cross-section is
sufficient at low masses to produce the necessary signal in the
channeling case.

There are four groups of scatters---off of Na and I, each with or
without channeling---that contribute to the modulation spectrum and
each of those groups will have a best fit to the data at different
regions of parameter space.  Whereas there were two ``preferred''
regions without channeling (\reffig{Binning}), there are now
potentially four.  Predominantly scattering off of Na and I, without
channeling, should again represent a good fit to the data around
10--15~GeV and 60--100~GeV, respectively.  The best fit regions to
channeled scatters off of Na is at even lower masses and is
apparent as the bump in the 2 bin region of \reffig{BinningIC} around
2--4~GeV.  However, as evident by the lack of 36 bin regions (only the
7$\sigma$ contour covers that mass), channeled events off of Na do not
actually provide a reasonably shaped spectrum at these light masses,
even though that is where their ``best'' spot is.  The only other
region is that due to channeled events off of I.  That coincidentally
also occurs around 10--15~GeV, the same as unchanneled Na events.  In
fact, the channeled I events provide a much larger contribution to the
modulation at those WIMP masses than the unchanneled Na events do,
as evidenced by the reduction in the best fit cross-section by over
an order of magnitude from \reffig{Binning} to \reffig{BinningIC}.
Thus, there are two best fit mass regions when channeling is included:
a $\sim$10--15~GeV low mass region due to predominantly channeled
scattering events off of I and a $\sim$60--100~GeV high mass region due
to predominantly unchanneled scattering events off of I.  Though the
best fit masses without including channeling are nearly the same
as when including this effect, that is mainly a coincidence: the low
mass region in the former case is due to scatters off of Na, while the
low mass region in the latter case is due to channeled scatters off of
I.

\textit{Migdal effect.}
We do not include the Migdal effect in our analysis at this
time, but we plan to include it in a future paper.

\subsection{DAMA Analysis Techniques}

\begin{figure}
  \insertfig{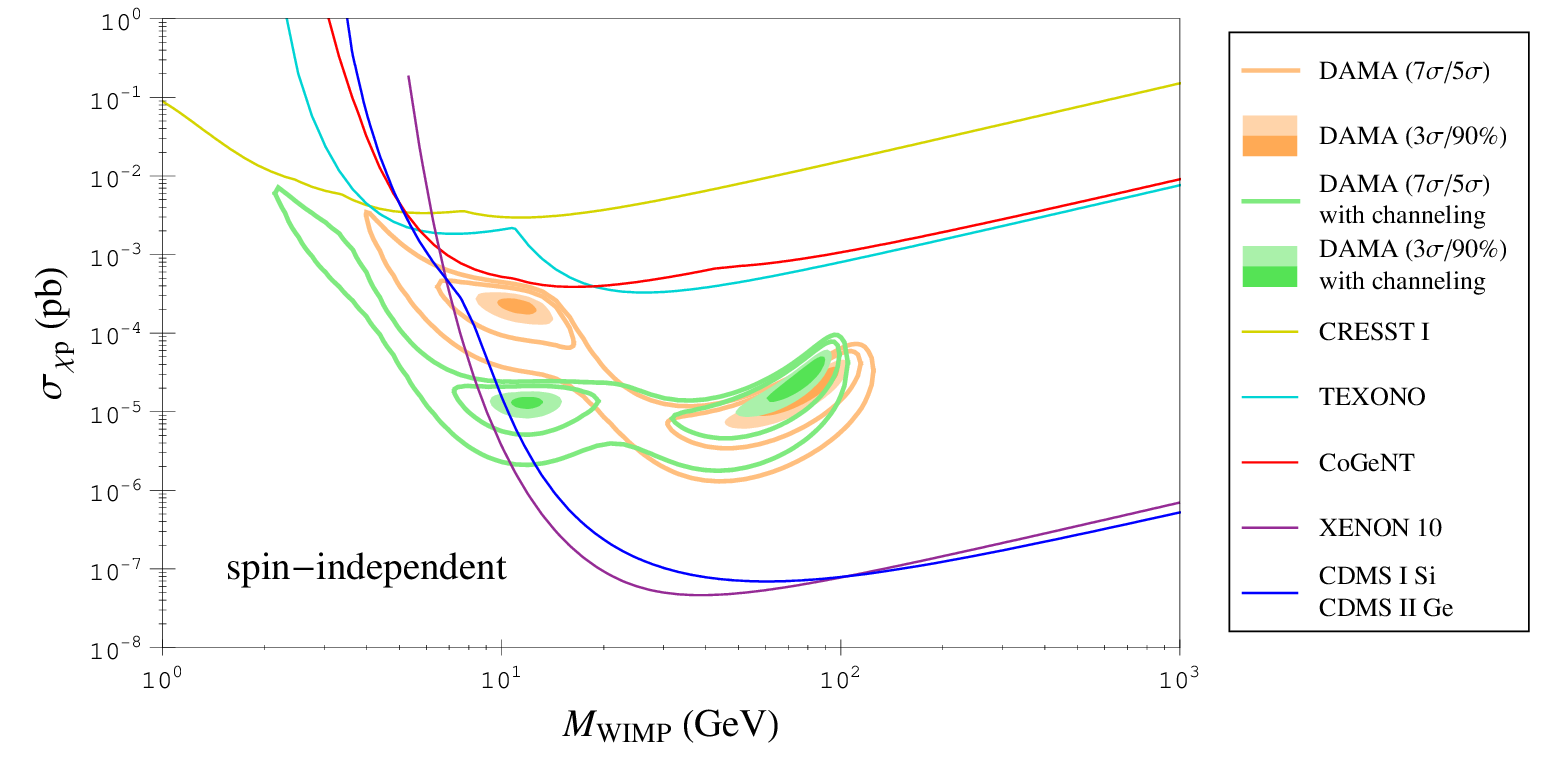}
  \caption{
    Experimental constraints and DAMA best fit parameters for SI
    only scattering.
    The DAMA best fit regions are determined using the likelihood
    ratio method with (green) and without (orange) the channeling
    effect.
    }
  \label{fig:SIp}
\end{figure}

\begin{figure}
  \insertfig{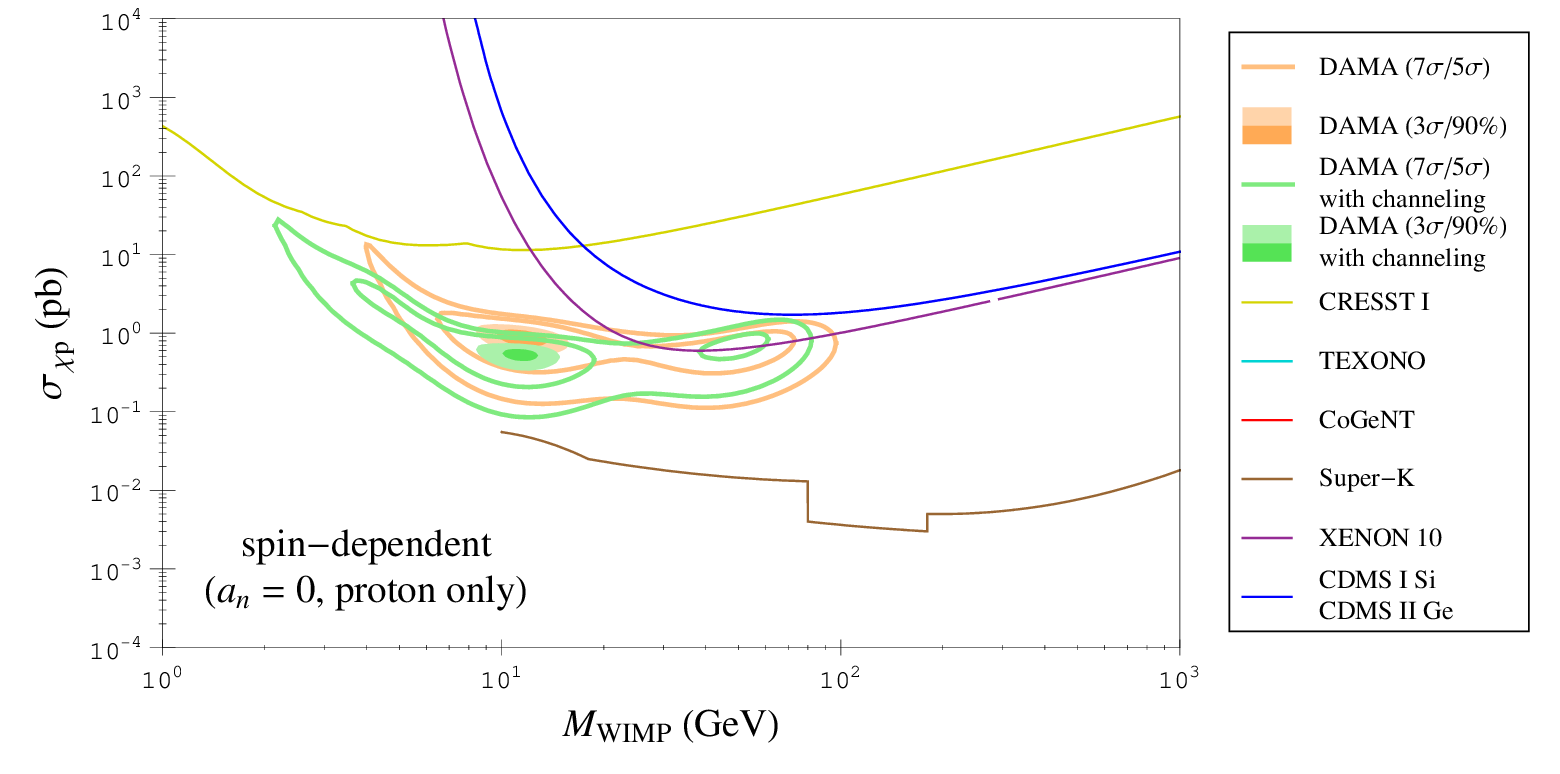}
  \caption{
    Experimental constraints and DAMA best fit parameters for SD
    proton-only scattering.
    The DAMA best fit regions are determined using the likelihood
    ratio method with (green) and without (orange) the channeling
    effect.
    The CoGeNT and TEXONO constraints are too weak to fall within the
    shown region.
    }
  \label{fig:SDp}
\end{figure}

\begin{figure}
  \insertfig{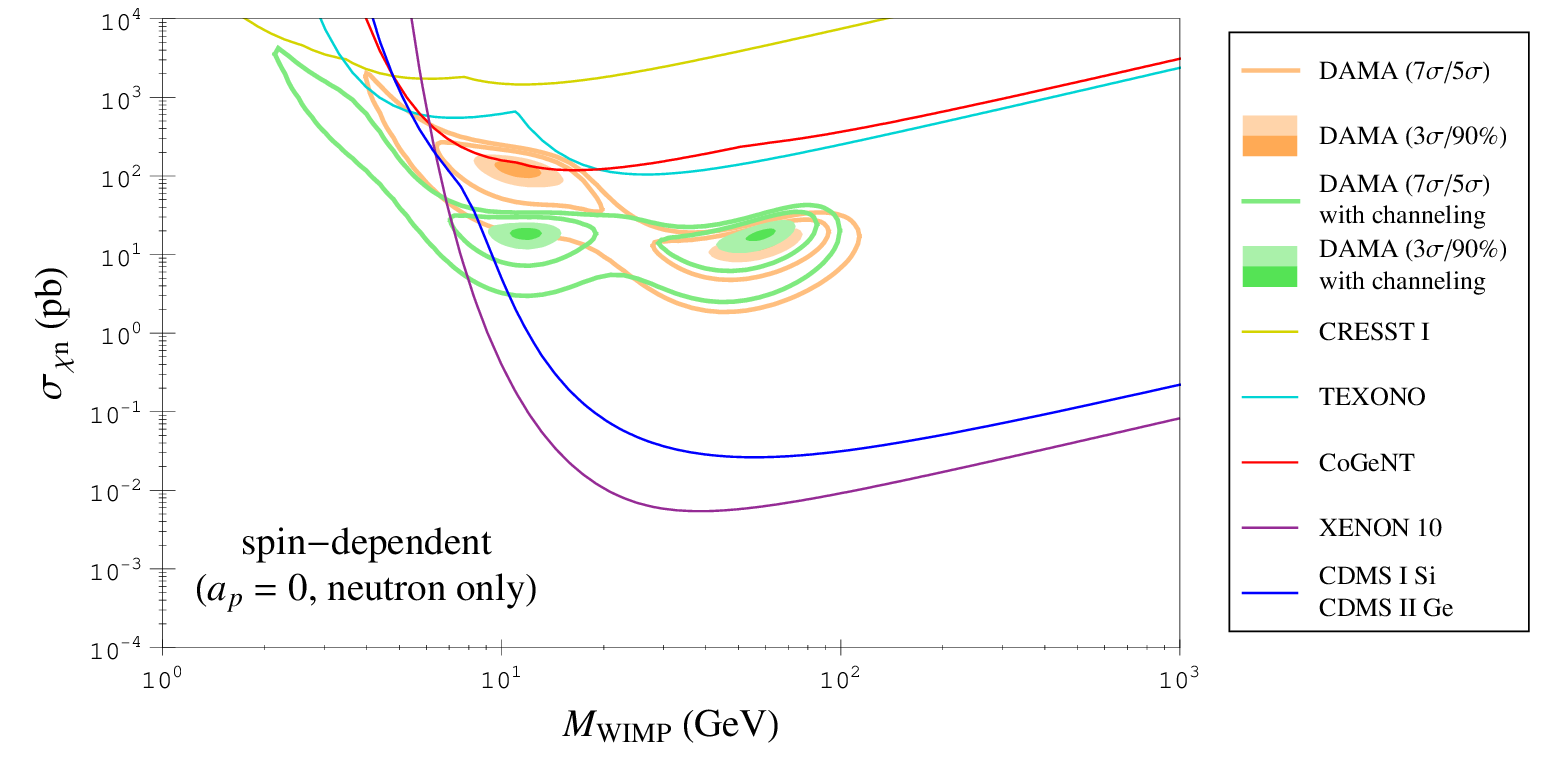}
  \caption{
    Experimental constraints and DAMA best fit parameters for SD
    neutron-only scattering.
    The DAMA best fit regions are determined using the likelihood
    ratio method with (green) and without (orange) the channeling
    effect.
    Super-Kamiokande (SuperK) provides no constraints in this case and
    is not shown.
    }
  \label{fig:SDn}
\end{figure}

The likelihood ratio analysis of DAMA yields a 4-dimensional confidence
region over the ($m,\sigmapSI,\apSD,\anSD$) parameter space.
In Figures~\ref{fig:SIp}, \ref{fig:SDp}, and~\ref{fig:SDn}, we show
the SI only ($\apSD = \anSD = 0$), SD proton-only
($\sigmapSI = \anSD = 0$), and SD neutron-only ($\sigmapSI = \apSD = 0$)
slices of this confidence region, respectively, with (green) and
without (orange) the channeling effect.  Later, we will show slices
in other, mixed coupling cases.  This analysis
indicates the parameters most likely to produce the DAMA signal.  If
some 2-dimensional plane in the full parameter space allows for only
a poor fit relative to the best fit point (the global $\chi^2$ minimum),
there will be no slice in that plane.  The fact that we have found
slices in all of the SI only, SD proton-only, SD neutron-only, and
mixed SD coupling cases is an indication that each of these cases is a
relatively reasonable possibility for producing the DAMA signal.  That
is, even though we allow for mixed SI and SD couplings, the DAMA signal
can still reasonably be produced by SI only scattering; the same may
be said of the other coupling possibilities.  One may see this from
the minimum $\chi^2$ given in \reftab{DAMAMin}: although the SI only
case provides a mildly lower $\chisqmin$ (27.1 without
channeling) than that the SD cases (29.3--31.0), they are all very
similar to the global minimum (27.1).

By finding a 4-dimensional confidence region, direct comparisons and
(to some degree) interpolations may be made between the DAMA regions
of different coupling cases.
Alternatively, we could have considered each case to be its own
2 parameter model in ($m$,$\sigma$), where $\sigma$ represents the type
of cross-section in that case (\eg\ $\sigmapSI$), and found the
corresponding 2-dimensional confidence regions.
However, each analysis would be based on a different theoretical
framework and would not be statistically correlated; one must then be
cautious when making comparisons between, \eg, the resulting SD
proton-only and SD neutron-only confidence regions.  In theory, such
2-dimensional confidence regions may be very different from the slices
of the 4-dimensional confidence region.  In practice, for the coupling
cases considered here, they turn out to be somewhat similar, which is a
consequence of the similar $\chisqmin$ in the different cases.

\begin{figure}
  \insertfig{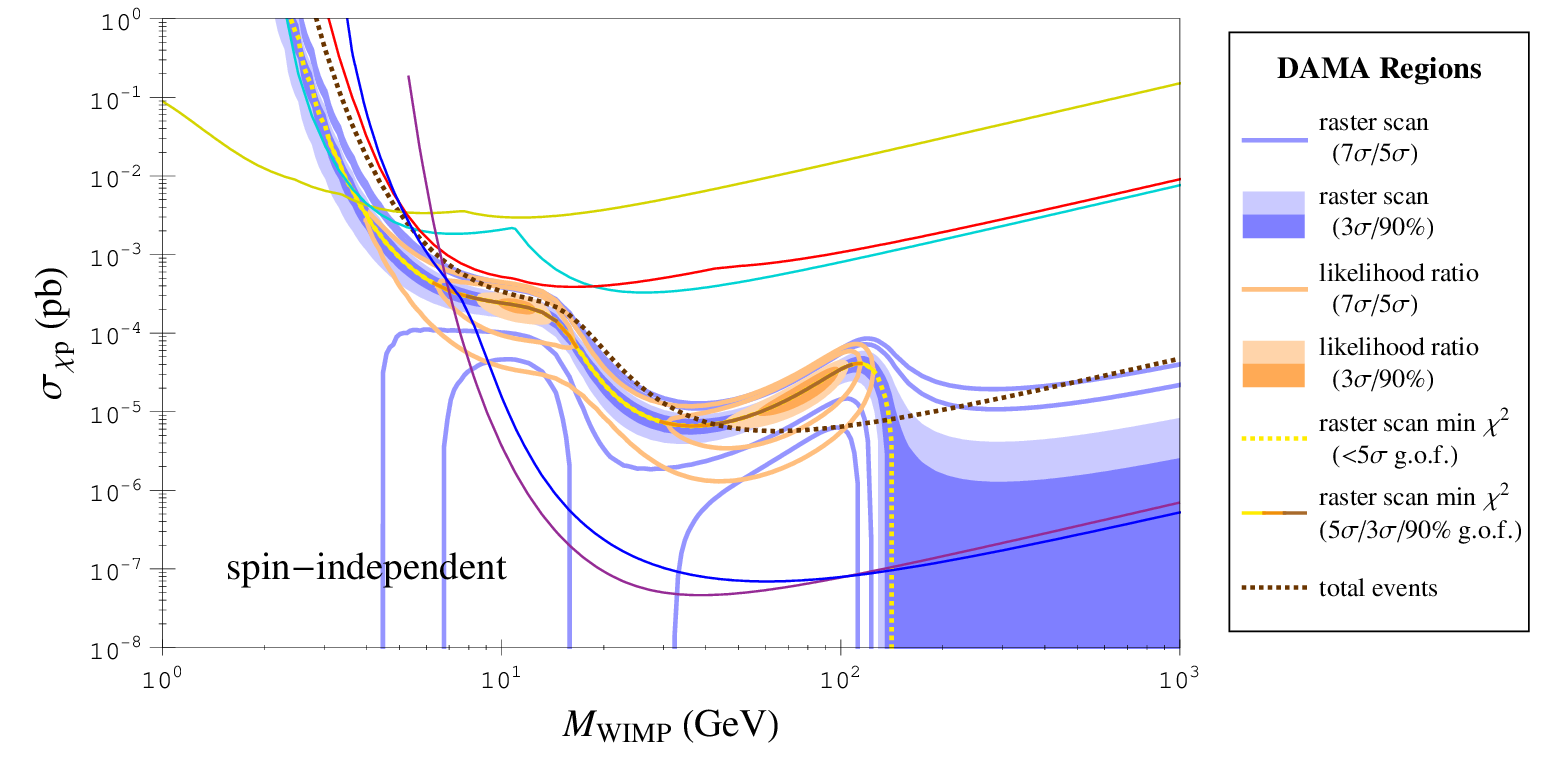}
  \caption{
    Two statistical analyses of DAMA (best fit and raster scan) vs.\
    null results of other experiments for SI scattering.
    DAMA best fit regions (likelihood analysis) are shown by the orange
    regions (90\%/3$\sigma$) and contours (5$\sigma$/7$\sigma$).
    [Technically, these regions are the 2D $\apSD = \anSD = 0$ slices
    of the full 4D confidence region as described in the text.]
    The blue bands and contours are raster scans over the WIMP mass,
    corresponding to the most likely cross-sections at each WIMP mass
    The data is not used to constrain the WIMP mass in a raster scan
    analysis; however, the brown/orange/yellow line shows the
    cross-section that minimizes the $\chi^2$ (best fit cross-section)
    at each mass, with the color indicating a
    goodness-of-fit within the 5$\sigma$/3$\sigma$/90\% level at those
    cross-sections (dotted yellow indicates a goodness-of-fit below
    5$\sigma$, a very poor fit).
    Null experiment constraints are as described in \reffig{SIp}.
    The constraint imposed by the total event rate in DAMA
    (as opposed to the modulation) is shown by the dark dotted curve.
    The likelihood ratio analysis is decoupled from a goodness-of-fit
    test and the lack of an overlap between the DAMA best fit regions
    (as determined by the likelihood ratio) and the allowed regions
    of other experiments does not, by itself, indicate an
    incompatibility between the experiments.
    Such comparisons, however, may be made using the $\chi^2$
    goodness-of-fit contours shown in \reffig{SIpMethodGOF}.
    The constraint due to the total event rate in DAMA (dotted
    line) shows that some parameters that yield the correct modulation
    amplitude predict too many total events and are therefore excluded.
    }
  \label{fig:SIpMethodRS}
\end{figure}

\begin{figure}
  \insertfig{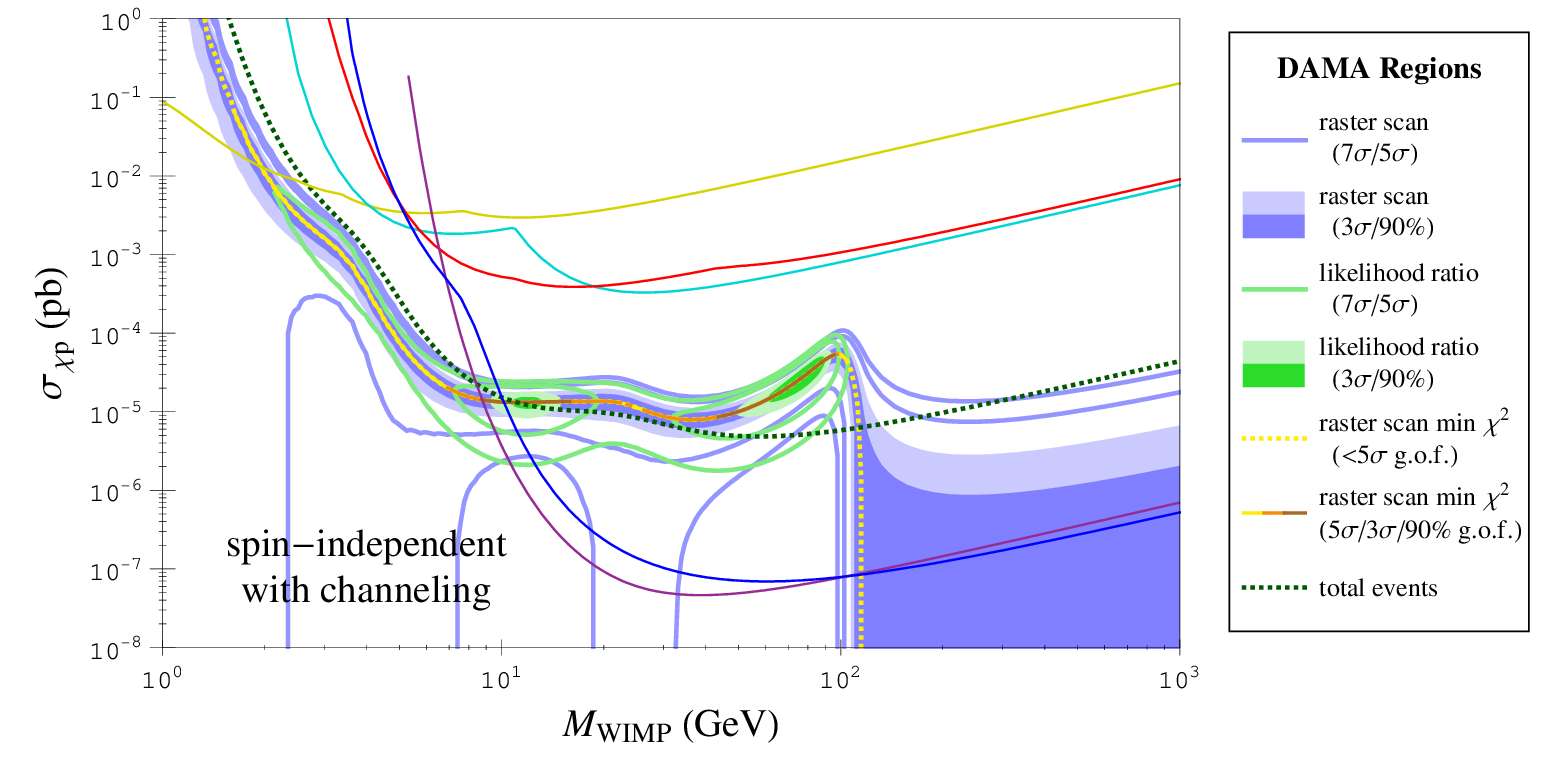}
  \caption{
    Same as \reffig{SIpMethodRS}, but including the channeling effect
    for DAMA.
    }
  \label{fig:SIpMethodICRS}
\end{figure}

As discussed previously, there are generally two preferred WIMP mass
ranges: a low mass range $\sim$10--15~GeV (due to scatters primarily
off of Na or channeled I scatters, when channeling is included) and
a high mass range $\sim$60--100~GeV (due to scatters primarily off
of I).  Low, high, and intermediate WIMP masses generally produce worse
fits (though not necessarily \textit{bad} fits).  Previous analyses
have found low WIMP masses, below these preferred ranges, to be of
interest.  In addition, the LHC or other (non-direct dectection)
experiments, if they observe \eg\ neutralinos, might be able to provide
the mass of the WIMP.  For these cases, it is useful to use the raster
scan, which does not use the data to make assumptions on the WIMP mass
and provides appropriate confidence intervals in the scattering
cross-section for a given (known) WIMP mass.

The raster scan is shown in \reffig{SIpMethodRS} (without channeling)
and \reffig{SIpMethodICRS} (with channeling) for SI only scattering.
Regions shown in dark/light blue are the confidence regions at
90\%/3$\sigma$ C.L., while the light blue contours indicate the
confidence regions at a 5$\sigma$/7$\sigma$ C.L. (the outer contour is
7$\sigma$).  For comparison, the DAMA likelihood ratio regions and
null experiment constraints are also shown; these regions and
constraints are equivalent to the ones shown in \reffig{SIp}.
The raster scan regions pass through the likelihood ratio regions
(as should be expected).  At masses where the raster scan
regions/contours extend to the bottom of the figure
(7$\sigma$/5$\sigma$ at low masses, 7$\sigma$ at intermediate masses,
and for all C.L.'s at high masses), the regions actually extend all the
way to a zero cross-section, \ie\ a cross-section of zero falls
within the corresponding C.L.\ at those masses.

The brown/orange/yellow line along the raster scan indicates the best
fit cross-sections (minimizing the $\chi^2$) at each WIMP mass.  At a
particular mass, the reasonableness of the best fit cross-section (and
corresponding confidence interval) may be tested against the $\chi^2$
distribution with 35 d.o.f.\ (36 bins with a fit to one parameter):
the brown/orange/yellow color of the line indicates the best fit
cross-section at that mass falls within a 90\%/3$\sigma$/5$\sigma$
probability level (dotted yellow is below 5$\sigma$).  Changes in the
color often occur near one of the likelihood ratio contours; this is
only coincidental as the two types of regions use different statistical
methods and are not in direct correspondence.

The most appropriate use of the raster scan is if one has an interest
in a particular WIMP mass (or specific range of WIMP masses): the
raster scan indicates the confidence interval (most preferred or best fit values)
in the cross-section at that mass and the color of the central line
indicates the goodness-of-fit of that particular model/interval
\footnote{A technical point: the raster scan confidence region is the
    band of confidence intervals over \textit{all} masses.  The
    goodness-of-fit test at the best fit cross-section at a given
    mass is only a test of the confidence interval at that mass.  If
    this test is used to ``cut'' the band at some WIMP masses,
    the resulting region would not formally be a confidence region.
    }.
If there is no interest in specific WIMP masses, the likelihood ratio
method should be used instead.

The regions determined using the likelihood ratio and raster scan
methods indicate the best fit parameters for producing the DAMA
signal, assuming that signal is actually produced by some particular
set of parameters in the assumed theoretical framework.  These
analyses, however, are decoupled from the goodness-of-fit: the
resulting regions do not indicate if those parameters are a good fit
to the data or a poor one (which may indicate the assumed theoretical
framework,\eg\ the SHM, is incorrect).
That is, the best fit parameter regions are not equivalent to the
regions that provide a good fit to the data.
Some parameters that fall outside of the best fit regions may actually
provide reasonably good fits to the data (just not as good as
parameters within the region), while some parameters inside the best fit
regions may provide poor fits to the data (just not as bad as
parameters outside the region).  In the likelihood ratio method, only
one of these two cases is possible, depending upon the value of
$\chisqmin$, \ie\ there will be either good fits outside the region or
bad fits within, but not both.

Because these two regions indicate best fit parameters, but not
the ``allowed'' parameters, they should not be used to determine the
compatibility with other experiments.  For example, in the SI case
with channeling shown in \reffig{SIp}, the XENON10 constraint
essentially rules out the 4$\sigma$ DAMA likelihood ratio best fit
parameter region (not shown, but falls midway between the 3$\sigma$
and 5$\sigma$ regions).  However, at about $m$ = 8~GeV and $\sigmapSI$
= 10$^{-5}$~pb,  the predicted recoil and modulation spectra are
compatible with XENON10 and DAMA at about the
90\% and 2$\sigma$ levels, respectively,
using a $\chi^2$ distribution for the latter.

\begin{figure}
  \insertfig{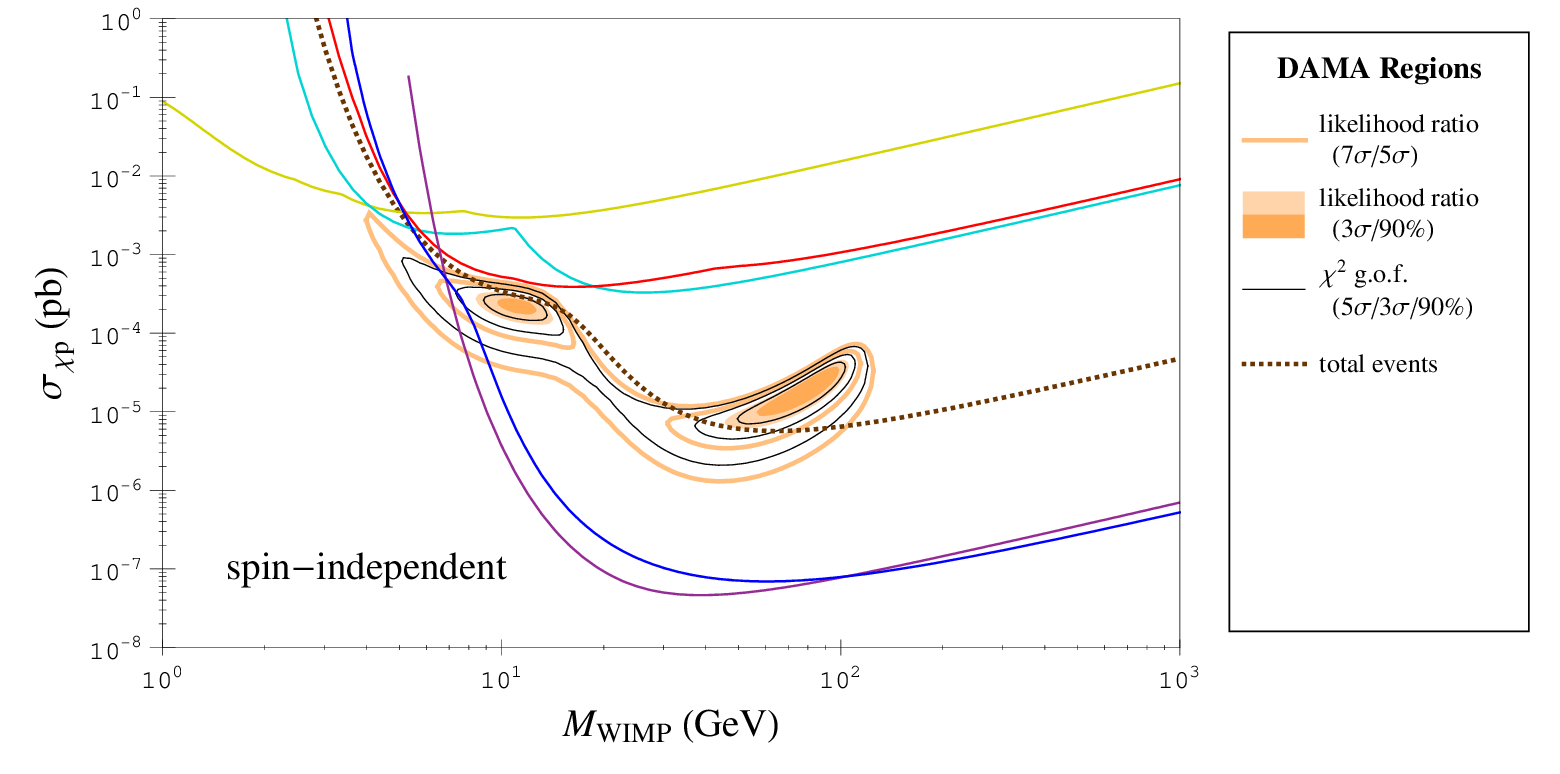}
  \caption{
    Two statistical analyses of DAMA (best fit and goodness-of-fit
    (g.o.f.)) vs.\ null results of other expeiments for SI scattering.
    The best comparison with other experiments is via the g.o.f.\ 
    regions shown here.
    The most likely parameters to produce the DAMA signal, as determined
    using the likelihood ratio method, are indicated by the orange
    regions (90\%/3$\sigma$) and contours (5$\sigma$/7$\sigma$).
    The black contours correspond to a conservative $\chi^2$
    g.o.f. test to the DAMA data.
    The constraint imposed by the total event rate in DAMA
    is shown by the dotted curve.
    Null experiment constraints are as described in \reffig{SIp}.
    Both DAMA regions shown are based on the $\chi^2$ value and thus
    have similar contours shapes, but they have different $\chi^2$
    cutoffs at a given confidence level and are not equivalent;
    \eg\ the 5$\sigma$ likelihood ratio contour is nearly the same as
    the 3$\sigma$ g.o.f.\ contour.
    Direct comparisons between DAMA and the other experiments should
    be done using the $\chi^2$ g.o.f.\ regions, not the likelihood
    ratio regions.
    }
  \label{fig:SIpMethodGOF}
\end{figure}

\begin{figure}
  \insertfig{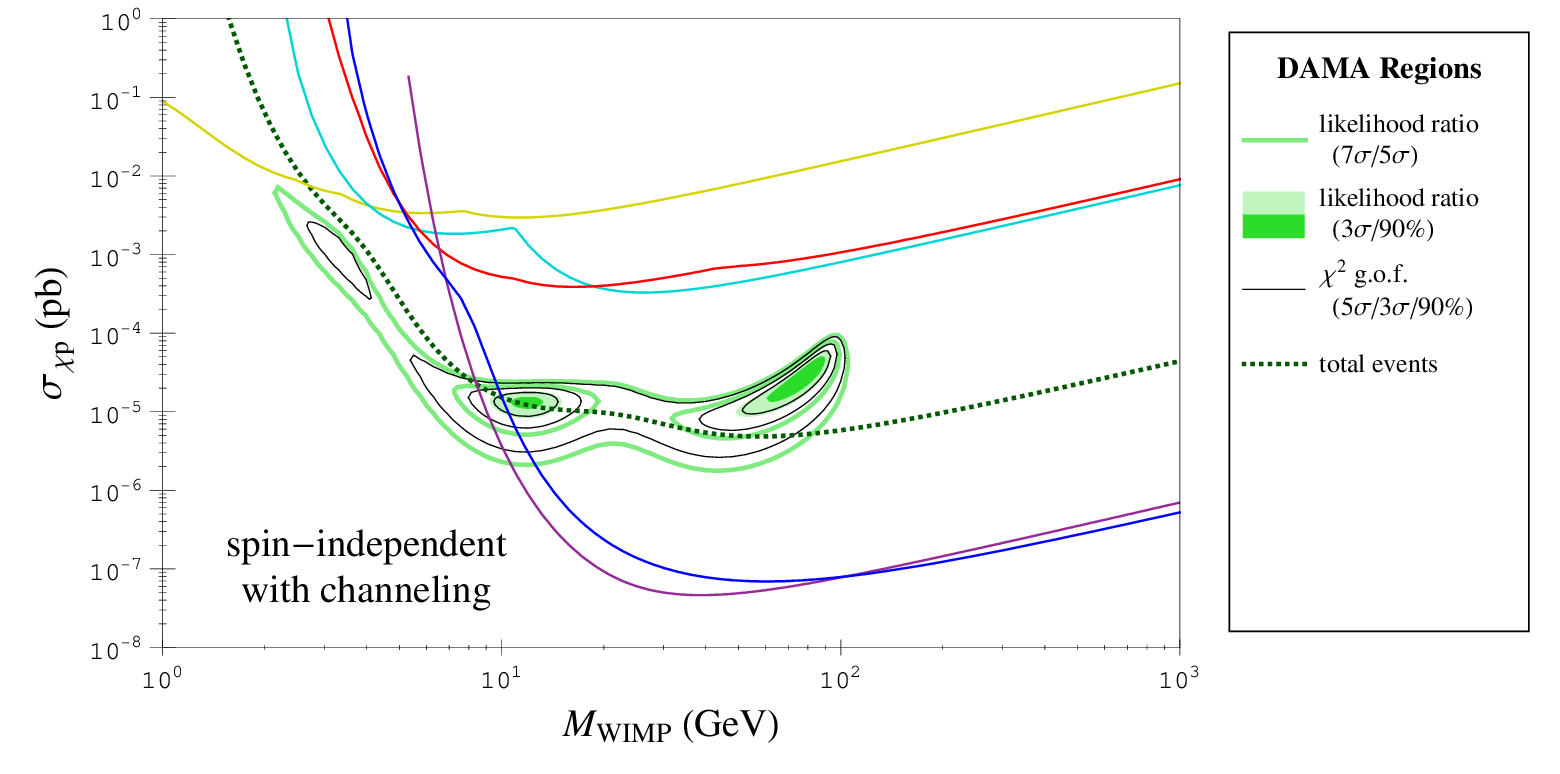}
  \caption{
    Same as \reffig{SIpMethodGOF}, but including the channeling effect
    for DAMA.
    }
  \label{fig:SIpMethodICGOF}
\end{figure}

The best way of determining the compatibility of the various
experiments is to do a combined analysis, such as a likelihood ratio
analysis, that takes into account all the available experimental
data \footnote{The best fit parameter region determined via the
    likelihood ratio method may shift significantly with these other
    experimental constraints added, yet still provide a reasonable fit
    to the data.}.
However, this is fairly difficult to do due to the wide range of
analyses used by the various experiments and the lack of background
subtraction makes it even more difficult (the likelihood is ill defined
in this case).  Rather, we will more conservatively compare experiments
using a $\chi^2$ goodness-of-fit (g.o.f.) test for DAMA for each set of
parameters (there is no fit to the parameters, so we use a $\chi^2$
distribution with 17 d.o.f., recalling the high energy bins have been
combined for the g.o.f.\ test).  The $\chi^2$ g.o.f.\ regions are shown
as black contours in \reffig{SIpMethodGOF} (without channeling) and
\reffig{SIpMethodICGOF} (with channeling) where the DAMA data is
compatible with the predicted spectrum at the 90\%, 3$\sigma$, and
5$\sigma$ level.  Parameters outside those contours are incompatible
with the data at at least the corresponding level; these contours
are constraints on the parameter space.
These contours are appropriate for comparing DAMA with the other
experiments since, if a set of parameters provides
a poor fit to the data of one experiment, including DAMA, it almost
certainly will be a poor fit in a combined analysis as well.
These $\chi^2$ g.o.f.\ contours are a conservative means of examining
the compatibility as a combined analysis could possibly provide
stronger constraints, but is unlikely to provide more lenient
constraints.

As they are both determined from the $\chi^2$ of the data, the contours
for the likelihood ratio and g.o.f.\ methods are similar in shape, as
can be seen in Figures~\ref{fig:SIpMethodGOF} \&~\ref{fig:SIpMethodICGOF}.
However, the $\chi^2$ corresponding to a given C.L.\ is different in
the two methods.
In particular, while some contours of the likelihood ratio analysis
seem to closely match contours of the $\chi^2$ g.o.f.\ analysis, they
are not of the same level.  For example, the 5$\sigma$ likelihood ratio
contour is very similar to the 3$\sigma$ g.o.f.\ contour.

To summarize, the choice of the DAMA analysis method to use should be
the following:
if the most likely parameters to produce the DAMA signal are desired,
use the likelihood ratio;
if interested in a \textit{particular} WIMP mass or range of masses,
use the raster scan; and
if direct comparisons (particularly, the compatibility) are to be made
with other experiments, use the $\chi^2$ g.o.f.\ analysis.

\subsection{DAMA Total Rate}

The total event rate in DAMA imposes constraints on the DAMA compatible
parameter space; this total rate constraint is shown as a dotted curve
in \reffig{SIpMethodRS} and the following figures.
Parts of parameter space that would produce the DAMA modulation signal,
such as the higher WIMP mass regions for the SI case shown in
\reffig{SIpMethodGOF} (without channeling) and
\reffig{SIpMethodICGOF} (with channeling), predict far more total
events in the DAMA detector than observed and are therefore ruled out
as an explanation for the DAMA signal.

In general, the total rate provides a strong constraint for the
higher WIMP mass region that is favorable for the modulation signal
in the various coupling cases, excluding the 90\% (without channeling)
and 3$\sigma$ (with channeling) g.o.f.\ regions.  Thus, high mass
WIMPs are disfavored as the source of the DAMA signal for the models
considered here.
The lower WIMP mass regions, however, are only minimally constrained
by the total rate in DAMA.

\subsection{Spin-Independent Scattering}

The DAMA best fit parameters and null experiment constraints for SI
only scattering are shown \reffig{SIp} both with (green) and without
(orange) channeling for DAMA.  Channeling greatly reduces the scattering
cross-sections necessary to produce the DAMA signal in the lower WIMP
mass region ($\sim$9--15~GeV) and opens up even lower mass regions
(down to $\sim$2~GeV), albeit only at a very low 7$\sigma$ C.L.
CDMS and XENON10 exclude all the 3$\sigma$ C.L.\ best fit regions.
The low exposure experiments---CoGenT, CRESST~I, and TEXONO---do not
quite constrain the DAMA best fit regions.  Both our older analyses
and those performed by some of the experimental groups indicate these
experiments \textit{do} constrain portions of the DAMA region, but
there are several differences between those analyses and our analysis
here that change the DAMA best fit region (aside from any potential
conservative assumptions on the experimental data sets as described
in \refsec{Experiments}):
(1) an increase in the number of DAMA data bins from 2 to 36;
(2) inclusion of DAMA's finite energy resolution, which increases the
number of scatters over threshold and reduces the necessary
cross-section, particularly at lower WIMP masses;
(3) inclusion of the channeling effect, where indicated; and
(4) no background subtraction.
Aside from shifting the DAMA best fit region downward, these changes
lead to the very low mass regions disappearing from the best fit
confidence regions, where the low threshold experiments provide the
strongest constraints.
SuperK provides relatively weak constraints for this SI case.

The best fit SI cross-sections at low WIMP masses are indicated
by the raster scan regions in \reffig{SIpMethodRS} (without channeling)
and \reffig{SIpMethodICRS} (with channeling).  Without channeling,
CRESST~I and, to some degree, TEXONO constrain the 3$\sigma$ best fit
cross-sections at masses below about 3~GeV.  The best fit
cross-section at these masses presents a poor fit to the DAMA data:
they are incompatible at at least the 5$\sigma$ level.
With channeling included, the best fit cross-sections shift downwards
at these low masses and are no longer constrained by TEXONO.
CRESST~I excludes the best fit cross-sections only below $\sim$1.5~GeV
in this case.
At high WIMP masses (above $\sim$100~GeV), with and without channeling,
the best fit cross-section is actually zero.  The reason for this is
that, for heavy WIMPs, the lowest energy bins will actually have a
negative modulation amplitude, whereas the observed amplitude is
positive.  The best fit cross-section would then be an unphysical
negative value; the best physically allowed value is zero.  As at
very low masses, the best fit cross-sections at these high masses
represent poor fits to the DAMA data (incompatible at $>5\sigma$).

To examine the compatibility of DAMA with other experiments, we use
the $\chi^2$ g.o.f.\ regions shown in \reffig{SIpMethodGOF} (without
channeling) and \reffig{SIpMethodICGOF} (with channeling).  Without
channeling, CDMS and XENON10 exclude all the parameter space that is
compatible with the DAMA data within the 3$\sigma$ level.
With
channeling, CDMS and XENON10 exclude all the parameter space that is
compatible with the DAMA data within the 90\% level.
There are
parameters at 8--9~GeV that are compatible with DAMA within the
3$\sigma$ level and still satisfy the other constraints.
Lower masses, while not constrained by
the null experiments or DAMA's total rate, are only compatible with
the DAMA data below the 3$\sigma$ level.

We note that these results are sensitive to the thresholds of the
two large exposure experiments, CDMS and XENON10.  In particular, the
DAMA 3$\sigma$ g.o.f.\ region with channeling would have been excluded
with the use of the
original $\mathcal{L}_{\mathrm{eff}}$ value of 0.19 for XENON10,
instead of the more recent measurement of 0.14 (which shifted the
XENON10 threshold from 4.6~keV to 6.1~keV).
The small window of compatibility at low WIMP masses is thus very
dependent on the calibrations of these experiments' detectors.
Given the closeness of the upward turn-off in the CDMS and XENON10
constraints (due to their thresholds) with the low mass DAMA region,
even small adjustments to the calibrations, which will shift the
turn-off in the constraints to slightly higher (or lower) WIMP masses,
may open up more of the DAMA low mass region (or exclude it completely).

\subsection{Spin-Dependent Scattering}

We here present results for the case of purely spin-dependent
interactions.  We study the cases of scattering off of protons only,
neutrons only, and mixed proton \& neutron scattering.

\subsubsection{Proton-Only}

\begin{figure}
  \insertfig{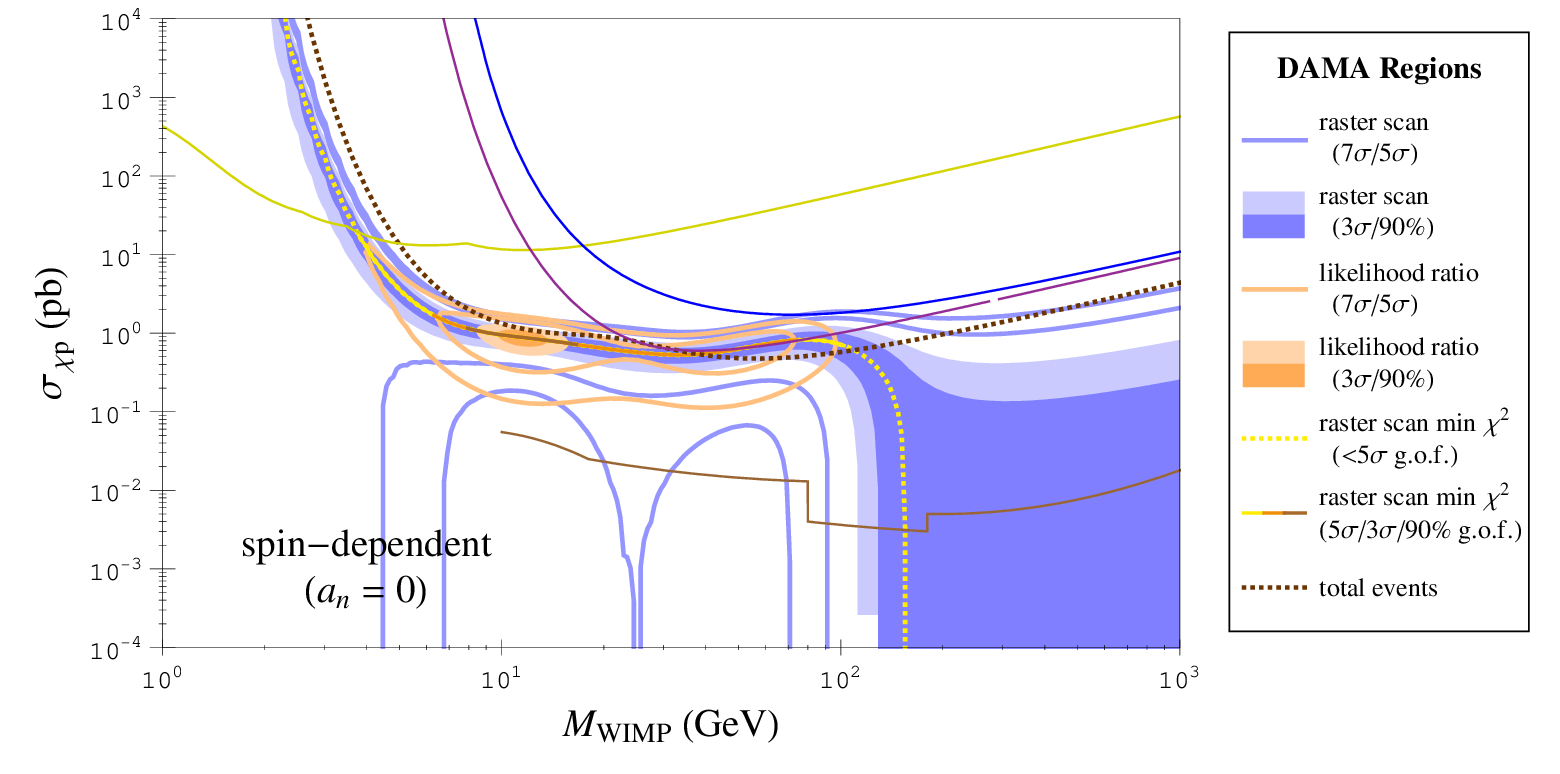}
  \caption{
    Experimental constraints and DAMA best fit parameters for
    spin-dependent (SD) proton-only scattering.
    The DAMA raster scan and likelihood ratio regions are as described
    in \reffig{SIpMethodRS}.
    Null experiment constraints are as given in \reffig{SDp}.
    }
  \label{fig:SDpMethodRS}
\end{figure}

\begin{figure}
  \insertfig{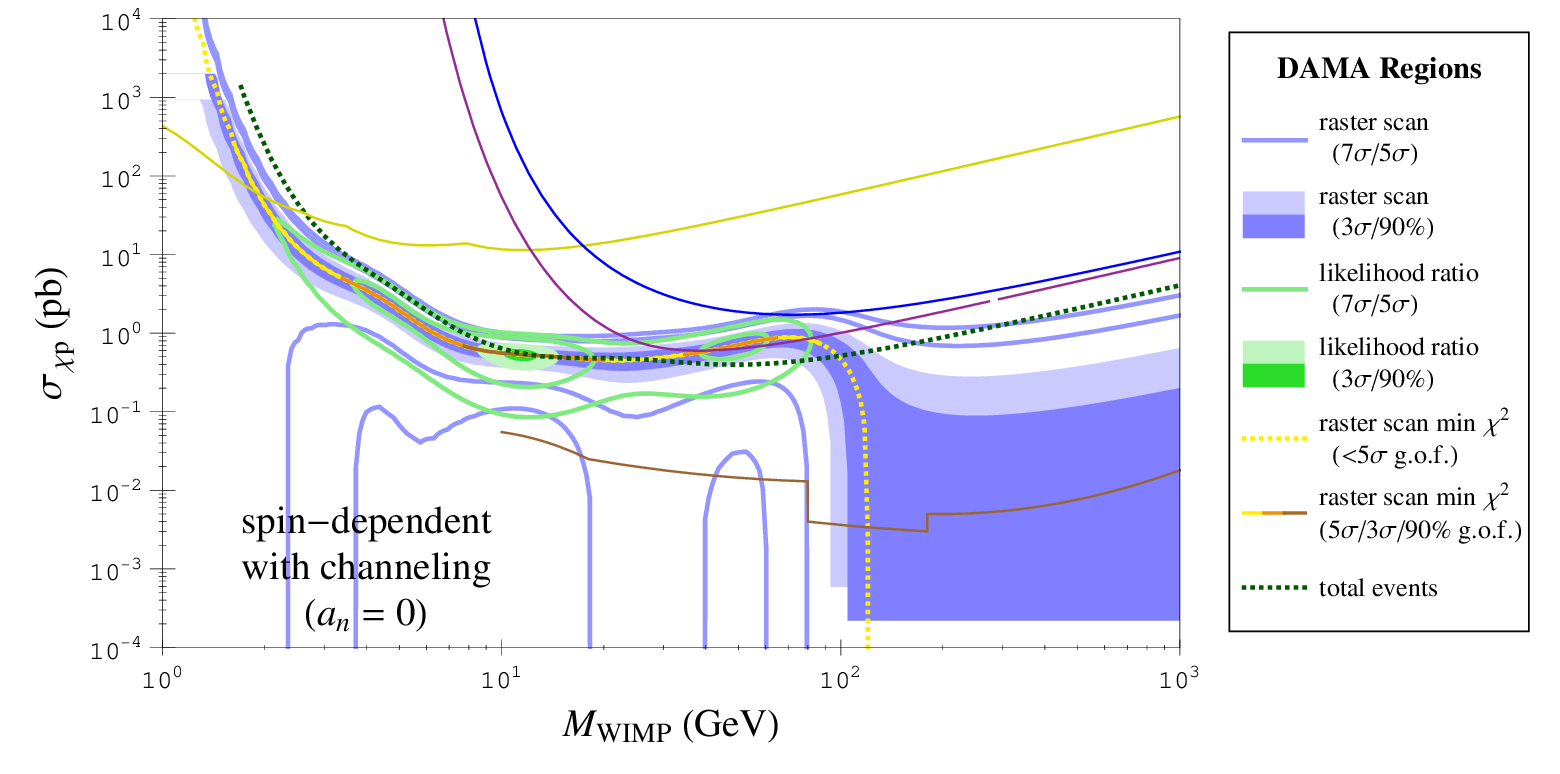}
  \caption{
    Same as \reffig{SDpMethodRS}, but including the channeling effect
    for DAMA.
    }
  \label{fig:SDpMethodICRS}
\end{figure}

\begin{figure}
  \insertfig{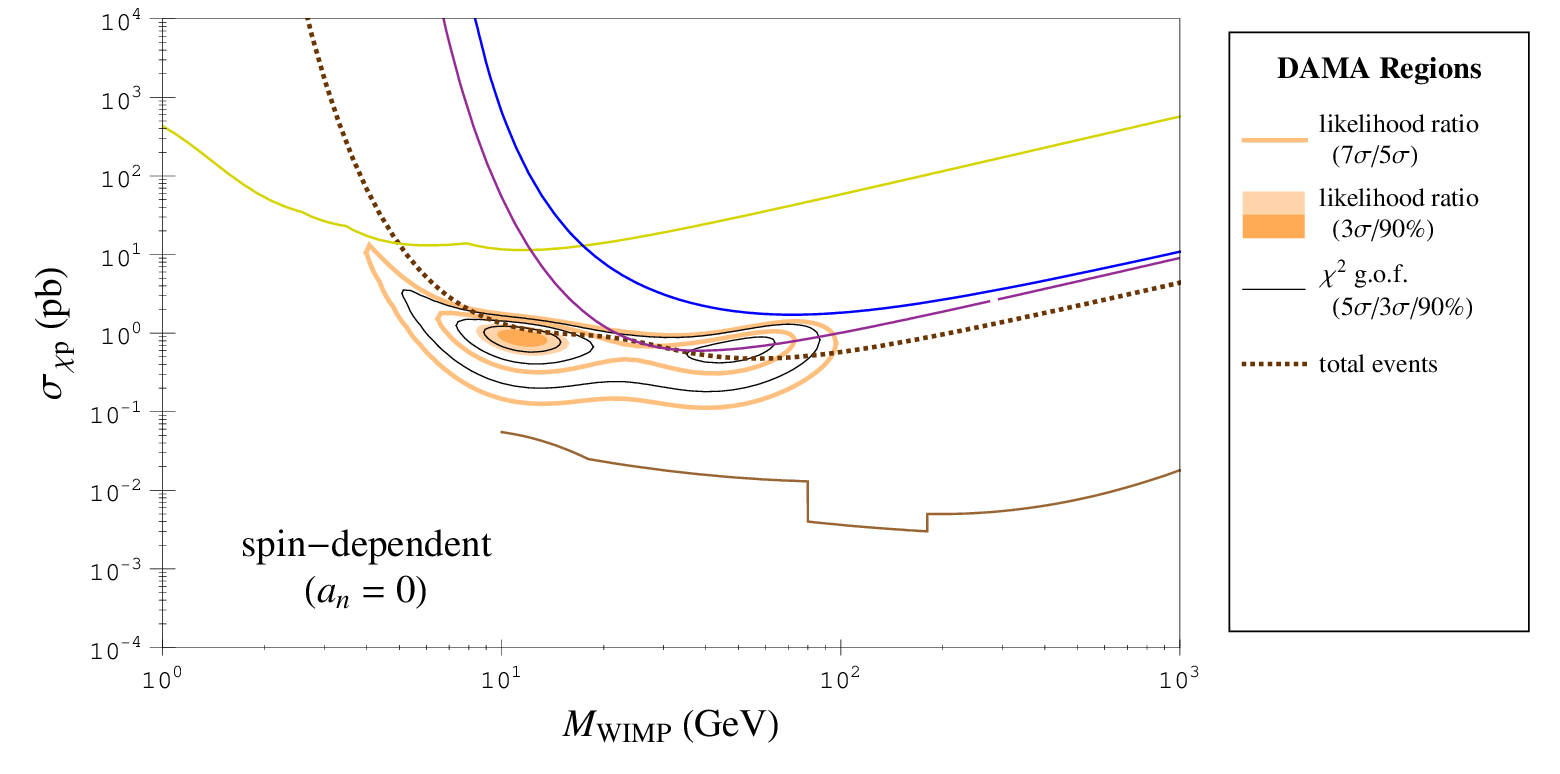}
  \caption{
    Experimental constraints and DAMA best fit parameters for
    spin-dependent (SD) proton-only scattering.
    The DAMA raster scan and $\chi^2$ g.o.f.\ regions are as described
    in \reffig{SIpMethodGOF}.
    Null experiment constraints are as given in \reffig{SDp}.
    }
  \label{fig:SDpMethodGOF}
\end{figure}

\begin{figure}
  \insertfig{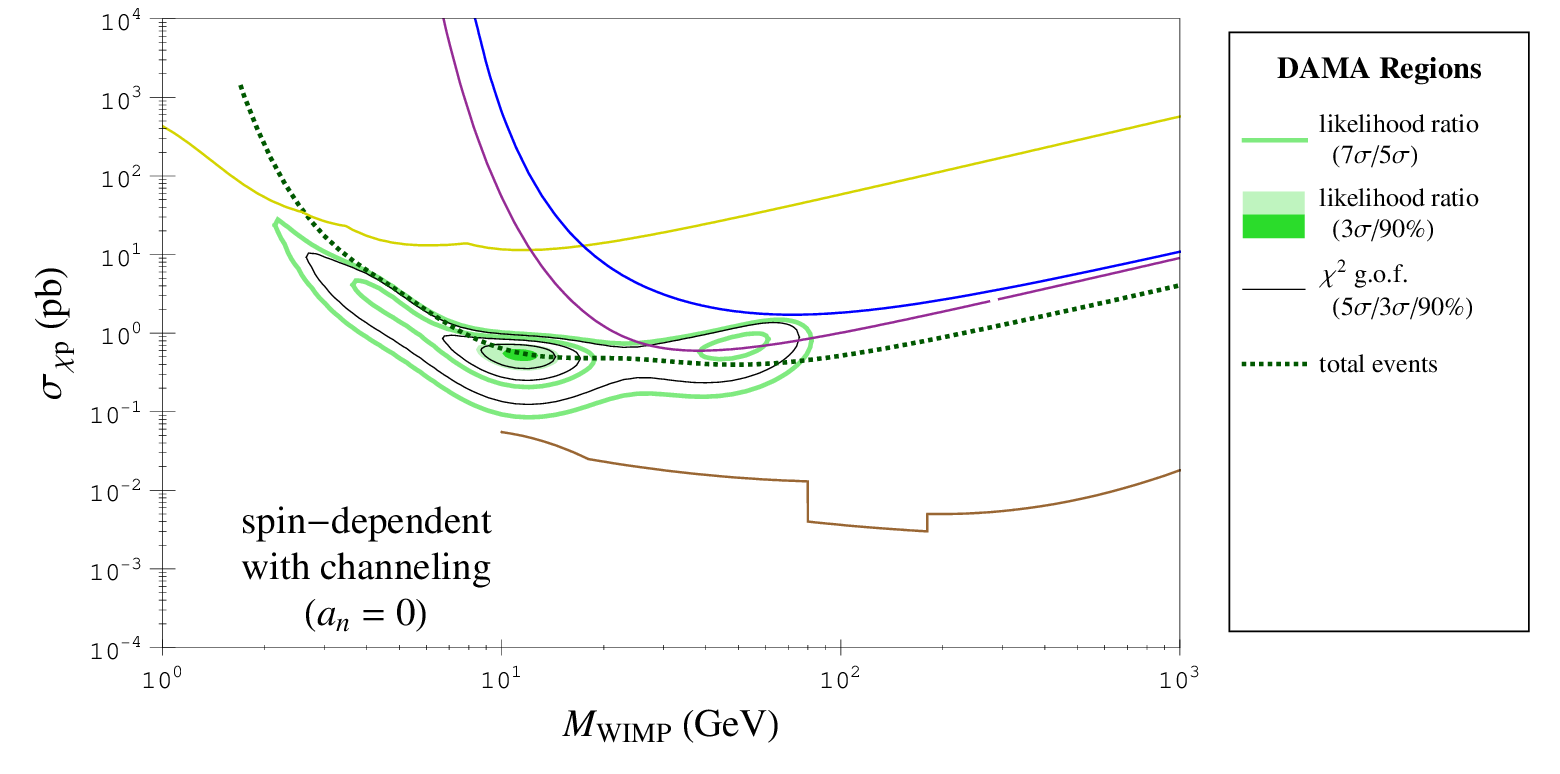}
  \caption{
    Same as \reffig{SDpMethodGOF}, but including the channeling effect
    for DAMA.
    }
  \label{fig:SDpMethodICGOF}
\end{figure}

\afterpage{\clearpage}

For the SD proton-only scattering case ($\anSD = 0$), the DAMA
best fit parameters and null experiment exclusion limits are shown in
\reffig{SDp}.  Unlike the SI case (and the SD neutron-only case to be
examined below), there is no high mass best fit region at better than
a 5$\sigma$ C.L.; the best fits are strictly found around 11~GeV.
The addition of channeling does not reduce the cross-sections of the
best fit region as much as in other cases.  These differences are a
result of the relatively small contribution from scatters off of I
in this case: where I presents the best fit to the DAMA data (the high
mass region), there are enough Na scatters (which do not match the
observed spectrum at those masses) to make the total spectrum a
relatively poor fit, thus eliminating the higher mass region found in
other cases at the 90\% \& 3$\sigma$ C.L.'s.  The relatively small
amount of I scatters means less of the channeled I scatters, which
contribute to the low mass region, so the low mass region does not
shift to much lower cross-sections.

Constraints from experiments with proton-even detector materials (\eg\ 
Si, Ge, and Xe) are significantly weakened as those elements have very
little spin in their proton groups for a WIMP to couple to.
Noteably, CoGeNT and TEXONO (both Ge) constraints are too weak to
appear in the various SD proton-only figures.  CDMS and XENON only
mildly constrain the 5$\sigma$ and 7$\sigma$ C.L.\ DAMA best fit
regions at higher WIMP masses.  CRESST~I, which uses proton-odd Al,
still has reasonably strong constraints, but these fall just outside
the 7$\sigma$ C.L.\ DAMA best fit region.  SuperK, on the other hand,
strongly constrains all of the DAMA best fit regions above their
analysis threshold of 10~GeV.  Only a small portion of the 3$\sigma$
C.L.\ region survives around 9--10~GeV, but would almost certainly
be excluded if SuperK extended their analysis to lower masses.  If
any of the assumptions required for analyzing indirect detection limits
from high energy solar neutrinos does not hold
(see \refsec{Experiments}), so that the SuperK constraint does not
apply, nearly all the the DAMA best fit parameter space survives in
this case.

Not shown in the figures are other proton-odd experiments such as
COUPP \cite{Behnke:2008zz,Collar:2008pc2} and KIMS \cite{Lee.:2007qn}
that can provide additional constraints in this SD proton-only
scattering case.  As noted in \refsec{Experiments}, COUPP does not
present their results in a way that allows us to perform our own analysis
and, hence, has not been included here.  Both COUPP and KIMS would
exclude the higher WIMP mass regions compatible with DAMA.  However,
KIMS has a relatively high threshold and cannot constrain the low WIMP
mass DAMA region.  COUPP could potentially constrain some of this low
mass region, although without a reanalysis it is unclear to what
degree (if it all) COUPP actually constrains the DAMA regions at these
low WIMP masses.

The best fit SD proton-only cross-sections at each mass are indicated
by the raster scans in \reffig{SDpMethodRS} (without channeling)
and \reffig{SDpMethodICRS} (with channeling).  CRESST~I strongly
constrains some of the raster scan band at low masses ($<$3~GeV)
without channeling and only mildly constrains that band at low masses
with channeling.  Again, the best fit cross-sections at those low
masses, as well as very high masses, are a poor fit to the data.

Parameters compatible with the DAMA data are indicated by the $\chi^2$
g.o.f.\ regions shown in \reffig{SDpMethodGOF} (without channeling) and
\reffig{SDpMethodICGOF} (with channeling).  Regions compatible with
each experiment at the 90\% level are found at 8--10~GeV both with and
without channeling \footnote{A cautionary note: saying parameters
    are consistent with each experiment at \eg\ the 90\% level is not
    equivalent to saying the experiments are consistent within the
    90\% level.
    The latter is a statement of a combined analysis, which is not
    performed here.}.
In addition, regions compatible with DAMA within
the 3$\sigma$ level can be found that satisfy the DAMA constraint at
7--10~GeV, or as low as 6--10~GeV when channeling is included.  Without
the SuperK constraint, parameters compatible with DAMA within the 90\% 
level can be found over 8--15~GeV.  Within the 3$\sigma$ level, that
range expands to 7--19 \& 35--55~GeV (no channeling) or
6--18~GeV (with channeling), where some of the compatible regions have
been limited by DAMA's total event rate.  However, the high mass
(35--55~GeV) DAMA compatible region without channeling would be
incompatible with the COUPP and KIMS experimental constraints.

\subsubsection{Neutron-Only}

\begin{figure}
  \insertfig{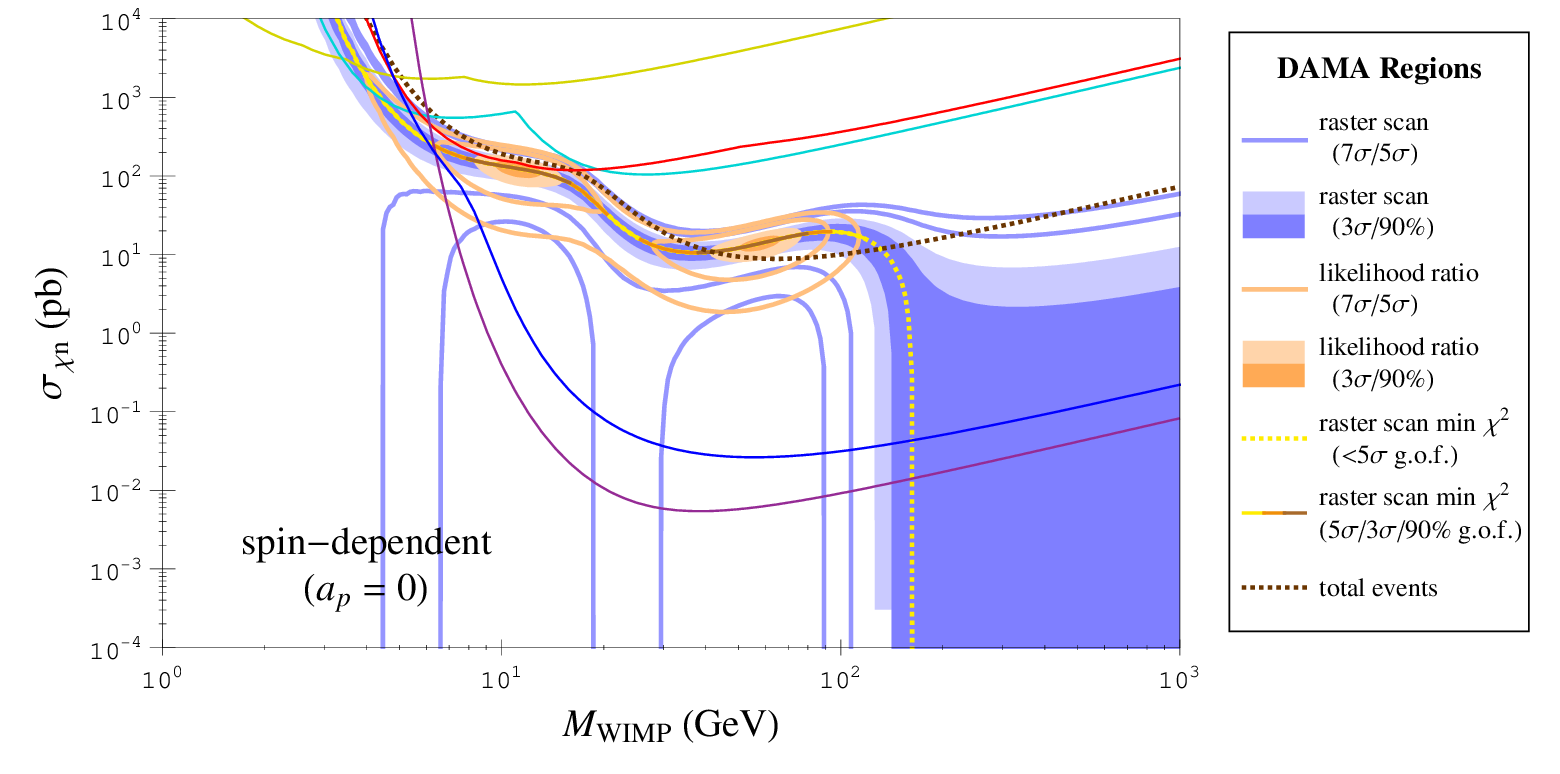}
  \caption{
    Experimental constraints and DAMA best fit parameters for
    spin-dependent (SD) neutron-only scattering.
    The DAMA raster scan and likelihood ratio regions are as described
    in \reffig{SIpMethodRS}.
    Null experiment constraints are as given in \reffig{SDn}.
    }
  \label{fig:SDnMethodRS}
\end{figure}

\begin{figure}
  \insertfig{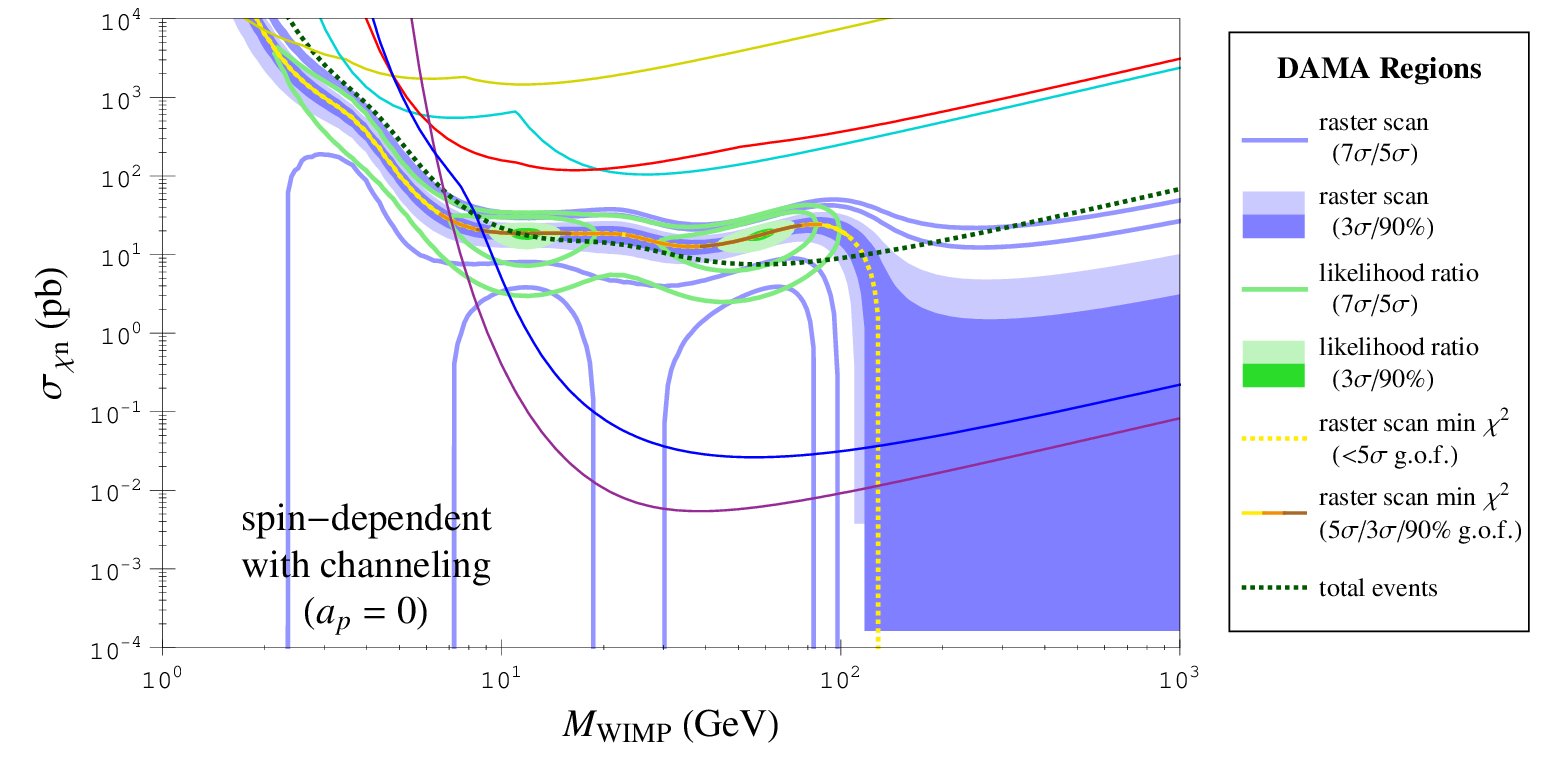}
  \caption{
    Same as \reffig{SDnMethodRS}, but including the channeling effect
    for DAMA.
    }
  \label{fig:SDnMethodICRS}
\end{figure}

In the SD neutron-only coupling case ($\apSD = 0$), there is again two
best fit mass regions, as shown in \reffig{SDn}.  The addition of the
channeling effect shifts the best fit cross-sections in the low mass
region to lower values.
For this neutron-only coupling case, experiments with the neutron-odd
isotopes (\eg\ Si-29, Ge-73, Xe-129, and Xe-131) are expected to gain
in sensitivity relative to DAMA, which has only neutron-even isotopes
(Na-23 and I-127).  The CDMS and XENON10 constraints exclude the
5$\sigma$ C.L.\ DAMA best fit region with and without scattering.
CoGeNT also constrains some of the 3$\sigma$ C.L.\ DAMA best fit
region.  Only DAMA best fit regions below a 5$\sigma$ C.L.\ survive
the other experimental constraints.
SuperK provides no constraints in this case.

Once again, the DAMA best fit SD neutron-only cross-sections at each
mass are indicated by the raster scans regions, shown in
\reffig{SDnMethodRS} (without channeling) and \reffig{SDnMethodICRS}
(with channeling).  The best fit cross-sections within a 3$\sigma$
C.L.\ are excluded by CRESST~I for masses below 3~GeV and by CDMS and
XENON10 for masses above 7~GeV; TEXONO constrains some of
the best fit cross-sections at WIMP masses in between.  For masses
of 3--7~GeV, the best fit cross-sections are consistent with the DAMA
data only within the 3$\sigma$--5$\sigma$ level.

\begin{figure}
  \insertfig{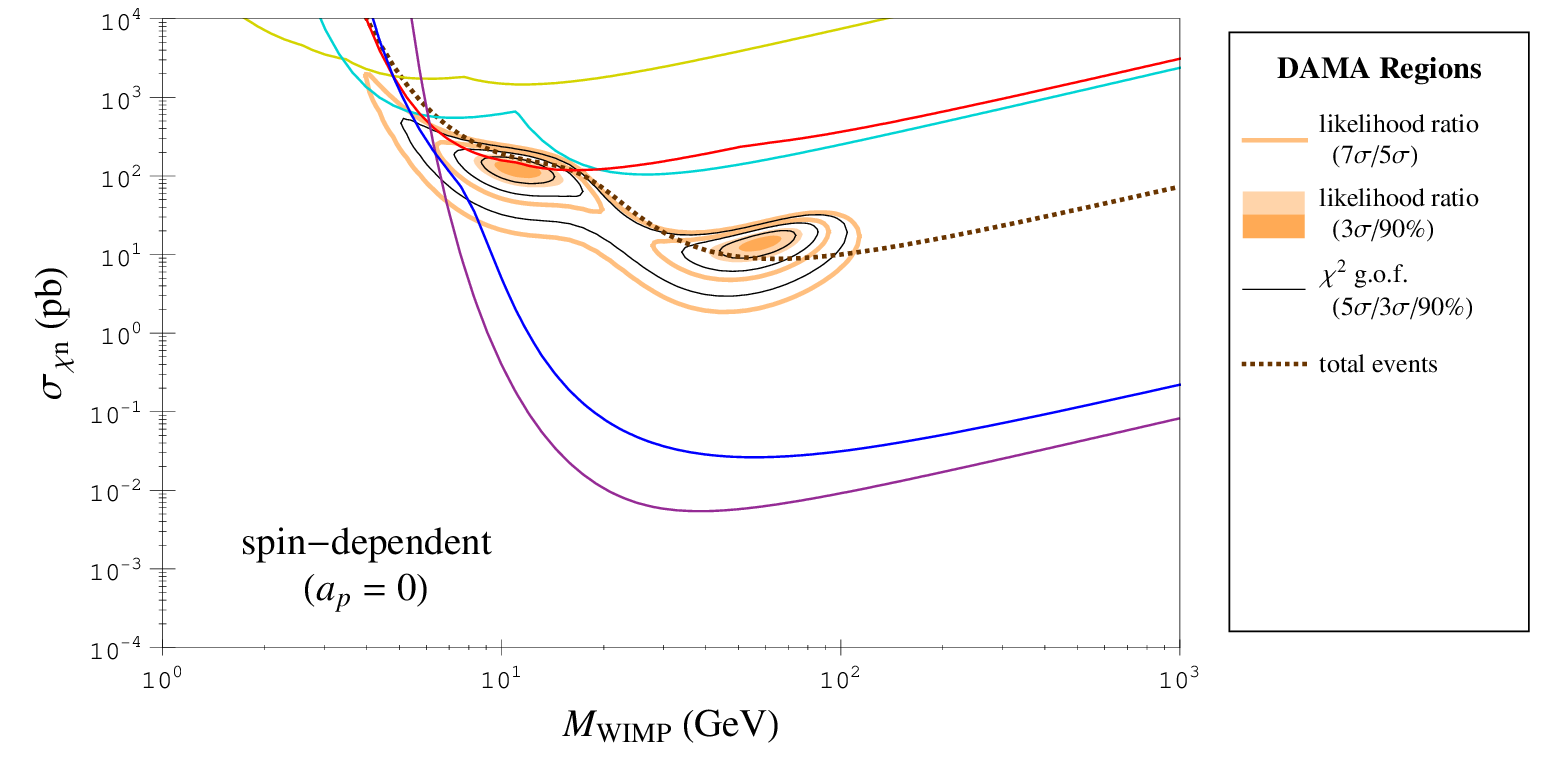}
  \caption{
    Experimental constraints and DAMA best fit parameters for
    spin-dependent (SD) neutron-only scattering.
    The DAMA raster scan and $\chi^2$ g.o.f.\ regions are as described
    in \reffig{SIpMethodGOF}.
    Null experiment constraints are as given in \reffig{SDn}.
    }
  \label{fig:SDnMethodGOF}
\end{figure}

\begin{figure}
  \insertfig{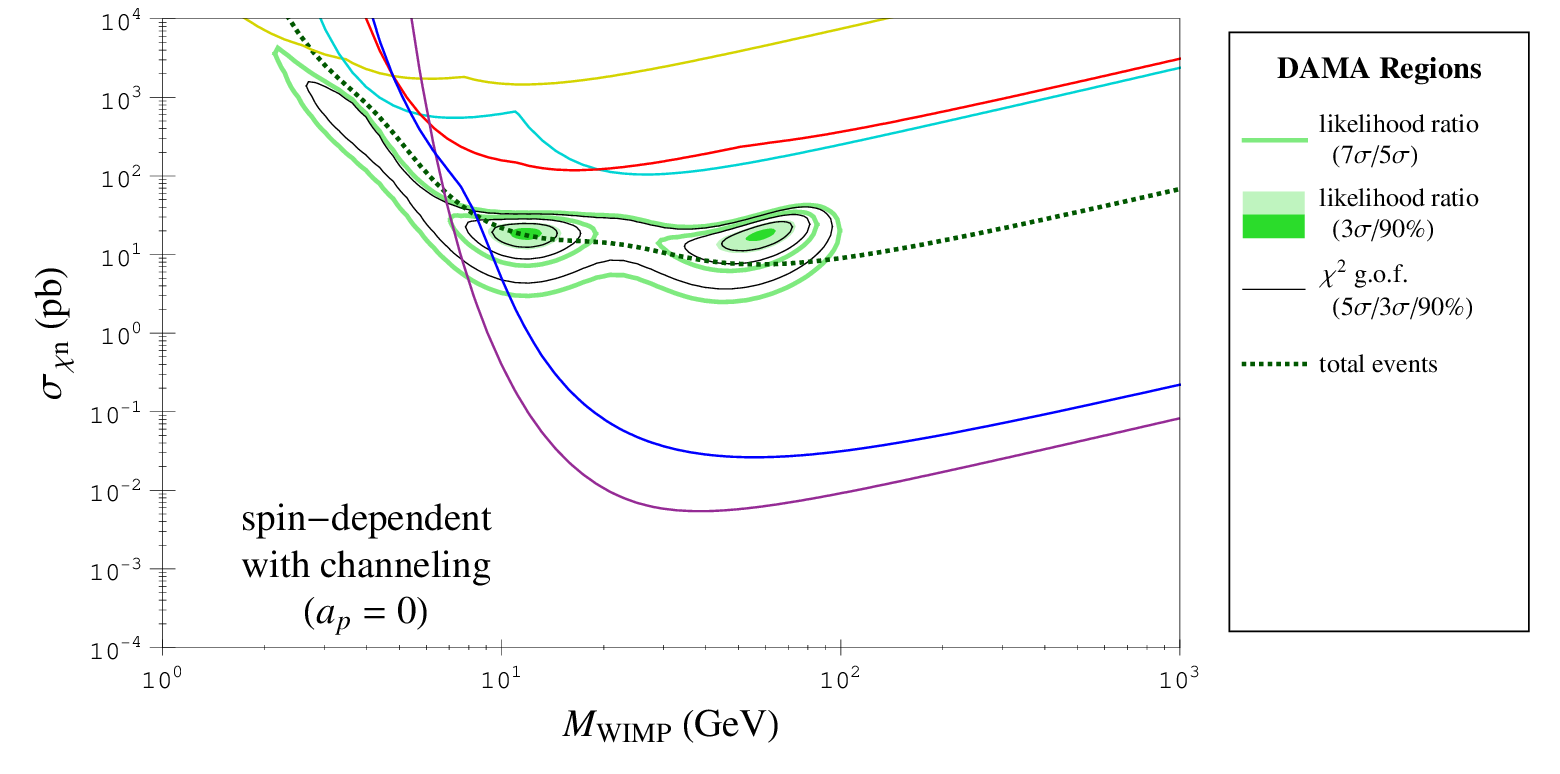}
  \caption{
    Same as \reffig{SDnMethodGOF}, but including the channeling effect
    for DAMA.
    }
  \label{fig:SDnMethodICGOF}
\end{figure}

Parameters compatible with the DAMA data are indicated by the $\chi^2$
g.o.f.\ regions shown in \reffig{SDnMethodGOF} (without channeling) and
\reffig{SDnMethodICGOF} (with channeling).  Parameters compatible
with the DAMA data within the 3$\sigma$ level are excluded by the
CDMS and XENON10 constraints both with and without channeling.

\afterpage{\clearpage}

\subsubsection{Mixed Couplings}

\begin{figure}
  \insertfig{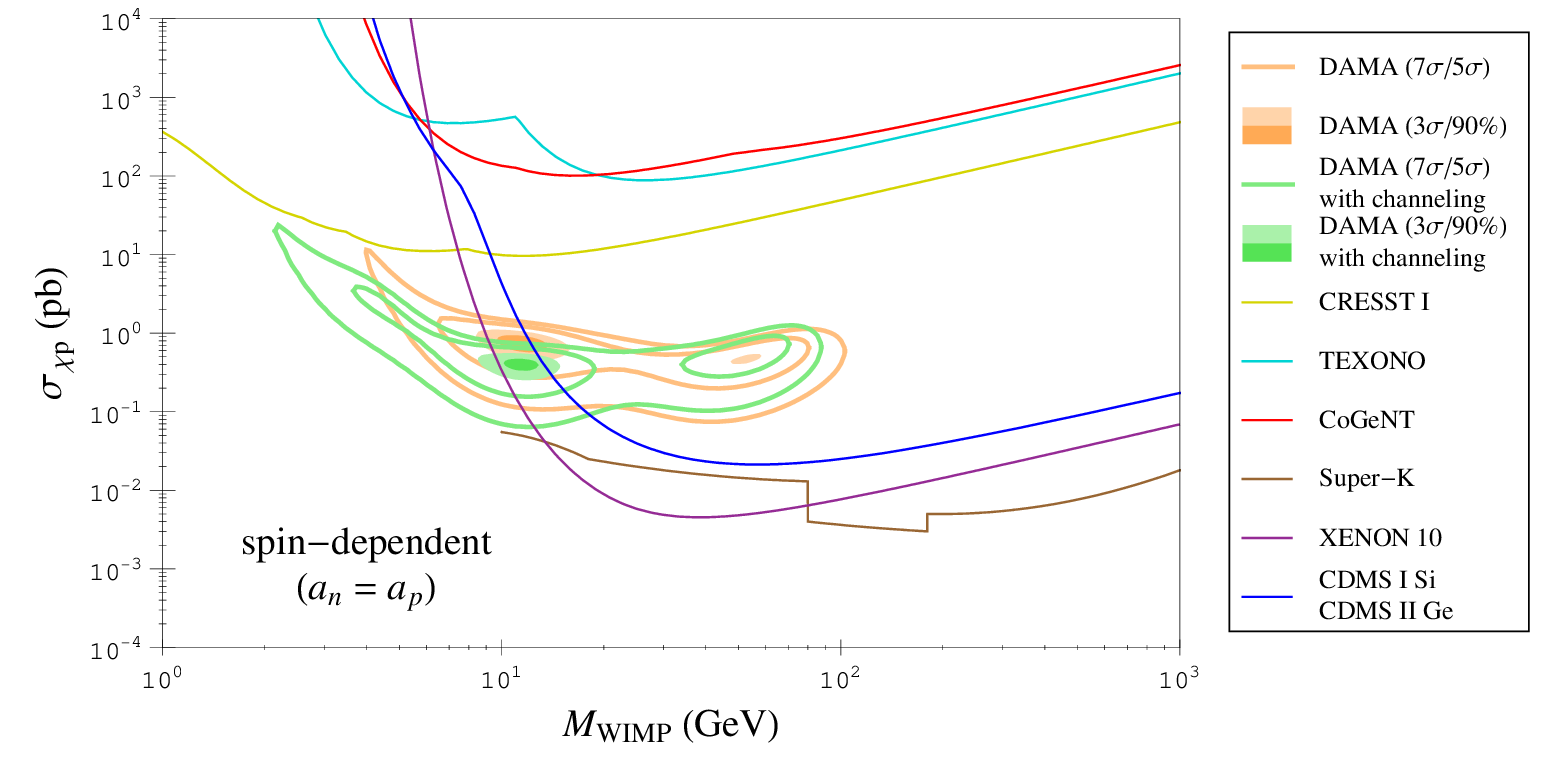}
  \caption{
    Experimental constraints and DAMA best fit parameters for SD
    only scattering in the case $\anSD = \apSD$.
    The DAMA best fit regions are determined using the likelihood
    ratio method with (green) and without (orange) the channeling
    effect.
    }
  \label{fig:SDpS}
\end{figure}

\begin{figure}
  \insertfig{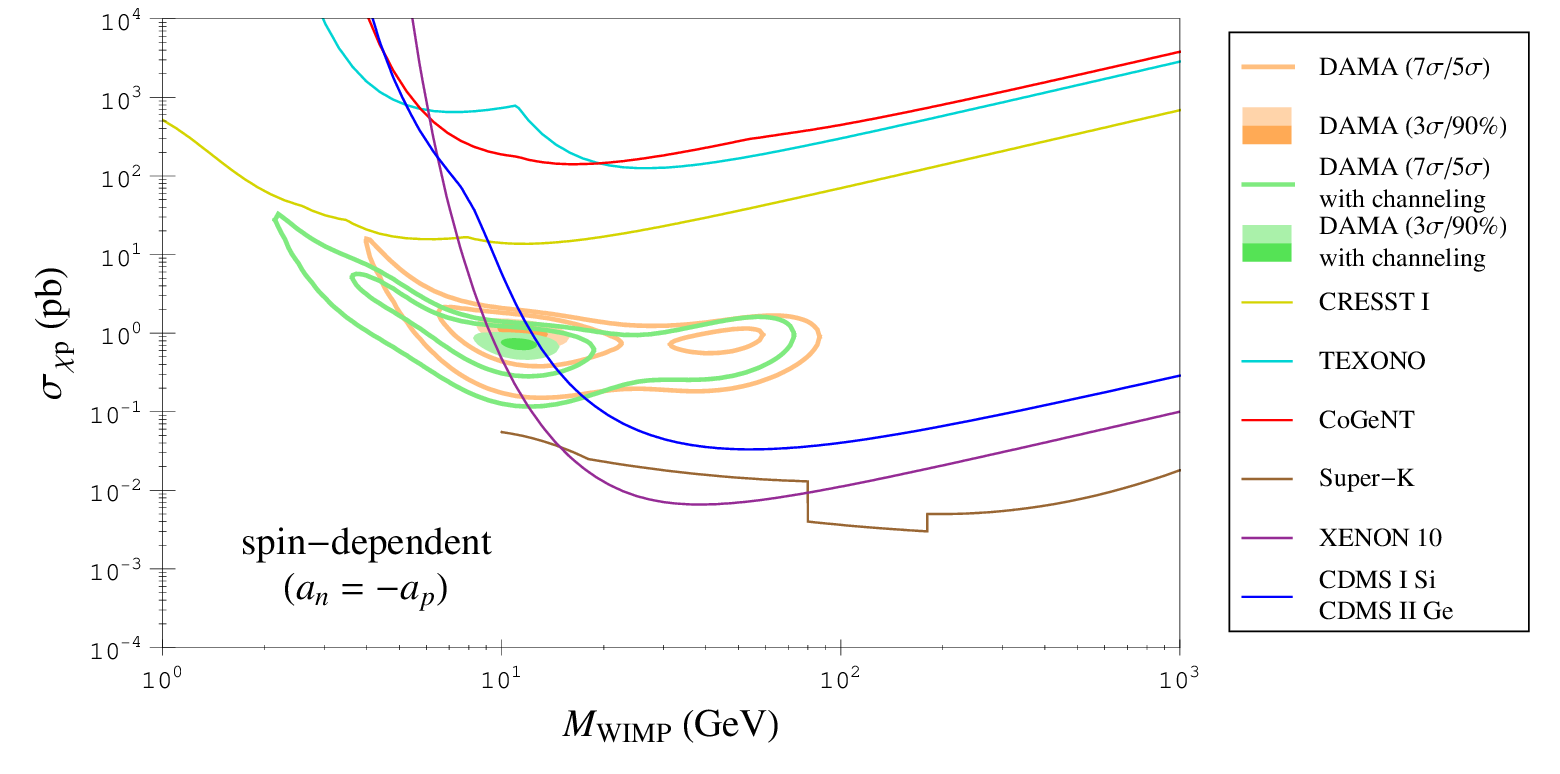}
  \caption{
    Experimental constraints and DAMA best fit parameters for SD
    only scattering in the case $\anSD = -\apSD$.
    The DAMA best fit regions are determined using the likelihood
    ratio method with (green) and without (orange) the channeling
    effect.
    }
  \label{fig:SDpO}
\end{figure}

While the previous two cases took one of the two SD couplings to be
zero, there is little reason to suspect one of the couplings to be
significantly smaller than the other.  Now we will consider the case
of arbitrary SD couplings, where both couplings may be non-zero.
We show in \reffig{SDpS} the experimental constraints and DAMA
best fit parameters (WIMP mass and SD WIMP-proton cross-section) in
the case of equal SD couplings ($\anSD = \apSD$) and in \reffig{SDpO}
for $\anSD = -\apSD$.  In both cases, the SD scattering cross-sections
off of the proton and neutron are equal.  The null experiment
constraints are nearly the same in both cases, but the DAMA best fit
regions have some differences.  In the $\anSD = - \apSD$ case, as
compared to the $\anSD = \apSD$ case, the inclusion of channeling
does not lower the best fit cross-sections in the low mass region by
as much and the the higher mass region falls under a weaker C.L.
As with the previous cases, we find DAMA best fit regions within
a 3$\sigma$ C.L.\ survive other constraints at WIMP masses of
$\sim$8--10~GeV.  Those regions fall within the 90\% $\chi^2$ g.o.f.\ 
DAMA compatibility region (not shown), so there are parameters in these
two cases that are compatible with each of the experiments within
the 90\% level.

\begin{figure}
  \insertfig{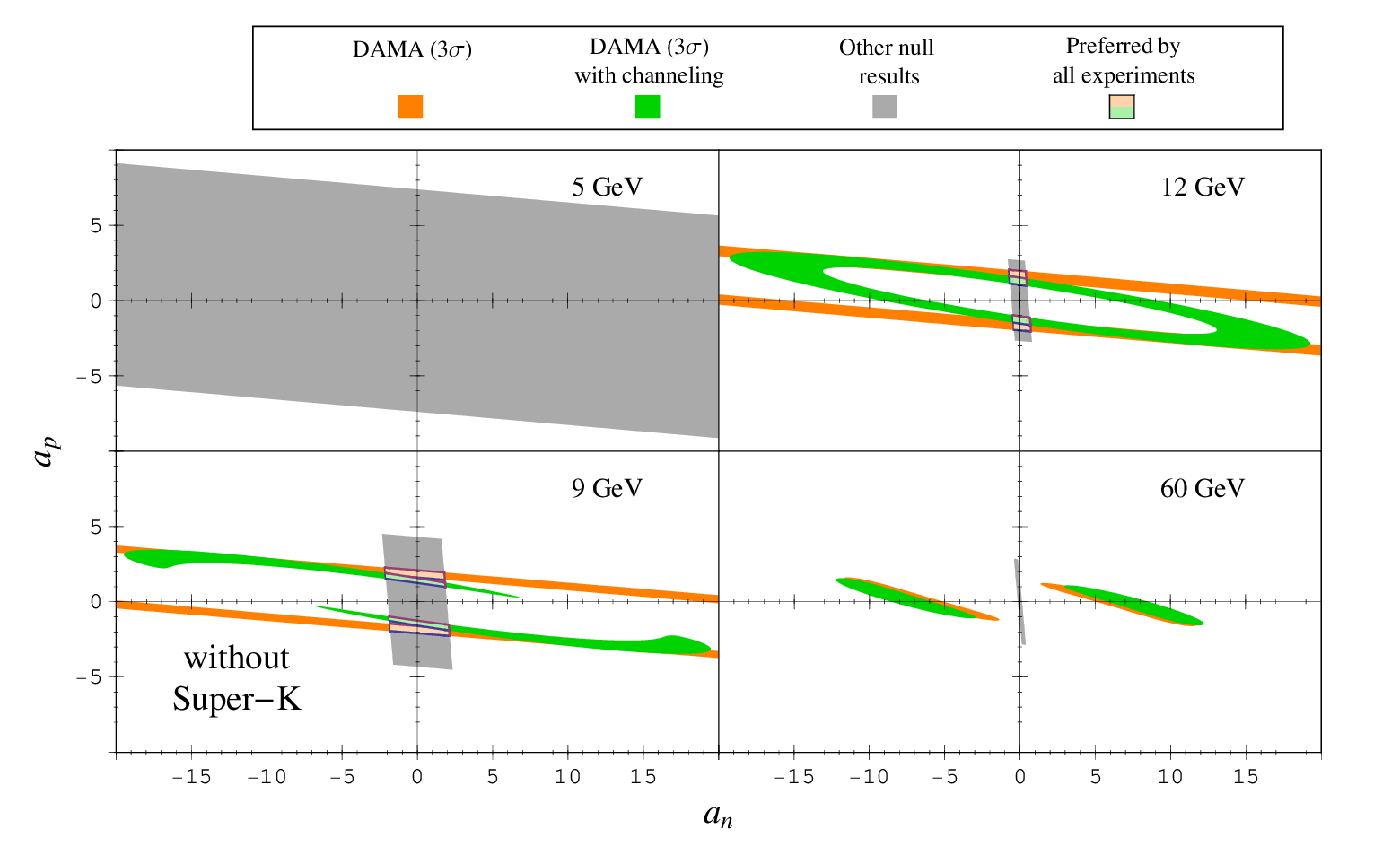}
  \caption{
    Experimental constraints and DAMA best fit parameters for SD only
    scattering in the $\apSD$-$\anSD$ plane at several different WIMP
    masses.
    The 3$\sigma$ best fit DAMA parameters are shown in orange
    (no channeling) and green (with channeling).  The regions allowed
    at 90\% by a combination of the null experiments, excluding
    Super-Kamiokande, is indicated by the gray regions.  Parameters
    in the overlapping regions (gray+orange or gray+green) are
    consistent with both the DAMA best fit regions and all null
    experiments.
    Super-Kamiokande constraints, if valid, would exclude all of the
    overlapping regions shown at 12~GeV (and all such regions with
    WIMP masses above 10~GeV).
    }
  \label{fig:SDaLR}
\end{figure}

In \reffig{SDaLR}, we show the DAMA best fit couplings in the
$\apSD$-$\anSD$ plane, along with null experiment constraints, at
several different WIMP masses; SuperK constraints have not been
included.  The gray regions indicate the regions allowed by the
various null experiments, while the green and orange regions indicate
the parameters most preferred (at a 3$\sigma$ C.L.) for producing the
DAMA signal with and without channeling, respectively (there is no
DAMA best fit region at a WIMP mass of 5~GeV).
The parameters in the overlapping regions (shown in lighter shades of
green and orange) are DAMA best fit parameters that satisfy the other
constraints.  Such regions can be found at masses of $\sim$8--15~GeV.
Above 10~GeV, these regions have $|\apSD| \gg |\anSD|$.  At masses of
8--10~GeV, $|\apSD| \gae |\anSD|$.  SuperK constraints exclude all the
overlapping regions above 10~GeV.

\begin{figure}
  \insertfig{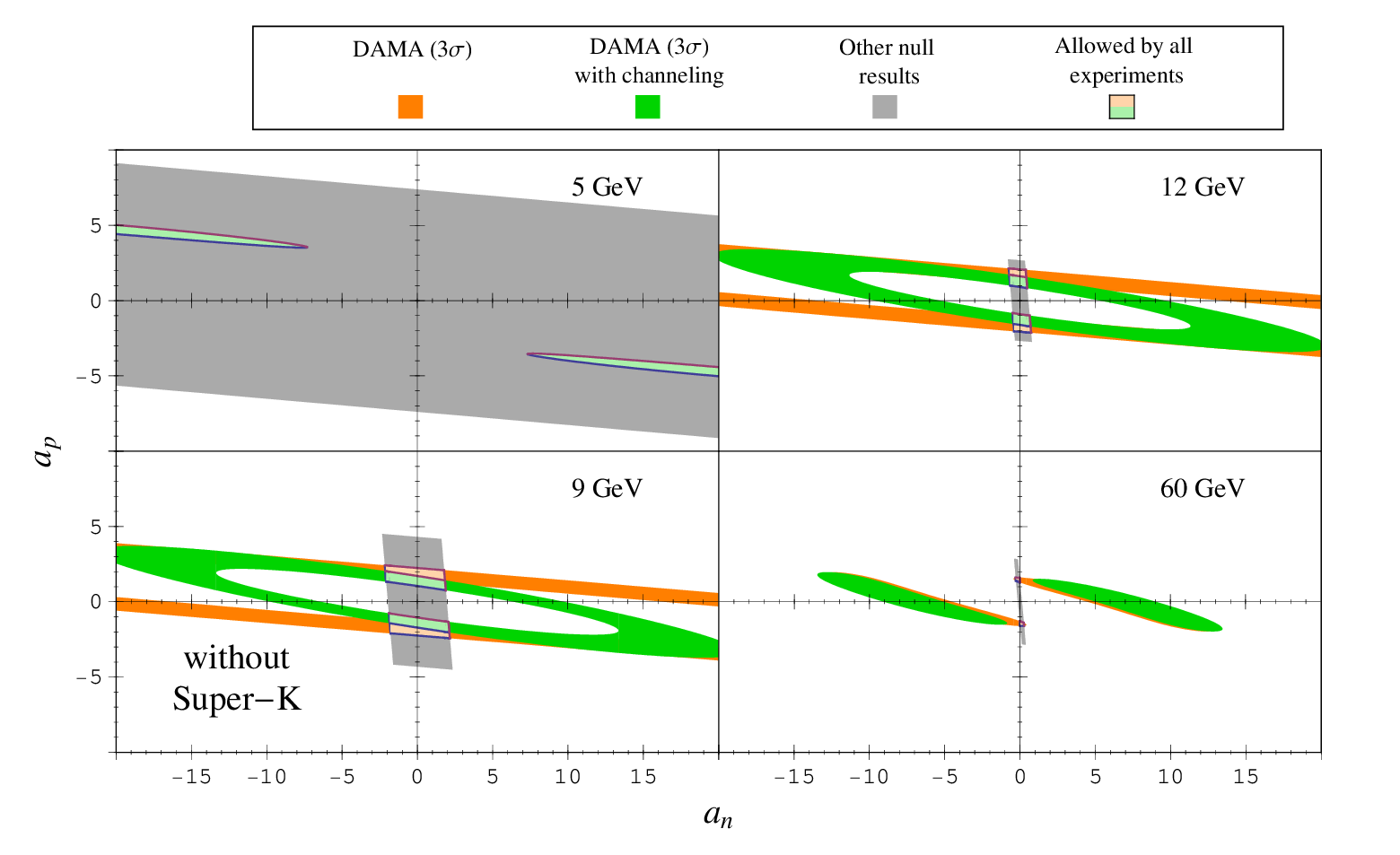}
  \caption{
    Null experiment and DAMA constraints for SD only
    scattering in the $\apSD$-$\anSD$ plane at several different WIMP
    masses.
    The 3$\sigma$ allowed DAMA parameters are shown in orange
    (no channeling) and green (with channeling).  The regions allowed
    at 90\% by a combination of the null experiments, excluding
    Super-Kamiokande, is indicated by the gray regions.  Parameters
    in the overlapping regions (gray+orange or gray+green) are
    consistent with all experiments.
    There is no DAMA region at 5~GeV without channeling that is
    compatible within the 3$\sigma$ level.
    Super-Kamiokande constraints, if valid, would exclude all of the
    overlapping regions shown at 12~GeV and 60~GeV (and all such
    regions with WIMP masses above 10~GeV).
    }
  \label{fig:SDaGOF}
\end{figure}

\Reffig{SDaGOF} similarly shows constraints in the $\apSD$-$\anSD$
plane, but shows the couplings compatible with the DAMA data within
the 3$\sigma$ level using a $\chi^2$ g.o.f.\ test.  Again, SuperK
constraints are not included.  Overlapping regions in this case
are the parameters that are compatible with all experiments (within
3$\sigma$ for DAMA and 90\% for the others).  Such regions are found
over $\sim$7--19 \& 35--55~GeV without channeling and $\sim$6--18~GeV
with channeling.  Inclusion of the SuperK constraint excludes
all the overlapping regions above 10~GeV.

\subsection{Mixed SI \& SD Scattering}

The SI only scattering case provides a slightly better fit to the DAMA
modulation data than SD only scattering, as can be seen in
\reftab{DAMAMin}, although the improvement is mild enough that each of
the SI and SD spectra provide reasonable fits to DAMA.  In the case of
no channeling, SI only scattering actually provides the best global
fit, but mixed SI and SD couplings can also provide reasonable fits
(only slightly worse than the SI only case).
With channeling, the best fit has a mix of SI and SD contributions.

Due to the linearity of the modulation amplitude and recoil events
in each of the SI and SD cross-sections, $\sigma = \sigmaSI + \sigmaSD$,
one may find approximate mixed SI and SD couplings by interpolating
between the SI and SD only regions given in the various figures.  For
example, at a WIMP mass of 10~GeV, an SI-only WIMP-proton cross-section
of $1.5 \times 10^{-5}$~pb and an SD proton-only WIMP-proton
cross-section of 1~pb each fall within the 90\% g.o.f.\ constraints of
their respective cases.  Then the case where $\sim$60\% of the signal
comes from SI scattering, with $\sigmapSI = 0.9 \times 10^{-5}$~pb
(60\% of the cross-section that gives the full signal in SI), and
the other $\sim$40\% comes from SD scattering, with
$\sigmapSD = 0.4$~pb (40\% of the cross-section that gives the full
signal in SD), should fall within (or very near) a 90\% g.o.f.\ 
constraint in the mixed SI \& SD coupling case.  A variety of mixed SI
\& SD couplings for WIMP masses around $\sim$8~GeV can be found that
reasonably fit the DAMA data as well as the various null results of
other experiments.

\section{\label{sec:Conclusion} Conclusions}

The latest DAMA/LIBRA annual modulation data, which may be interpreted
as a signal for the existence of weakly interacting dark matter
(WIMPs) in our Galactic Halo, are examined in light of null results
from other experiments.  We investigate consequences for both
spin-independent (SI) and spin-dependent (SD) WIMP couplings as well
as mixed SI \& SD couplings.  The positive
signal for annual modulation is compared to negative results from
CDMS, CoGeNT, CRESST~I, TEXONO, XENON10, and Super-Kamiokande (SuperK).
In addition to the new experimental results, two new ingredients are
included compared to our previous studies: 1) the energy spectrum of
DAMA/LIBRA data is now given in 36 bins, and 2) the effect of
channeling is taken into account, which lowers the nuclear recoil
energy threshold and increases sensitivity to low WIMP masses.
Several statistical tools are implemented in our study: best fit
regions are found; goodness-of-fit (g.o.f.); and raster scans over the
WIMP mass.
These approaches allow us to differentiate between the most preferred
(best fit) and allowed (g.o.f.) parameters.
It is hard to find WIMP masses and couplings consistent with all
existing data sets; the surviving regions of parameter space are found
here.  For SI interactions, the best fit DAMA regions of
any WIMP mass are ruled out to 3$\sigma$, even with channeling taken
into account.  However, some parameters outside the best fit
regions still yield a moderately reasonable fit to the DAMA data: we find
parameters in the SI case at masses of $\sim$8--9~GeV that
are compatible with each of the experiments at better than the
3$\sigma$ level when channeling is included.
For SD interactions, we find no neutron-only couplings that are
compatible with the DAMA data to within the 3$\sigma$ level and satisfy
other experimental constraints, with or without channeling included.
For SD proton-only couplings, a range of masses below 10~GeV is
compatible with DAMA and other experiments, with and without
channeling, when SuperK indirect detection constraints are included;
without the SuperK constraints,
masses as high as $\sim$20~GeV are compatible.
Mixed couplings with $\anSD = \pm \apSD$ are found to be consistent
with all experiments at $\sim$8~GeV.
The general case of mixed SD proton and neutron couplings are examined
as well.
In short, there are surviving regions at low mass for both SI and SD
interactions (although only moderately so for the SI case) and at high
masses for the case of SD proton-only couplings if indirect detection
limits are relaxed.

While in the process of preparing this paper, several papers
were presented that also examine the compatibility of the DAMA
signal with other negative experimental results, always assuming that
the dark matter consists of WIMPs \cite{Bottino:2008mf,
Petriello:2008jj,Chang:2008xa,Chang:2008gd,Fairbairn:2008gz,
Hooper:2008cf} (although other non-WIMP candidates were studied as
well, such as mirror~\cite{Foot:2008nw}, composite~\cite{Khlopov}, and
WIMPless~\cite{Feng:2008dz} dark matter). Our results
are compatible with the results of these other studies of WIMPs,
although our conclusions sometimes differ. In all of them channeling
is taken into account.  Ref.~\cite{Bottino:2008mf} concludes that the
recent CDMS results are compatible with the DAMA regions as derived by
the DAMA collaboration, for WIMPs with spin independent interactions
with mass in the 7 to 10 GeV range, a results similar to ours.
Ref.~ \cite{Petriello:2008jj} used only the raster scan and the DAMA
data binned into two bins to study WIMPs with spin independent
interactions (while here we use 36 data bins with this method) and find
light WIMPs compatible with all experimental results, when channelling
is included.  Refs.~\cite{Chang:2008xa} and~\cite{Fairbairn:2008gz}
both studied WIMPs with spin independent interactions using global
fits with two parameters and the data binned into many bins, thus
including the spectral modulation amplitude information.  Their results
are similar to those of our global fits, but they conclude that light
WIMPs with spin independent interactions ``cannot account for the
data''~\cite{Chang:2008xa} or ``are strongly
disfavoured''~\cite{Fairbairn:2008gz} within the standard halo model
(they considered also the limits imposed by the DAMA total spectrum,
not included here, which, however, affect heavier WIMPs).
Ref.~\cite{Hooper:2008cf} studies indirect detection bounds which may
be very important for light WIMPs, but depend on the WIMP-annihilation
channels.


\begin{acknowledgments}
  K.F.\ acknowledges the support of the DOE and the Michigan
  Center for Theoretical Physics via the University of Michigan.
  G.G.\ was supported in part by the US DOE grant DE-FG03-91ER40662 
  Task C.
  P.G.\ was partially supported by NSF grant PHY-0756962 at the
  University of Utah.
  C.S.\ acknowledges the support of the William I.\ Fine Theoretical
  Physics Institute at the University of Minnesota and thanks V.\ Mandic
  for useful conversations.
\end{acknowledgments}




\end{document}